\documentclass[conference]{IEEEtran}
\IEEEoverridecommandlockouts
\AtBeginDocument{%
  }
\usepackage{amsmath,amsfonts}
\usepackage{algorithmic}
\usepackage{graphicx}
\usepackage{textcomp}
\usepackage{xcolor}
\usepackage{algorithmic}
\usepackage{graphicx}
\usepackage{textcomp}
\usepackage[table]{xcolor}
\usepackage{wrapfig}
\usepackage{comment}
\usepackage{booktabs}
\usepackage{hyperref}
\hypersetup{
    colorlinks=true,
    citecolor=blue,
    linkcolor=blue,
    urlcolor=blue,
}
\usepackage{amsmath}
\usepackage{cleveref}
\usepackage{multirow}
\usepackage{listings}
\usepackage{subcaption}
\usepackage{enumitem}
\usepackage{circuitikz}
\usepackage{float}
\usepackage[sort]{cite}
\usepackage{tabularx}
\newcolumntype{Y}{>{\centering\arraybackslash}X}

\usepackage{tikz}
\usetikzlibrary{arrows.meta}

\newcommand{\pname}{\mbox{{\scshape BoostedSOSA}}}
\newcommand{\stannic}{\mbox{{\scshape Stannic}}}

\newcommand{\schedfpga}{Scheduling\ FPGA}

\newcommand{\bem}[1]{{\bf\em #1}}

\newcommand{\doubted}[1]{\textcolor{magenta}{#1}}

\newcommand{\eg}{\mbox{{\em e.g.}}}
\newcommand{\ie}{\mbox{{\em i.e.}}}
\newcommand{\cf}{\mbox{{c.f.}}}

\newtheorem{definition}{\bf Definition}

\definecolor{myNewColorOne}{HTML}{9933FF}

\DeclareRobustCommand{\stannic}{\mbox{{\scshape Stannic}}}

\definecolor{codegreen}{rgb}{0,0.6,0}
\definecolor{codegray}{rgb}{0.5,0.5,0.5}
\definecolor{codepurple}{rgb}{0.58,0,0.82}
\definecolor{backcolour}{rgb}{0.95,0.95,0.92}

\lstdefinestyle{mystyle}{
    backgroundcolor=\color{backcolour},   
    commentstyle=\color{codegreen},
    keywordstyle=\color{blue},
    numberstyle=\tiny\color{codegray},
    stringstyle=\color{codepurple},
    basicstyle=\ttfamily\footnotesize,
    breakatwhitespace=false,         
    breaklines=true,                 
    captionpos=b,                    
    keepspaces=true,                 
    numbers=left,                    
    numbersep=5pt,                  
    showspaces=false,                
    showstringspaces=false,
    showtabs=false,                  
    tabsize=2
}

\usepackage{orcidlink}

\begin{document}

% \title{Conference Paper Title*\\

\title{
$\pname$: Accelerated Inferencing for Low Variance Stochastic Online Scheduling %Acceleration
}
% \and
% \and
% \and
% \and
% \and

\author{\IEEEauthorblockN{
Adam H. Ross\IEEEauthorrefmark{1}\orcidlink{0009-0004-0120-7852}, 
Riccardo Revalor\IEEEauthorrefmark{2}\orcidlink{0009-0004-0120-7852},~\IEEEmembership{Graduate Student Member,~IEEE},
Aryan Singh\IEEEauthorrefmark{4}\orcidlink{0009-0003-1116-0325}, \\
Ayush Jain\IEEEauthorrefmark{5}\orcidlink{0009-0009-1475-3250}, and
Debjit Pal\IEEEauthorrefmark{3}\orcidlink{0000-0003-3722-5126},~\IEEEmembership{Member,~IEEE}}
\IEEEauthorblockA{
Email: \{aross50\IEEEauthorrefmark{1},
rreva\IEEEauthorrefmark{2},
dpal2\IEEEauthorrefmark{3}\}@uic.edu
, \{asing271\IEEEauthorrefmark{4}, ayushj\IEEEauthorrefmark{5}\}@illinois.edu}
\thanks{Adam H. Ross, Riccardo Revalor, and Debjit Pal are with the Department of Electrical and Computer Engineering, 
  University of Illinois Chicago, USA.}\thanks{Aryan Singh and Ayush Jain are with the Department of Electrical and Computer Engineering, 
  University of Illinois Urbana-Champaign, USA.}
}

\maketitle

%%%% DP: To be read.
\begin{abstract}
Heterogeneous scheduling in stochastic, online environments, such as high-performance computing (HPC) systems, presents a significant challenge. Stochastic Online Scheduling Accelerators (SOSAs) offer a promising solution, but their effectiveness is compromised by a reliance on runtime estimates provided by users. These estimates introduce substantial variance into the scheduling process (mean MAE in hundreds of %of $1{,}295$ 
Core-Days), %with a $\pm 1{,}368$ spread in our evaluation
%with large variance
%), 
thereby %explicitly 
%harming the %scheduling performance bounding guarantees of the Stochastic Online Scheduling algorithm.
%competitiveness of Stochastic Online Scheduling algorithms, whose competitive-ratio bound, $(3+\sqrt{5})(2+\Delta)$, worsens as runtime variability $\Delta$ increases.
weakening the competitiveness of Stochastic Online Scheduling algorithms as their competitive-ratio bound %, $(3+\sqrt{5})(2+\Delta)$, 
increases with runtime variability. %$\Delta$.

To address this limitation, we introduce $\pname$, a dual-FPGA ML-assisted Scheduling architecture that integrates a %processing time %inference model 
Machine Learning predictor for expected processing times, with a novel temporal-aware training policy. %This model predicts expected job runtimes using standard SLURM-based job management %arguments 
%parameters as input features, making it compatible with existing HPC systems. 
The predictor estimates job runtimes using only scheduler parameters available at submission time, enabling its use in existing HPC systems.
Using historical real-world HPC job data (from the Argonne Leadership Computing Facility, MIT Supercloud and UIUC Blue Waters workload datasets), %we demonstrate that $\pname$ reduces both the error and variance of runtime estimates relative to conventional human estimates, remains robust under temporal workload shifts through additive model updates, and achieves an average $13\times$ throughput increase over an AVX-optimized software scheduler.
we show that the predictor reduces MAE by up to $63.85\%$ compared to user runtime estimates, %runtime estimates provided by users,
and the additive training policy reduces MAE by up to $71.88\%$ compared to a static model. % under severe temporal shifts. 
End-to-end, $\pname$ achieves an average $17\times$ speedup over an AVX-optimized software baseline and processes up to %approximately 
$1{,}711$ jobs/seconds.

\end{abstract}


%moved for clean copy paste:
%achieves significant reductions in runtime estimate variance, marking a key improvement in automated heterogeneous scheduling, while also demonstrating an average of $13\times$ increase in throughput over an AVX optimized software scheduler.
\section{Introduction
%\assign{Adam}
}\label{sec:intro}

Efficient workload scheduling in %stochastic 
high-performance computing (HPC) systems is challenging, %a complex challenge, 
because it requires navigating a multi-dimensional decision space, characterized by unpredictable job arrivals, variable execution times, and resource contention among others 
\cite{mualem2001utilization}. To address this challenge, %complexity 
%without incurring major computational overhead, %hardware accelerators for 
%and to minimize computational overhead,
Stochastic Online Scheduling Accelerators (SOSAs) have emerged as an effective solution to minimize computational overhead \cite{kohutka2023new, palaniappan2025herculeshardwareacceleratorstochastic, stannic_2026}. %\doubted{since it's double blind review, we should mask hercules ref}. 
Unlike conventional computationally expensive and sequential software-based schedulers~\cite{slurm}, %which are often sequential and computationally expensive, 
SOSAs {\em exploit parallelism and spatial locality}, %architectures, 
to maintain and update job %task %virtual 
schedules continuously. By offloading multi-dimensional scheduling from the host CPU to dedicated hardware accelerator, SOSAs enable near-real-time load balancing and job %task 
scheduling %routing 
while significantly reducing queue latency and maintaining high throughput~\cite{fusco2022hardware}.

However, the effectiveness of hardware accelerators is fundamentally dependent on the availability of accurate expected processing times (EPT) 
for incoming jobs \cite{saldanha2020probabilistic}. Within stochastic scheduling algorithms, EPT serves as the primary mathematical metric for calculating assignment costs and determining execution priority \cite{pinedo2016scheduling}. In practice, accurate EPTs are hard to obtain. Conventional schedulers %usually 
rely on user-provided estimates or past averages. However, %, but 
such sources introduce significant variance between EPT and real processing times, \cite{liang2024resourceallocation}, leading to %. Additionally, This dependency creates a 
sub-optimal scheduling. Furthermore, such variance is heavily reliant on temporal %factors
conditions, such as high utilization periods or shared resource contention \cite{li2019effect}. Additionally, if the EPT estimation is performed purely \emph{in software}, the latency of the prediction phase can negate the speedup gained by the \emph{hardware} accelerator \cite{fusco2022hardware,patel2025racer}. 
Existing hardware-based approaches often ignore estimation overhead by assuming that EPTs are already available at the hardware interface \cite{fusco2022hardware, kohutka2023new}.  
In %dynamic 
real-world HPC environments, however, jobs %tasks 
may arrive only microseconds apart, leaving little time for %complex 
software-based profiling~\cite{reiss2012heterogeneity} resulting in throttled throughput. 

To address these %estimation 
challenges, Machine Learning (ML) models have been deployed %researchers have started to use Machine Learning (ML) models 
to capture patterns in HPC telemetry distributions. Among these, gradient-boosted decision trees algorithms, \eg, XGBoost, have demonstrated superior accuracy~\cite{ChenGuestrin2016}. XGBoost is particularly effective due to its %for this domain, because it is 
strong performance on tabular data~\cite{ChenGuestrin2016} and %can handle the 
unusual runtime patterns common in cluster logs, \eg, %sudden 
short interactive runs. Furthermore, the multi-tree structure lends itself well to parallelization for fast inferencing, such as in the Field-Programmable Gate Array (FPGA)-based Forest Processing Unit (FPU)~\cite{conifer}.  %\doubted{Its tree-based design also makes parallel inference easier, since the same input can be processed by many trees at once}. 

% However, deploying this model on a hardware Forest Processing Unit (FPU)~\cite{conifer} has challenges. Software uses flexible floating-point arithmetic, while hardware tree engines require uniform fixed-point formats. As result, \doubted{raw HPC features can overflow and distort the model, and the FPU's high-precision outputs do not naturally match the quantized inputs expected by systolic scheduling cores}.
%Bridging these gaps requires a \doubted{design approach} that \bem{reconciles the variability of HPC telemetry to the strict numerical and timing constraints of a %real-time realistic hardware scheduling pipeline}. 

To \bem{rapidly bridge the EPT variability correction and accelerated EPT inferencing}, %scheduling}, 
we propose
%To address these challenges, we propose our vision 
$\pname$, a dual-FPGA ML-assisted Scheduling pipeline %and our vision 
that shifts the heterogeneous scheduling paradigm from relying on %static 
user estimates to %\emph{holistic predictive intelligence}. 
using hardware-accelerated EPT predictions. For each incoming job, we %predict its EPT, %using only submission-time features, 
%and 
accelerate its EPT prediction on an %dedicated 
{\em Inference FPGA}. Then, we schedule the job with predicted EPT using a {\em Scheduling FPGA}. This separation keeps EPT prediction and scheduling as independent stages, and allows each component to be {\em swapped} %modified 
without redesigning the other. As workload behavior changes over time, we %update 
adapt the prediction model to temporal shifts %with newly completed jobs 
using an %our 
additive training policy. % to adapt it to temporal shifts.

Achieving this vision requires %exploration and 
efficient and 
effective solutions to the following %three 
research questions (RQ): \textbf{RQ1:} How can a software-based ensemble model be adapted to a %rigid 
fixed-point hardware architecture to ensure logical integrity and prevent overflow when processing high-variance, stochastic telemetry? \textbf{RQ2:} Can job runtime on an HPC cluster be accurately predicted using only scheduling parameters %features 
available at %the moment of 
submission ($t=0$), without relying on post-submission system state information? \textbf{RQ3:} How can we make such a system adaptable to temporal changes to runtime activity?

Our technical contributions are as follows. 

\begin{enumerate}[ leftmargin=*, parsep=0cm, itemindent=1.2em, itemsep=0.2em, topsep=0.1em]
 \item {\bf Variable-Specific Normalization and Hardware Translation}: To address RQ1, we scale each training feature independently so its $99.9^{th}$ percentile fits within the FPU's $32$-bit integer range, maximizing precision while avoiding overflow.

 \item {\bf Submission-Time EPT Prediction}: To address RQ2, we remove post-submission %and outcome-derived 
 telemetry from historical traces and retain only scheduler-request fields available when a job arrives. In this setting, we show that employing the XGBoost algorithm improves Mean Absolute Error (MAE) by $\mathbf{63.85\%}$ compared to user-provided estimates.

 \item {\bf FPU-Scheduler Integration via Additive Training}: To address RQ3, we design a distributed architecture composed of two AMD Alveo U55C FPGAs. %with one acting 
 One FPGA acts as an inference engine by leveraging %implemented using 
 a Conifer %Feature Processing Unit (FPU) 
FPU. %\cite{conifer}. 
The other FPGA acts as a scheduler, hosting an augmented %the 
version of the $\stannic$ SOSA \cite{stannic_2026}. To our knowledge, $\pname$ is the {\em first} system to couple a hardware-accelerated gradient-boosted inference engine %directly 
with a SOSA to eliminate  %eliminating 
the %software-side 
prediction latency %inherent to 
of prior predictive scheduling approaches \cite{BrownGibbBelikovNash2022,fusco2022hardware}. For inference, we develop %create 
an additive training policy to track temporal workload shifts without retraining the model from scratch on streaming task data of evolving characteristics, and show that it reduces MAE by $\mathbf{71.88\%}$ compared to a static baseline under severe temporal shifts.

\item {\bf End-to-End Evaluation}: We evaluate $\pname$ on $3.97$M historical jobs from the Argonne Leadership Computing Facility (ALCF)~\cite{LiuEtAl2020,ArgonneALCF}. We further test $\pname$ on MIT Supercloud \cite{MITSamsi} and UIUC Blue Waters \cite{UIUC_data} %workload 
datasets to test generalization. %to test its generalizability across independent environments with different schedulers, hardware configurations, and workload distributions. 
%which 
%spanning different schedulers, hardware configurations, and workload distributions. %End-to-end, 
$\pname$ achieves an average $\mathbf{17}\times$ speedup over an AVX-optimized software baseline, processing $\sim$1,700 jobs/second.
\end{enumerate}

The %rest of the 
paper is organized as follows.~\Cref{sec:background} discusses necessary background,~\Cref{sec:methodology} discusses ML %our employed Machine Learning 
predictor and training methodology, %and preliminaries,
~\Cref{sec:hw_design} details the $\pname$ architecture, %for this work,
~\Cref{sec:exp_setup,sec:exp_results} %and~\Cref{} 
detail our experimental setup and experimental results, respectively.~\Cref{sec:related_work} discusses related work and positions our %current 
research,~\Cref{sec:limitations,sec:validity_threats} %summarizes 
discuss limitations and validity threats of $\pname$, respectively, and~\Cref{sec:conclusion} concludes the paper.

\section{Background and Preliminaries}\label{sec:background}
In this section, we define the formal abstractions for %used in 
stochastic online scheduling (SOS) and briefly review the hardware constraints of deploying tree-based ML inference on FPGAs. 

\subsection{Stochastic Scheduling Abstractions}\label{sec:background_definitions}

The core objective of %Stochastic Online Scheduling 
SOS %algorithm 
is to dynamically map streaming %continuously arriving 
jobs to %heterogeneous machines 
machines containing heterogeneous compute resources ($\eg$, CPUs, GPUs) without prior knowledge of future workloads or compute resource availability (thereby introducing uncertainty) while minimizing the weighted expected processing %completion 
time \cite{GuptaEtAl2020, Jager2023} (WEPT).
To manage this uncertainty, the scheduler relies on specific abstractions for jobs and schedules. %Therefore, we introduce the following definitions.

\begin{definition}\label{def:job}
    A \bem{Job} $J$ is an abstraction of a program $P$ with uncertain execution time, 
    represented as a quadruple
    $J = \langle W, \hat{\epsilon}, \mathcal{P}, ID \rangle$ \cite{stannic_2026},
    where $W$ is the \emph{weight} (priority) of $J$, $\hat{\epsilon}$ is a list of \emph{Expected Processing Times} (EPTs) ($\ie$,  how long program $P$ is expected to run on the machine), with $|\hat{\epsilon}| = N$ for $N$ machines, $\mathcal{P}$ denotes the nature of the program, with $\mathcal{P} \in [\mathrm{Compute}, \mathrm{Memory}, \mathrm{Mixed}]$, and $ID \in \mathbb{Z}^{+}$ is a unique job identifier \cite{stannic_2026}. %represented as a tuple $J = \langle W, \hat{\epsilon} \rangle$, %. Here, 
    %where $W$ is the {\em weight} (priority) of $J$, and $\hat{\epsilon}$ is the {\em Expected Processing Time} (EPT) on a given machine. We can then compute the \bem{Weighted Shortest Processing Time} (WSPT) ratio as $T^J = J.W / \hat{\epsilon}$. 
    For the $k^{th}$ machine, we compute the \bem{Weighted Shortest Processing Time} (WSPT) ratio as
    $T_k^J = J.W / \hat{\epsilon}_k$, where $\hat{\epsilon}_k \in \hat{\epsilon}$ \cite{stannic_2026}.
    Because exact execution times are unknown at the moment of submission due to stochastic variance ($\eg$, shared resource contention or data movement overhead), EPT is used as a predictive estimate of the job's wall-clock runtime, and can differ from Real Processing Time (RPT), %which is 
    only known after the job is processed by a machine.
\end{definition}

\begin{definition}\label{def:virtual_schedule}
    A \bem{Virtual Schedule} (VS) $V_i$ for machine $M_i$ is a partial order for the execution of an uncommitted set of jobs $\{K\}$, sorted continuously by their relative WSPT ratios. %It is an interim schedule that is updated at regular intervals. 
    In an online scheduling context, VS %this 
    allows the scheduler to have an interim schedule that can be adjusted on the fly. % can be updated.
\end{definition}

Hardware accelerators for SOS %, known as SOSAs, 
use systolic arrays to maintain %these 
Virtual Schedules ($V_i$s) in near-real-time \cite{fusco2022hardware,stannic_2026}. By mapping each $V_i$ to a one-dimensional array of Processing Elements (PEs), the hardware can physically reorder jobs and perform cost calculations by leveraging spatial data locality. \bem{However, the mathematical integrity of this systolic sorting is entirely dependent on the availability and accuracy of the EPT ($\hat{\epsilon}$) provided at the exact moment of job insertion} \cite{KossenWeijland1990}. %Specifically, in \cite{Jager2023}, the core SOS algorithm used for SOSAs~\cite{palaniappan2025herculeshardwareacceleratorstochastic,stannic_2026}, J\"{a}ger %the authors 
%mathematically proved that the \bem{competitiveness} bounds of a schedule produced by SOS is directly correlated to the variance between input EPT and RPT. 
Specifically, J\"{a}ger~\cite{Jager2023} mathematically proved that the \bem{competitiveness} bound of schedules produced by the core SOS algorithm adopted by SOSAs~\cite{palaniappan2025herculeshardwareacceleratorstochastic,stannic_2026} is directly correlated with the variance between the input EPT and RPT.
As such, \bem{minimizing %input 
EPT variance directly 
results in better guaranteed schedule %quality 
resulting in enhanced performance}.

%\doubted{rreva: continue from here, carefully re-read jager's paper}
Let $P_i^J$ denote the stochastic processing time of job $J$ on
machine $M_i$. %($\ie$,  the random variable whose expectation is the
%job's true EPT and whose realized value after execution is its RPT).
Its expectation, $\mathbb{E}[P_i^J]$, is the job's
true EPT. %while 
%We denote by 
$p_i^J$ denotes a realized value of $P_i^J$,
\ie, the job's RPT observed post execution. Formally, J\"{a}ger showed that the competitive ratio of the SOS %policy 
is bounded by $(3+\sqrt{5})(2+\Delta)$ for a deterministic greedy policy, \ie, the expected scheduling cost is at most $(3+\sqrt{5})(2+\Delta)$ times that of the offline reference policy~\cite{Jager2023}. Here, $\Delta$ is an upper bound on the squared coefficient of variation (SCV) of the %processing times
RPTs \cite{Jager2023}. In our setting, $\Delta$ is governed directly by the variance between a job's EPT and RPT introduced in Definition 1: for a job $J$ on machine $M_i$, %$\Delta\leq\frac{Var(J.\hat{\epsilon}_i)}{\mathbb{E}[J.\hat{\epsilon}_i]^2}$. 
$
\mathrm{SCV}(P_i^J)
    := \frac{\operatorname{Var}[P_i^J]}
            {\mathbb{E}[P_i^J]^2}
    \leq \Delta
$ \cite{Jager2023}.
Thus, smaller $\Delta$ indicates lower variability between RPT and EPT.
%of the realized processing times (RPTs) around their expected value (EPT).

%This inequality makes the mechanism above explicit; because 
As $\Delta$ scales with the variance-to-mean-squared ratio of the EPT, any reduction in EPT variance %directly 
tightens both competitive ratios, independent of the particular scheduling implementation. The guarantee further depends on maintaining low prediction error as the underlying workload evolves: {\em a predictor whose error grows over time yields a growing $\Delta$ and a corresponding weaker guarantee}. This dependence is what {\em motivates} the temporal-aware training policy %introduced in, which is 
designed explicitly to keep prediction error low as the workload distribution drifts (\cf,~\Cref{sec:additive_training}).

\subsection{$\stannic$ FPGA-Assisted Scheduler}\label{sec:stannic_background}
To %dynamically 
map incoming jobs {\em on-the-fly} to the available heterogeneous machines, the generated EPTs must be fed into a \emph{scheduler}. SOSAs %Stochastic Online Scheduling Accelerators (SOSAs) 
bridge the gap  between complex scheduling heuristics and the strict low-latency requirements of heterogeneous %high-performance computing.
HPC. $\stannic$ %SOSA 
implements a SOS algorithm \cite{Jager2023} from a Virtual {\em schedule-centric} perspective. It uses a one-dimensional systolic array per machine to maintain WSPT-ordered Virtual Schedules ($V_i$) and mitigates routing overhead \cite{stannic_2026}. To minimize software overhead and effectively leverage spatial parallelism, $\stannic$ is deployed on an %Field-Programmable Gate Array (FPGA)
FPGA. Unlike task-centric approaches~\cite{palaniappan2025herculeshardwareacceleratorstochastic}, $\stannic$ organizes the %Virtual Schedule ($V_i$) 
$V_i$ of each machine into a centralized Systolic Memory Management Unit (SMMU). %The assignment of an incoming job $J$ to a machine $M_i$ relies on calculating an expected delay cost, which is split into two components based on the relative WSPT priority of $J$ compared to the existing jobs $K \in V_i$. 
It assigns an incoming job $J$ to a machine $M_i$ based on an expected delay cost %with two components, 
determined by the WSPT %priority 
of $J$ relative to the existing jobs $K \in V_i$. %: $cost^{H}$ and $cost^{L}$. %$ cost^{H} = (J.W) \cdot \left(J.\hat{\epsilon}_i + \sum_{K \in V_i, T_i^K \ge T_i^J} \left(K.\hat{\epsilon}_i - n_K(t_J)\right)\right)
% %$ and $cost^{L} = J.\hat{\epsilon}_i \cdot \sum_{K \in V_i, T_i^K < T_i^J} \left(K.W - n_K(t_J)\frac{K.W}{K.\hat{\epsilon}_i}\right)$. 
% Here, $cost^{H}$ accumulates the expected delay imposed on $J$ by existing jobs with a higher or equal priority, while $cost^{L}$ calculates the delay $J$ imposes onto jobs with a lower priority \cite{stannic_2026}.  Within these %summations
% sums, the assigned weights (absolute priorities) of jobs $J$ and $K$ are denoted by $J.W$ and $K.W$. Their %Expected Processing Times (EPTs) 
% EPTs on machine $M_i$ are given by $J.\hat{\epsilon}_i$ and $K.\hat{\epsilon}_i$. The variables $T_i^J$ and $T_i^K$ denote their WSPT ratios ($W / \hat{\epsilon}_i$), used as internal sorting metric. $n_K(t_J)$ is the discretized virtual work, %representing 
% and represents the %exact 
% number of cycles job $K$ has spent at the head of $V_i$ by the time $J$ arrives at $t_J$.
To compute the cost %se costs 
in near real time, the systolic array keeps the PEs in strict WSPT order. Each PE stores one job $K$ and compares it to the incoming job using the broadcasted $T_i^J$. %, setting $C = 0$ if $T_i^K \ge T_i^J$ and $C = 1$ otherwise. 
This enables the PEs to identify the boundary between the high- and low-priority sets locally. Because each PE stores the cost, %cumulative sums ($sum^{HI}$ and $sum^{LO}$), 
the boundary PEs can return the cost via a single-cycle parallel lookup $\left(\mathcal{O}(1)\right)$  %directly, reducing an 
instead of $\mathcal{O}(N)$, where $N$ %
is the maximum number of jobs in $V_i$. %) sum to .
\bem{Due to $\stannic$'s near real-time scheduling speeds, it is imperative that prediction latency is minimized}. 

\subsection{Gradient Boosted Inference}\label{sec:xgboost_bkgr}
To generate high-fidelity EPTs for incoming jobs, gradient-boosted decision tree ensembles, specifically XGBoost, are highly effective \cite{MenearEtAl2023}. XGBoost is %particularly 
suited for HPC telemetry because it natively handles tabular data with extreme variance, sparse features, and bimodal distributions ($\eg$, immediate job crashes resulting in zero-runtime spikes).
%Mathematically, 
XGBoost formulates the predictive task as an additive expansion of $Q$ trees \cite{ChenGuestrin2016}. For a given job's feature vector $x_i$, the final ensemble prediction $\hat{y}_i$ is the accumulated sum of the outputs from individual trees:
$\hat{y}_i = \sum_{q=1}^Q f_p(x_i),  f_q \in \mathcal{F}$
where $\mathcal{F}$ represents the functional space of all possible trees, and each $f_q$ is an independent tree. The algorithm trains these trees sequentially. At iteration $j$, the model learns a new tree $f_j$ designed to minimize a regularized objective function $\mathcal{L}^{(j)}$ based on the residual errors of the previous $j-1$ trees.
%: $\mathcal{L}^{(j)} = \sum_{i=1}^n l\left(y_i, \hat{y}_i^{(j-1)} + f_j(x_i)\right) + \Omega(f_j)$,
%where $l$ is a differentiable convex loss function measuring the variance between the true execution time $y_i$ and the current prediction. The regularization term $\Omega(f_j)$ %= \gamma T + \frac{1}{2} \lambda \|w\|^2$ 
% penalizes model complexity by constraining the number of leaves ${\cal T}$ and the magnitude of the leaf weights $w$, with the goal of mitigating overfitting on stochastic outliers. XGBoost optimizes ${\cal L}$ %this objective 
% using a second-order Taylor expansion based on the first- and second-order gradients of the loss.
% :
% $
% \mathcal{L}^{(t)} \approx \sum_{i=1}^n \left[ g_i f_t(x_i) + \frac{1}{2} h_i f_t^2(x_i) \right] + \Omega(f_t)
% $
At inference, each input \(x_i\) is routed through Boolean threshold tests in the trees until it reaches a leaf weight \(w\). \bem{The final prediction is %then 
the sum of these independent leaf weights, making %which makes 
the method well-suited for spatial hardware acceleration through adder-tree cascades}. %\ar{Wanted to flag this a bit more for readers down the line, but maybe it's clunky.} 
Furthermore, the iterative training %method 
makes XGBoost %a 
suitable %fit 
to %for 
warm-start partial training schemes~\cite{xgboost_warm_start}. We leverage this %warm-start 
capability in \Cref{sec:async_xgboost} to adapt predictors to temporality %temporal changes in 
of HPC runtime conditions. %to keep the predictors up to date and accurate.

%\dpal{Should we have a subsection on STANNIC? Bring~\Cref{sec:stannic} here.}

\section{ML Technique for EPT Prediction
%Methodology
}\label{sec:methodology}

In this section, we explain the ML strategy employed to predict EPTs, the data translation techniques %required 
to bridge %high-level 
software prediction models with strict hardware scheduling constraints, and detail %motivations behind our 
the hardware architectural components. %choices. 

%\dpal{We need fancy figures.}

\subsection{Data Sanitization and Feature Selection}\label{sec:data_sanitization}
%Before training the XGBoost model, the historical HPC dataset must be  sanitized to ensure the predictive model learns valid, generalizable scheduling patterns. %rather than system anomalies or mathematically impossible artifacts.
%To ensure that the stream carries valid information, the predictive ML model must learn valid scheduling patterns. %, 
%To ensure high quality predictions, the predictive ML model must learn valid scheduling patterns.
To learn valid and generalizable scheduling patterns, we must ensure that the training data is free of %no 
faulty records (\eg, failed runs) to avoid skewing the predicted values. %necessitating sanitization of %. 
As such, the 
historical HPC dataset must be sanitized.

\subsubsection{Filtering Bimodal Zero-Value Spikes}\label{sec:filtering_bimodal}

To prevent the ML model from learning to predict immediate system crashes or scheduler cancellations, we apply a filtering step to the training data. %Our 
Initial analysis of the %historical %Argonne Leadership Computing Facility (ALCF) 
ALCF %workload 
dataset~\cite{alcf_data} %workload dataset 
revealed high bimodal spikes where the jobs' run times were recorded at exactly zero. These are likely failed jobs, immediate crashes, or scheduler cancellations. These records do not represent meaningful job execution. %as i
Including them could bias EPT predictions toward artificially short runtimes. %To prevent the model from learning to predict immediate system crashes, 
Therefore, we apply a pre-processing step that purges all rows involving zero or negative (erroneous record) runtime. %(immediate crashes/failed jobs) or negative runtime (erroneous record). %where the runtime is %exactly zero (``0''). \ar{I thought we did negatives too, as that was another error indicator (or just faulty data).}
%less of equal than zero.

\subsubsection{Mitigating Temporal Data Leakage and Input Features}\label{sec:temp_mitigating}

A critical requirement for %Stochastic Online Scheduling (SOS) 
SOS is that the EPT must be predicted at the %exact 
moment of job submission ($t=0$), for immediate routing decisions \cite{GuptaEtAl2020, kohutka2023new, Jager2023}. This must be done without introducing temporal data leakage, where the model uses information that becomes available only after job submission, which can potentially bias its predictions and overstate its real-world accuracy. In fact, features such as queue waiting time or delays caused by dependencies may correlate with runtime in historical data, but they are consequences of system congestion and are \bem{unavailable} at $t=0$. Furthermore, training the model on such future telemetry allows it to %artificially 
``cheat'' during the validation phase. To prevent this behavior, we explicitly remove all post-submission telemetry from the dataset. 
Consequently, we %explicitly drop 
remove these temporal features and define a restricted feature set where categorical identifiers represent the \bem{target machine and job queue identifiers}, resource requirements account for the \bem{requested number of nodes and cores}, and user expectations capture the \bem{total core-hours requested}, %by the user, 
which are all common features for standard {\scshape Slurm} and {\scshape Torque} requests in academic and industrial HPC systems~\cite{NCSASlurm,UIUC_data}. 

\subsection{Categorical Encoding and Feature Scaling}\label{sec:categorical_encoding}

%\ar{Rewriting for here, but we could also choose to move it? We can discuss. I think we could add it to the end of section four though, or maybe even cut it entirely? On the one hand, it's what we did, but it seems like pretty standard things, nothing ground breaking and it doesn't really come up anywhere else?}

%XGBoost's decision trees utilize comparison thresholds to determine tree traversal. Standard software implementations often support direct utilization of strings inside decision nodes. 
XGBoost supports features with different representations, such as numerical and categorical features ($\eg$, requested node counts and queue identifiers, respectively)~\cite{ChenGuestrin2016}. In hardware accelerated implementations such as Conifer's FPU \cite{conifer}, %the hardware must be designed to utilize a consistent typing for all features. 
these features must use fixed-width numerical types ($\eg$, \texttt{ap\_fixed<W,I>}, where \texttt{W} is total bit width and \texttt{I} is number of integer bits, including the sign bit). This imposes specific range and precision constraints, as insufficient integer bits can cause overflow, while insufficient fractional precision can introduce quantization or truncation errors. Therefore, to %properly 
consistently use %utilize 
certain %string based 
categorical features across models in various contexts and retraining (\cf ~\Cref{sec:async_xgboost,sec:hw_design}), we %must 
establish an %consistent
encoding for each of these features with %. This forced homogeneity in feature typing also has 
potential precision repercussions. During inference, the hardware %FPGA 
FPU applies a uniform fixed-point data type to evaluate branching logic across all features ($\eg$, \texttt{ap\_fixed<32,32>}). Since the features span substantially different numerical ranges, using the same type to represent their values can either exceed the representable range or barely use it. %One option would be to use large internal types to ensure everything can fit natively. 
One option would be to increase the bit width to accommodate larger values, but this would also increase hardware resource requirements.
%Features are never compared to each other, only a threshold trained to match the input feature during training. %This hardware constraint forces the tree thresholds to act as strict 32-bit integers. 
%Because features vary significantly, %wildly in scale from requested node counts (which may range into the thousands) to raw decimal metrics 
%they can't all fit %nicely 
%in the same unsigned fixed integer value range without scaling. % us scaling type of each node. Instead we 
%We perform feature scaling for %the sake of 
%resource efficiency and numerical precision %, getting each value 
%to fit within our designated value range. 
Instead, we scale each feature to fit within the available fixed-point range while preserving as much numerical precision as possible. Since each feature is compared only against thresholds defined for %that %same feature
itself, each feature can be scaled independently without changing the tree's comparison logic.

% \subsubsection{Categorical Encoding}
% To enable the FPGA to process string-based categorical data, we implement a \emph{translation ledger} on the Host CPU. In fact, 
% XGBoost can natively handle categorical features ($\eg$, features represented with strings). FPGAs, however, operate strictly on numerical data. %To bridge this, the Host CPU maintains a \emph{translation ledger} that 
% The ledger dynamically maps categorical strings-—such as target machine identifiers and queue names—-into integer IDs before packing them into the feature vector.

% \subsubsection{Host-Side Input Scaling}

\subsection{Data Normalization and Target Formulation}\label{sec:data_norm}

To align the XGBoost output (core hours) with the scheduler's input %logic 
(a job $J$'s EPT $J.\hat{\epsilon}_i$ on machine $M_i$), %runtime), 
we translate %we formulate the \bem{target} as wall-clock days.
%In fact, a key methodological challenge is translating 
XGBoost predictions into hardware-compatible formats %without mismatches in the scheduling logic. To achieve this, we convert XGBoost EPT predictions from raw core hours to wall-clock days, 
%computed as \texttt{USED\_CORE\_HOURS} $\div$ \texttt{CORES\_REQUESTED} 
computed as (\emph{consumed core-hours} $\div$ (24 $\times$ \emph{requested cores}))
($\ie$, the actual calendar time a job occupies a machine, regardless of how many cores it uses). Compared to raw core-hours, wall-clock days stay much smaller numerically and fit the fixed-point budget more safely. For example, a job running on 1,000 cores for 2 hours has %\texttt{USED\_CORE\_HOURS} $= 2{,}000$, 
consumed %but 
$2{,}000$ core hours, its wall-clock runtime is only about $0.083$ days. %This also makes the FPU output consistent with $\stannic$’s machine-level scheduling logic, mapping it directly to job $J$'s EPT, $J.\hat{\epsilon}_i$, on machine $M_i$.

%\doubted{The reason is that $\stannic$'s cost 
%function is expressed in terms of $J.\hat{\epsilon}_i$, how long a job occupies machine $M_i$.} %-- to compute WSPT ratios, track Virtual Work, and determine $\alpha_J$ release points.} 
%Instead, core-hours conflate core count with elapsed time. In fact, a 1,000-core job running for 2 hours and a 1-core job running for 2,000 hours have the same core-hours, but they occupy machines for completely different amounts of time. As result, feeding core-hours to $\stannic$ would create a unit mismatch and corrupt its WSPT ratios and Virtual Work tracking.

%\input{pic/data_to_tree_block}

\subsection{Asynchronous XGBoost Model Training}\label{sec:async_xgboost}
%\doubted{PUT HIST KNOWLEDGE ASSUMPTION HERE}

\subsubsection{Additive Training}\label{sec:additive_training}

To predict EPTs of jobs in %for 
evolving workloads, %adapt to changing workloads and reflect the current cluster state, 
we use XGBoost as our primary %main 
ML %predictive algorithm. %Because the Inference FPGA has strict spatial constraints, the total ensemble capacity is bounded by a maximum tree budget, $N_T$. 
predictor.
As %scheduled jobs finish on the simulated machines, 
jobs finish, %complete 
their %true 
execution data becomes available. %, especially the actual wall-clock runtime. This %ground-truth 
%information is used to keep EPT predictions accurate. 
These newly observed ground-truth values provide the feedback needed to keep EPT predictions accurate over time.
%We resort to these new ground-truth data, because of the inherently non-stationary nature of HPC workloads. 
We incorporate this feedback because HPC workloads are inherently non-stationary.
Relying on a static predictive model can lead to temporal degradation, as the scheduler is not aware of real-time cluster conditions, such as a sudden shifts in the runtime distributions. By iteratively feeding the execution results of recently completed jobs back into the model, we establish a \emph{%closed 
feedback loop}, transforming %This transforms 
$\pname$ into a dynamic, ``situation-aware'' scheduling architecture. However, retraining the entire predictive model from scratch on an %evolving % continuously growing 
streaming dataset to maintain the %this 
situation %real-time 
awareness is \bem{computationally prohibitive, and can introduce significant overheads}. 

To solve this issue, we implement \emph{additive training}. In gradient boosted ensembles like XGBoost, additive training allows us 
to append new decision trees to an existing model to %, which 
iteratively correct predictor error. %for 
%earlier trees %error, 
as described in \Cref{sec:xgboost_bkgr}. This enables the ensemble to learn from new data without discarding previously learned patterns and avoid %structures 
%or requiring 
a computationally expensive %full 
retraining. %cycle. %, which can be computationally expensive. 
%To execute the additive training, we %must 
%define and 
%develop %implement 
%an \emph{update policy} on %to manage 
%how new trees are appended. 
Although $\pname$ uses XGBoost as its primary predictor, we additionally evaluate the additive training policy with Random Forest, showing that the proposed adaptation strategy is not specific to XGBoost.

\begin{figure}
    \centering
    \includegraphics[scale=0.25]{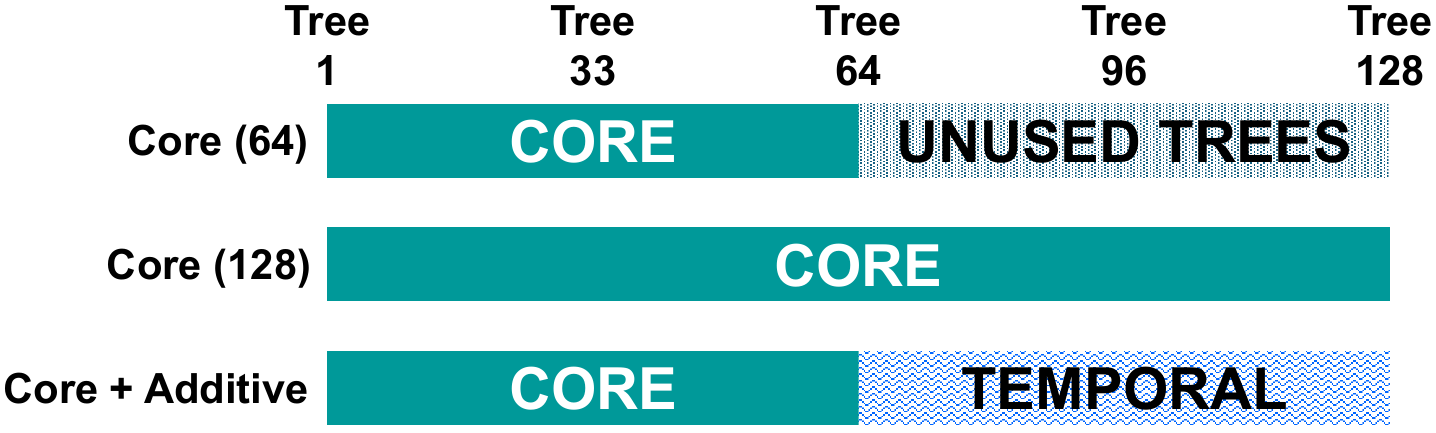}
    \caption{Data to tree allocation during model training.}
    \label{fig:data_to_tree}
    \vspace{-6mm}
\end{figure}

\subsubsection{Bifurcated Core + Additive Policy}\label{sec:bifurcated}

While multiple policies could be defined, we designed a strictly \emph{bifurcated Core + Additive} policy (illustrated in \Cref{fig:data_to_tree}). This policy partitions the learning capacity %and the dataset 
into two distinct phases. \textbf{1. Primer Phase:} Before the system goes live, we split a historical workload dataset (detailed in \Cref{sec:exp_setup}) into two halves. Because the inference hardware %Inference FPGA 
has strict %spatial 
capacity constraints (\eg, limited on-device storage capacity), the total ensemble capacity is bounded by a maximum tree budget, $N_T$. The first half of $N_T$, designated as the \emph{CORE model}, is used to train the initial model using exactly 50\% of the $N_T$ tree estimators. The CORE model serves as the base knowledge for the model, and captures the long-term, stable statistical behavior of the HPC cluster. \textbf{2. Update Phase:} Once the whole system is online and running, the continuous stream of newly completed jobs is accumulated into discrete, sequential \emph{temporal buckets}. %Refer to \Cref{sec:xgboost_rationale} %of different sizes. 
%for our experimental evaluation to find the optimal size of these buckets and validate the policy's robustness against temporal workload drift.
When a temporal bucket is filled, the asynchronous Trainer thread (\cf,~\Cref{sec:host_cpu}) on host takes the frozen CORE model and warm-starts a new training session. It then appends the remaining 50\% of the tree budget trained exclusively on the %latest 
temporal bucket.
There are two fundamental motivations of \bem{why} we choose this bifurcated policy. \emph{First}, it allows the model to capture new trends without overwriting established ones. The frozen core model keeps the cluster’s long-term baseline behavior, while the additive trees adapt to short-term changes such as queue time shifts or temporary hardware slowdowns. \emph{Second}, this static partitioning keeps our inference model consistently sized, allowing us to always %utilize 
use as much of our inferencing hardware as possible. {\em A fixed tree budget keeps resource usage predictable, bounds Host-side training cost, and simplifies additive model updates}.

To empirically validate this specific allocation, we performed a sensitivity analysis using the datasets described in \Cref{sec:exp_setup}, over the CORE/Additive tree-budget split, sweeping five %interior 
configurations (25/75, 37.5/62.5, 50/50, 62.5/37.5, 75/25) as well as the two degenerate configurations (100/0, \ie, CORE only, and 0/100, \ie, Additive only) under increasing synthetic temporal drift. %(\Cref{sec:async_xgboost}). 
MAE is largely insensitive to the exact allocation as long as both components receive a nonzero share of the tree budget: %interior 
splits perform comparably to one another, while the 100/0 configuration exhibits a marked degradation (+$8\%$ MAE on average). from the loss of drift-tracking capacity, and the 0/100 configuration remains competitive (-$3\%$ MAE on average), but forfeits the stable long-term prior that the frozen CORE %component 
provides. We adopt the 50/50 split because %it balances these two failure modes: 
it preserves a substantial frozen baseline while allocating sufficient capacity to the additive component to track distributional shifts, making it a natural regularization point between under- and over-committing to either behavior.

%this static partitioning makes the design \emph{portable}, in the sense that it becomes easier to scale to FPGA platforms with different resource limits. A fixed tree budget keeps resource usage predictable, bounds Host-side training cost, and simplifies additive model updates.}

\begin{figure}
    \centering
    \includegraphics[scale=0.25]{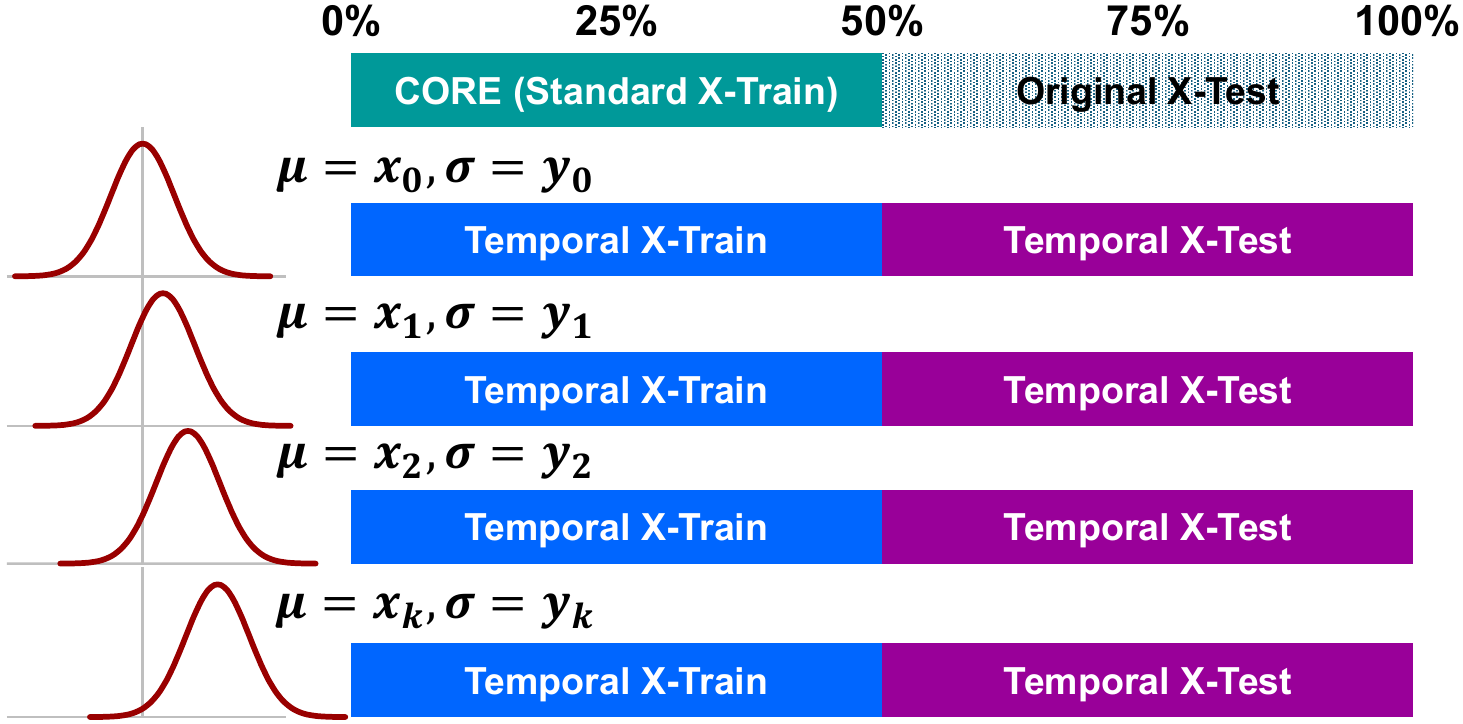}
    \caption{Dataset split and temporal dataset creation.}
    \label{fig:temporal_dataset}
    \vspace{-4mm}
\end{figure}

\begin{figure}
    \centering
    \includegraphics[width=0.85\linewidth]{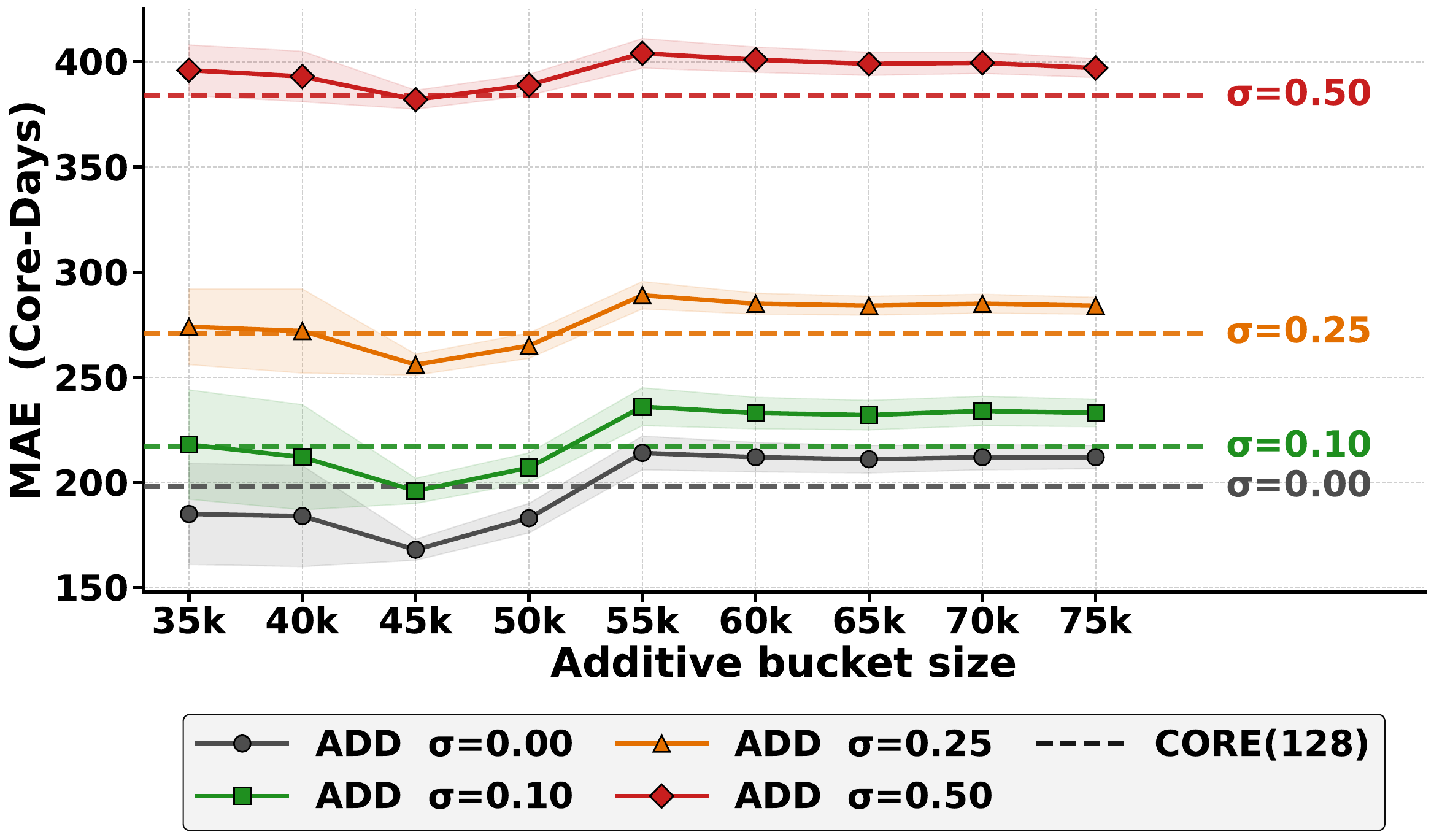}
    \caption{MAE of the additive XGBoost model versus static baselines under increasing synthetic temporal drift ($\sigma$). Performance is best at an additive bucket size of 45,000 jobs, where the additive model (ADD) consistently outperforms the baseline (CORE) across all drift ($\sigma$) levels, showing robust adaptation to temporal workload shifts.}
    \label{fig:drift_rmse_methodology}
    \vspace{-6mm}
\end{figure}

\subsubsection{XGBoost Rationale%-- ALCF Calibration Example
}\label{sec:xgboost_rationale}

%We use the ALCF data %workload 
%as a running example to illustrate how the temporal bucket size is calibrated for a target deployment.
To evaluate the robustness of XGBoost against temporal workload shifts, we simulate temporal drift by applying a Gaussian scaling factor, $N(1, \sigma^2)$, to job runtimes in ALCF data across four severity levels ($\sigma \in [0.00, 0.50]$) %ranging from 0.00 to 0.50
(\cf,~\Cref{fig:temporal_dataset}). We compare a static baseline %(BA) 
model -- trained on half the historical dataset using %64 and 128 
trees (CORE) -- against approaches based on the  \emph{bifurcated Core + Additive} policy (ADD) defined in \Cref{sec:bifurcated}. %, with different bucket sizes. 
The additive models warm-start from the 64-tree CORE baseline model and sequentially train 64 new trees exclusively on a recent bucket of drifted jobs. %\Cref{fig:drift_rmse_methodology} illustrates this comparison, measuring the Root Mean Square Error (MAE) across different tested additive bucket sizes. 
To ensure statistical robustness and mitigate evaluation bias, we evaluate these models using a $K$-fold cross-validation ($K=5$). %\cite{cross_validation_1995}. 
We %also 
partition the buckets using \emph{stratified sampling} \cite{Neyman1992}, to ensure that each fold %receives 
has a proportional and representative share of extreme temporal outliers jobs from the original dataset.

\Cref{fig:drift_rmse_methodology} shows the comparison and reports the mean MAE and its shaded %$\pm 1\sigma$ variance bands 
$\pm 1$ standard deviation bands ($\sigma_{\text{MAE}}$) computed %across
across different %tested 
additive bucket sizes. %\ar{Hmmm, lets be careful here, we don't want to confuse this variance $\sigma$ with the variance $\sigma$ we use in our temporal data creation. Can we be more specific here? Differentiate them somehow? Use $\sigma_{MAE}$ or something?}
The cross-validated evaluation reveals an %distinct 
optimal bucket size of \bem{45,000 jobs} for the ALCF %workload 
dataset. At this granularity, ADD
%the additive model 
consistently outperforms %both static baselines 
CORE across all drift severities ($\sigma$). Under medium drift ($\sigma = 0.25$), %the static 64-tree baseline 
CORE degrades to an MAE of approximately %4,650, 
$285$, while the $45$K additive model maintains a %superior 
lower MAE of approximately %, near %3,880. 
$255$. %\ar{Check these values, I think these are still the RMSE numbers, but now we use MAE. Also, I think we should use the 128 core numbers if we later say "rather than simply increasing the overall tree count."} 
This confirms that the performance gains are driven by active drift tracking within the leftover data rather than simply increasing the overall tree count. While prediction accuracy gradually converges back toward the baseline at larger bucket sizes ($> 60,000$) due to over-correcting on bucket-specific distributions, the additive model remains stable and avoids catastrophic degradation even under severe drift.
%\bem{These results confirm that additive training —-specifically when calibrated to the 45,000-job sweet spot—- is robust against temporal drift in HPC workloads. This validates XGBoost as a highly adaptive predictive engine, and fulfills our core goal of delivering continuous and robust runtime estimates to our hardware-accelerated scheduling pipeline.}
%{\em These results show that additive training, especially near the 45K-job niche %45{,}000-job 
%sweet spot, 
%can be robust to temporal drift in HPC workloads}. \bem{This validates XGBoost as an adaptive predictor for continuous runtime estimation in our hardware-accelerated scheduling pipeline}.
\bem{These results suggest that, when appropriately calibrated to the target workload, additive training can improve XGBoost's robustness to temporal drift.} %This supports its use for adaptive runtime estimation in our hardware-accelerated scheduling pipeline.

\smallskip

\noindent \bem{Remark}. The $45$K %,000-
job bucket size should not be interpreted as a universal parameter of $\pname$, rather %but as 
an ALCF-specific calibration. %example. 
The appropriate bucket size depends on the workload's temporal characteristics, \eg, job wall-times. %As shown in \doubted{ref to exp result where we compare mit vs uiuc}, shorter jobs provide execution feedback more rapidly and can therefore support larger bucket sizes over a comparable time horizon. 
For a new deployment, we do not assume a %complete 
cold start. Existing traces from the target system, or representative historical HPC %workload 
datasets such as those used in this work, can provide an initial CORE model and bucket-size estimate. Once $\pname$ is live, this initial configuration can be refined using the continuously observed wall-clock times of completed jobs.

\section{Hardware Architecture %Design
of $\pname$
%\dpal{START HERE}
}\label{sec:hw_design}

%\dpal{We need a figure for the architecture and likely its components like we did for HERCULES. START HERE.}

\begin{figure}
    \centering
    \includegraphics[width=\columnwidth]{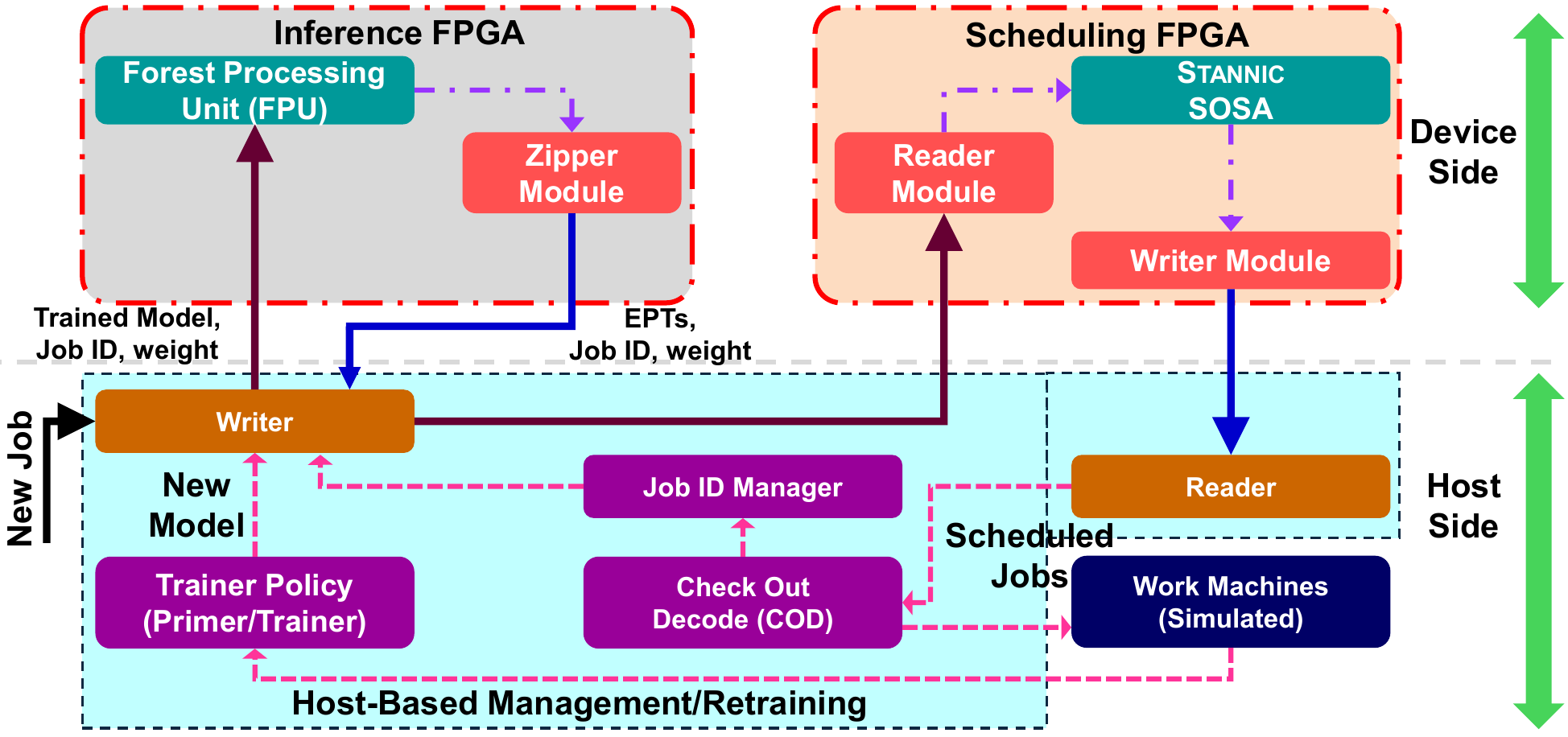}
    \caption{Top-Level %Block 
    diagram of $\pname$ hardware architecture. $\textcolor[HTML]{9933FF}{\rightarrow}$: Data movement inside FPGA. $\textcolor[HTML]{FF3399}{\rightarrow}$: Data movement inside host. $\textcolor[HTML]{660033}{\rightarrow}$: Data movement across host and FPGA.
    %\dpal{Will edit the figure further. Initial Placeholder.}
    }
    \label{fig:overall_boostedsosa}
    \vspace{-6mm}
\end{figure}

%The proposed 
$\pname$ %framework 
is a multi-stage pipeline running across a \emph{Host CPU} and two connected %Alveo U55C -> ALVEO WILL GO TO EXP SETUP
FPGAs, as shown %seen 
in \Cref{fig:overall_boostedsosa}. To separate %divide 
concerns and reduce delay, %preprocessing and model training run on the host, 
inference runs on the Conifer FPU \cite{conifer} based \emph{Inference FPGA} and systolic array based scheduler %scheduling 
runs on the \emph{Scheduling FPGA} using %the 
$\stannic$ scheduler \cite{stannic_2026}. The host is responsible for preprocessing and model training. The modules, along with their integration, are described in the following subsections. %The next subsections describe the hardware and software components of this co-design.

\begin{comment}
%THIS GOES INTO METHODOLOGY
\subsection{Motivation: The Transition from Batch to Streaming I/O}
Prior architectures, including the baseline STANNIC architecture \cite{stannic_2026}, relied heavily on batch-loading I/O between the Host CPU and the $\schedfpga$. The Host CPU had to accumulate rigid batches of incoming jobs before executing a bulk transfer to the hardware. While this strategy was sufficient for virtual simulations, it introduced severe limitations, that hindered its deployment in real-world, live scenarios. \emph{First}, it does not reflect how online scheduling works, where job arrivals are sporadic and unpredictable. Having to wait until the batch is completely filled introduces undesired latencies. \emph{Second}, it is space-inefficient, because it requires  reservations of idle hardware memory on both the Host and the FPGA to stage the batch data. To achieve true SOS, the architecture must be capable of processing jobs instantaneously as they arrive. Therefore, we propose a \bem{continuous streaming architecture} built as a co-design of multiple interconnected modules that communicate with each other \doubted{using a PCIe shared memory}. %The advantage of 
By using this modular architecture, % is that 
we enforce \emph{separation of concerns} (each component executes its own tasks and treats the others as black box layers), we reduce starvation and overhead (each component works in asynchronous mode), and we mitigate %having the 
risk of a \emph{single point of failure.} 
\end{comment}

\subsection{The Host CPU}\label{sec:host_cpu}

%\dpal{I think we should write 2-3 lines discussing about various threads here. Otherwise, the threads are appearing all of a sudden in~\Cref{sec:host_side,sec:model_integration}. We need figures as well.}
\bem{The Host CPU %has the goals of orchestrating 
orchestrates the pipeline, %runtime, 
and coordinates job management and model maintenance during continuous execution}. %In particular, 
All data %intermediate results 
exchanged by the two FPGAs are written to and retrieved from Host RAM through PCIe, making the Host CPU the central coordination point for data movement across the pipeline. The Host CPU operates through an asynchronous multi-threaded architecture. This design parallelizes communication, scheduling control, and model maintenance to hide PCIe latency and %avoid 
mitigate host-side bottlenecks during continuous execution. Five threads are used: \emph{Job ID Manager}, \emph{Writer}, \emph{Reader}, \emph{Checkout Decode}, and \emph{Trainer}.

\subsubsection{Host Side Job Orchestration}\label{sec:host_side}

The pipeline begins at the Host CPU. To reduce latency and separate concerns inside the Host CPU, we resort to multi-threaded architecture that enables asynchronous execution.  
When a job is submitted to the cluster by an external user, %($t=0$)
it is managed by the \emph{Writer}. %thread. 
%This thread 
Writer receives a unique, quantized, Job ID from the \emph{Job ID Manager} (JIM), %thread, 
and applies it to the current job. The JIM acts as a \emph{translation ledger} %, because it updates a queue of available Job IDs, 
between the software cluster and the hardware accelerator. Real-world HPC jobs often use long, complex external identifiers ($\eg$, UUIDs). %or string hashes). 
The JIM maps each external ID to a compact, hardware-friendly internal ID. It tracks these internal IDs through a simple queue. When a new job arrives, JIM %the Manager 
assigns the next free internal ID and stores the mapping from internal to external ID in a lookup table. Then, the Writer writes this information into the shared memory to be streamed via PCIe to the Inference FPGA, and %forwarding 
forwards the results to the Scheduling FPGA when ready. %Operating asynchronously from the Writer, 
The \emph{Reader} %thread 
continuously monitors the return PCIe channels. Once the {\em Scheduling FPGA} finalizes a scheduling decision, the Reader ingests $\stannic$'s output and passes it to the \emph{Checkout Decode}. %thread. %module. 
This thread executes two operations -- \bem{First}, it asks the JIM to convert the hardware's internal ID back to the original external ID, thereby releasing the internal ID for reuse and \bem{Second}, it sends the scheduled job to the simulated machines.

\subsubsection{Model Integration and Payload Deployment}\label{sec:model_integration}

\begin{figure}
    \centering
    \includegraphics[width=\columnwidth]{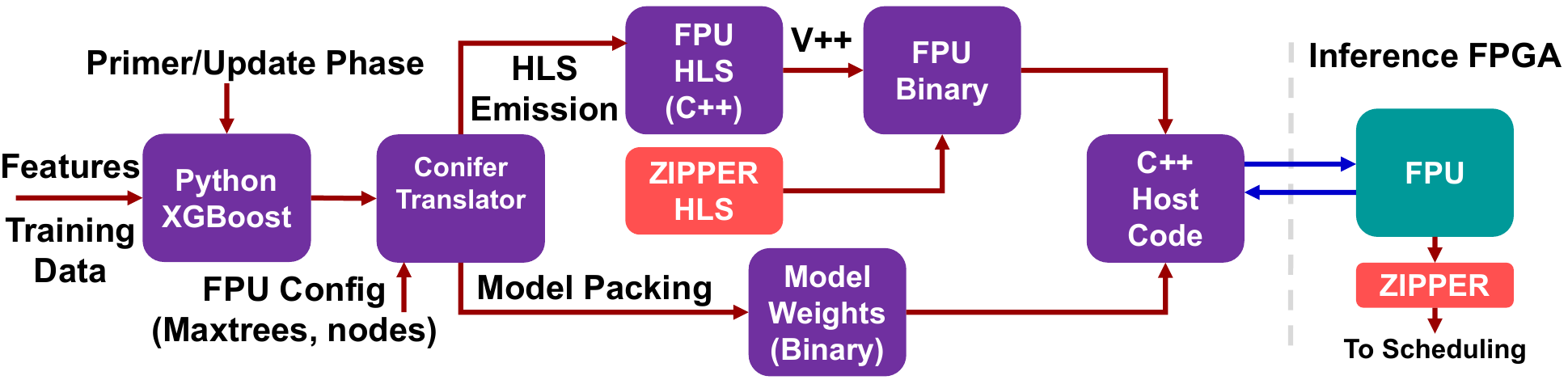}
    \caption{XGBoost trained model to FPGA kernel bitstream and weight %conversion 
    via Conifer translator~\cite{conifer}. $\textcolor[HTML]{990000}{\rightarrow}$: Data movement between Python XGBoost model and Confier translator. $\textcolor[HTML]{0000CC}{\rightarrow}$: Data movement between host code and FPU kernel module.
    %\dpal{Will edit the figure. Initial Placeholder.}
    %\ar{Should we modify to show that we have hand modified the HLS/Use additional separate HLS (zipper) to create our binary?}
    }
    \label{fig:xgboost_conifer}
    \vspace{-7mm}
\end{figure}

%The framework 
$\pname$ includes a dedicated %asynchronous 
\emph{Trainer} %thread 
that updates the XGBoost predictive model (\cf,~\Cref{sec:methodology}). %the training phase is detailed in Section \ref{sec:methodology}). 
% \begin{comment}
% Once an updated model is ready, it is compiled by \emph{Conifer} into two deployable payloads. The first is a synthesized C++ HLS kernel encoding the inference data path and fixed-point arithmetic (\texttt{ap\_fixed<32,32>}), compiled into an FPGA bitstream via Vitis (\texttt{v++}). The second is the \emph{Model Nodes Binary}, which serializes the numerical content of the trained model. For each incoming job, the Host creates a batch of $N$ feature vectors (one per target machine) and streams them to the Inference FPGA.
% \end{comment}
Once training is complete, {\em Conifer}~\cite{conifer}
 compiles the updated model %is compiled 
into two FPGA-deployable %distinct hardware 
payloads,  %by \emph{Conifer}~\cite{conifer},
%a framework that 
%Conifer converts XGBoost ensembles into FPGA-deployable representations, 
taking as additional inputs 
the \emph{FPU configuration} --- (i) number of tree engines%($50$)
, (ii) nodes per engine%($512$)
, (iii) feature width%($8$, padded)
, and (iv) fixed-point precision. %(\texttt{ap\_fixed<32,32>}). 
\Cref{fig:xgboost_conifer} shows an overview %summary 
of the %subset 
%%of the 
{\em Conifer} flow %that we utilize 
in \pname. %\ can be seen in . 
The compilation produces two outputs. The {\em first} is a synthesized C++ HLS kernel that encodes the inference data path, such as the tree traversal logic, fixed-point arithmetic and pipeline structure which Vitis (\texttt{v++}) compiles %it 
into an FPGA bitstream 
targeting the Inference FPGA~\cite{vitis}. %Alveo U55C. 
The {\em second} is the 
\emph{Model Nodes Binary}, which serializes the numerical 
content of the trained ensemble, \eg, a node's feature index, split threshold, and leaf score. This binary is %not compiled into hardware directly, and can instead be 
loaded into the hardware at runtime. %\doubted{Unlike the bitstream, these binaries are not synthesized into hardware}. \doubted{Instead, they are loaded at runtime by the Host Code into the Inference FPGA's on-chip memory to populate the decision tree lookup structures.} %. This separation is what makes live 
%model updates non-disruptive, because upon each retraining cycle, only nodes and scales binaries are regenerated and hot-swapped into FPGA memory, with no re-synthesis or FPGA reprogramming required.}
For each incoming job, the Host creates a batch of $N-1$ features (all but the Machine ID), and sends them to the Inference FPGA.

\subsection{The Inference FPGA}\label{sec:inf_fpga}

\bem{The Inference FPGA %accelerates 
%has the goal of accelerating 
accelerates the XGBoost inference phase} (\cf,~\Cref{fig:xgboost_conifer}). %and 
It has two modules: the \emph{FPU}, and the \emph{Zipper}. %XRT OpenCL kernel 
%and the \emph{Reader}. 
%The Inference FPGA %(Alveo U55C) 
%has the goal of accelerating the XGBoost inference phase. %and hosts the hardware XGBoost inference engine, programmed onto the Alveo U55C 
%and is programmed from the bitstream created by Conifer. 
The \emph{FPU} is %\fixme{\emph{Forest Processing Unit} (FPU)}, 
a 
hardware-based decision tree engine synthesized by Vitis into an FPGA \emph{bitstream} %($\ie$, the binary configuration used to program the FPGA's hardware logic used to execute decision trees),
from the {\em Conifer}-emitted 
HLS C++ \cite{conifer}. This is purely the ``architecture'' of the predictor models, and is completely segregated %divorced 
from individual weights within each tree.
%, which is synthesized by Vitis into an FPGA \emph{bitstream}.
At runtime, the Host code loads the Model Nodes Binary weights into the FPU's on-chip memory, populating the 128 independent \emph{tree engine} units, one per tree in the ensemble, each backed by a dedicated node memory of up to $64$ entries. %This separation is what makes live 
The separation between the bitstream and the model weights makes live updates 
%model updates 
non-disruptive, because upon each retraining cycle, only node weight binaries are regenerated and hot-swapped into FPGA memory, with no re-synthesis or FPGA reprogramming required. 
The FPU reads the $N-1$ feature vector, then iterates over the machine indices, %$0-5$, 
running a pipelined batch inference for all machines. %processes all $N$ vectors loaded from the Host CPU in a single batched inference pass.  %across its $50$ parallel tree engines, producing the full EPT vector $J.\hat{\epsilon}$, with one predicted runtime per machine. This vector is then streamed to the $\schedfpga$ as input to the $\stannic$ scheduler.
%When a feature vector arrives over PCIe from the Host CPU, 
All $128$ tree engines %evaluate it 
perform inference \emph{concurrently}. 
Then, their leaf scores are accumulated to produce the final EPT %prediction.  
vector, $J.\hat{\epsilon}$, with one predicted runtime per machine. %This vector is then streamed to the $\schedfpga$ as input to the $\stannic$ scheduler. 
The \emph{FPU} has been modified to stream it's data %out 
to %our 
\emph{Zipper} module internally. 
%Before transmission to the $\schedfpga$, this vector is passed to the \emph{Zipper} % XRT OpenCL kernel, which sits
%kernel that sits between the FPU and the PCIe stream. 
The \emph{Zipper} packages the EPT vector together with the corresponding job metadata retrieved from the Host (such as Internal Job ID and job weight), and writes the resulting packet to Host RAM over PCIe, which is then forwarded to the {\em Scheduling FPGA}.
Additionally, the Zipper has a precision-correction role, %which is 
detailed in \Cref{sec:inf}. 
% The Inference FPGA is connected to its respective Reader and Writer modules. 
% The Reader asynchronously listens on the PCIe channel for incoming feature vectors from the Host. The Writer forwards the %computed and integrated 
% Zipper-formatted EPT predictions over PCIe to the $\schedfpga$. 
%the Host must extract its metadata to construct the feature vector for downstream Expected Processing Time (EPT) prediction. In this first step, the pipeline reduces the dataset to only the variables needed for scheduling. The host then cleans the data by removing incomplete records and invalid entries, such as jobs with zero requested resources or negative wait times. This ensures a consistent dataset before %normalization and 
%model training.

\subsection{The Scheduling FPGA %$\schedfpga$
}\label{sec:scheduling_fpga}

\bem{The Scheduling FPGA %has the goal of executing 
executes the scheduling % phase of the 
pipeline}. It %has these moodules: 
hosts the $\stannic$, %scheduler, 
and its own \emph{Reader} and \emph{Writer} modules.

\subsubsection{$\stannic$ Scheduler}\label{sec:stannic}

%\dpal{This is where I am a bit concerned. It should be in the Preliminaries and Background section. We can mention the $\schedfpga$ data flow and the Free running memory part, but it feels very weird to include $\stannic$ here since this is NOT a contribution of this paper.}
The {\em Scheduling FPGA} runs %(which is an Alveo U55C, like the Inference FPGA) 
the $\stannic$ SOSA %, whose micro-architecture is described in details in \cite{stannic_2026}. 
(\Cref{sec:stannic_background}). 
%$\stannic$ implements the Stochastic Online Scheduling algorithm \cite{Jager2023} from a Virtual Schedule-centric perspective, using a one-dimensional systolic array per machine to maintain WSPT-ordered Virtual Schedules with minimal routing overhead. 
In this work, we treat $\stannic$ as a \emph{black-box} scheduling engine and integrate it into the streaming pipeline without major modifications, apart from %the following 
minimal %numeric 
interface-level adaptations. %We make data types adjustments to accommodate the fixed-point precision
%of the %Argonne dataset
In particular, we align its input data types with the upstream fixed-point pipeline, enabling packet-based ingestion through the {\em Scheduling FPGA}'s \emph{Reader}, and return scheduling decisions through its \emph{Writer}.

\subsubsection{%$\schedfpga$ 
Scheduling FPGA Data Flow}\label{sec:kernel_data_flow}

%\dpal{Where do we talk about the {\em Zipper} module?}

%The $\schedfpga$ receives %inputs from two sources. The 
%two inputs, from two sources. %\emph{Mem Block} module buffers the EPT vector $J.\hat{\epsilon}$ streamed directly from the Inference FPGA. %The \emph{Zipper} module receives the corresponding job metadata, such as the Internal Job ID, from the Host Writer thread over PCIe. 
%The Zipper module, located on the Inference FPGA, forwards jobs metadata, such as the Internal Job ID, to the Kernal FPGA's Reader thread via PCIe.
%The $\schedfpga$ receives two inputs, the EPT vector $J.\hat{\epsilon}$ from the \emph{Mem Block}, which buffers the stream from the Inference FPGA, \doubted{and the job metadata, such as the Internal Job ID, from the \emph{Reader}, which receives it over PCIe %from the Zipper on the Inference FPGA.
%from the Host CPU's Writer.}
\emph{Reader} %module 
receives the full packet produced by the \emph{Zipper} ($\cf$, \Cref{sec:inf_fpga}), retrieving it from Host RAM via PCIe.
%Both information are 
%This packet is fed into 
$\stannic$ consumes this packet to compute the scheduling decision. Once $\stannic$ assigns the job to a machine, its \emph{Writer} module streams the result back to the Host CPU over PCIe, where the Host's Reader thread reads it for checkout and dispatch.

\subsection{Integration of Inference and  Scheduling FPGAs 
%Integration
}\label{sec:inf}

To integrate the two FPGAs, a key challenge is the precision mismatch between the FPU's output and $\stannic$'s input. %constraints. 
%This creates a tradeoff: 
Preserving the FPU's full numerical precision can require wider data additional hardware resources in $\stannic$, whereas reducing the bit width can introduce quantization or truncation errors. %To overcome this, we 
We therefore propose a multi-stage data type strategy that %maximizes 
%preserves mathematical precision within the FPU while adhering to $\stannic$'s lightweight ingestion constraints.
%strategy that 
maximizes numerical precision within the constraints imposed by $\stannic$'s lightweight integer %input 
representation.

The FPU operates at high precision, using \texttt{ap\_fixed<32,32>} for tree threshold navigation and \texttt{ap\_fixed<32,16>} for leaf score accumulation. %As result, 
% \begin{comment}
% %THIS PART IS MORE HW AGNOSTIC AND GOES TO METHODOLOGY
% EPT predictions are expressed in wall-clock days, computed as  
% \texttt{USED\_CORE\_HOURS} $\div$ \texttt{CORES\_REQUESTED} ($\ie$, the actual calendar time a job occupies a machine, regardless of how many cores it uses). Compared to raw core-hours, wall-clock days stay much smaller numerically and fit the fixed-point budget more safely. For example, a job running on 1,000 cores for 2 hours has \texttt{USED\_CORE\_HOURS} $= 2{,}000$, but its wall-clock runtime is only about $0.083$ days. This also makes the FPU output consistent with $\stannic$’s machine-level scheduling logic. \doubted{The reason is that $\stannic$'s cost 
% function is expressed in terms of $J.\hat{\epsilon}_i$, how long a job occupies machine $M_i$.} %-- to compute WSPT ratios, track Virtual Work, and determine $\alpha_J$ release points.} 
% Instead, core-hours conflate core count with elapsed time. In fact, a 1,000-core job running for 2 hours and a 1-core job running for 2,000 hours have the same core-hours, but they occupy machines for completely different amounts of time. As result, feeding core-hours to $\stannic$ would create a unit mismatch and corrupt its WSPT ratios and Virtual Work tracking.
% \end{comment}
% However, $\stannic$ does not directly accept 
% fixed-point %days as input
% inputs, because it operates on lightweight integer formats \cite{stannic_2026}. 
The \emph{Zipper} kernel acts %solves this issue by acting 
as a precision bridge between the two FPGAs. It multiplies the FPU's day-expressed prediction by $96$, since 
$1~\text{day} = 24~\text{hours} \times 4~\text{blocks/}{\text{hour}} = 96~\text{blocks}$ of 15 minutes each. It then casts the result to \texttt{uint16\_t}: $\text{EPT}\_{\text{blocks}} = \left\lfloor \text{EPT}\_{\text{days}} \times 96 \right\rfloor$. The \texttt{uint16\_t} range $[0; 65{,}535]$ %blocks 
can represent up to about 682 days, which is far beyond the runtime of any realistic HPC job.  %(\doubted{in practice, the vast majority of HPC jobs in the Argonne dataset 
%fall well within $72$ hours of wall-clock runtime.}) 
%\doubted{At the same time, a 15-minute unit is precise enough for $\stannic$’s scheduling cost calculations.}
%Because of this design, we decide to 
Hence, we disable the {\em Conifer} dynamic scaler which otherwise %. Had it been enabled, the 
%scaler 
would re-normalize the FPU output back to raw floating-point before returning it over PCIe. This would %forcing 
force the Zipper to parse, multiply, and re-cast the FPU output, introducing unnecessary overhead and  %Also, the scaler itself consumes additional DSP slices to perform floating-point affine transformations %$(y=(x-offset)×scaley = (x - \text{offset}) \times \text{scale} y=(x-offset)×scale)$
%, which require multipliers. 
%Multiple type and precision castings would also introduce an unnecessary 
precision penalty at both the entry and exit points of the pipeline. Instead, the Host CPU does the feature scaling %is done by the Host CPU 
before transmission, to ensure the data reaches the \texttt{ap\_fixed<32,32>} boundary as integers. The Zipper then applies a $\times 96$ scaling to the final score values, quantizing from \texttt{ap\_fixed<32,16>} to \texttt{uint16} for $\stannic$. 
%%% @Riccardo: I reread this. The following text is just paraphrasing the previous text. So I commented it out.
%\doubted{
% In other words, if the dynamic scaler was enabled, the FPU output would first be re-scaled to its floating-point representation, then parsed, multiplied by $96$, and cast again to \texttt{uint16}, introducing conversion overhead and potential loss of precision. %loss. 
% We chose to disable it, and let the Host pre-scale the input features, and the Zipper directly apply the $\times 96$ conversion to the final \texttt{ap\_fixed<32,16>} prediction before casting to \texttt{uint16}, to minimize intermediate conversions.
%}.

\subsection{%Motivation: The 
Transition from Batch to Streaming I/O
}\label{sec:streaming}

%To enable real-time responsiveness in stochastic environments, we propose a \bem{continuous streaming architecture} that replaces traditional batch-loading I/O.
The hardware architecture of $\pname$ %organization described above 
is \bem{motivated} by the need to accommodate processing jobs as they arrive (streaming), %replace traditional batch-loading I/O with a 
%continuous streaming jobs, %design, 
to enable real-time responsiveness in HPC %stochastic 
environments.
Prior architectures, including %the %baseline 
%original 
$\stannic$, %architecture \cite{stannic_2026}, 
relied heavily on batch-loading incoming jobs %I/O 
between the Host CPU and the Scheduling FPGA. %$\schedfpga$. 
% The Host CPU had to accumulate %rigid 
% batches of incoming jobs before executing a bulk transfer to the hardware. 
While this strategy was sufficient for virtual simulations, it introduced %severe 
%practical 
limitations, %that 
hindering its real-world usability. %deployment. %in real-world, live scenarios. 
\emph{First}, it does not reflect realistic %how 
online scheduling %works, 
where job arrivals are sporadic and unpredictable. %Having 
To wait until the batch is completely filled would introduce undesired latencies and %potentially 
a degradation %significant drop 
in quality-of-service. \emph{Second}, it is space-inefficient, %because it 
requiring reservations of %idle 
hardware memory on %both the 
Host and the FPGA for data staging. %to stage the batch data. 
%To achieve true SOS, the architecture must be capable of processing jobs instantaneously as they arrive (stream into the scheduler). 

Therefore, we propose a \bem{continuous streaming architecture} built as a co-design of multiple interconnected modules that communicate with each other using a PCIe shared memory (\cf,~\Cref{fig:overall_boostedsosa}). %This architecture is visible in \Cref{fig:overall_boostedsosa}. %The advantage of 
%By 
By introducing %these 
separate \emph{Reader} and \emph{Writer} modules, 
the core $\stannic$ scheduler can be turned into a free-running module. All three elements are connected internally by FIFO streams, with all reads being non-blocking. In doing so, the scheduler's operation is fully decoupled from job arrivals. This, when combined with the host-side job orchestration (\cf,~\Cref{sec:host_side}), allows for asynchronous job dispatching to and from both FPGAs. 
Using this modular architecture, % is that 
we enforce \emph{separation of concerns} (each component executes its own tasks and treats the others as black box layers), and we reduce starvation and overhead. %(each component works in asynchronous mode). 

\section{Experimental Setup}\label{sec:exp_setup}

\noindent \textbf{HPC Datasets}. %To train and evaluate our predictive scheduling architecture, we use historical HPC workload datasets from the \emph{ALCF} public data catalog \cite{PatelEtAl2020,alcf_data}. Specifically, we aggregate the \emph{DIM\_JOB\_COMPOSITE} datasets spanning several years across multiple major ALCF supercomputers, including Cooley, Polaris, Aurora, Theta, Mira, and Intrepid~\cite{ArgonneALCF}. These composite datasets provide detailed historical logs of user job submissions, requested hardware allocations, and final execution outcomes. %Following \Cref{sec:methodology}, we restrict the XGBoost input vector to six explicit submission-time features: \texttt{MACHINE\_NAME}, \texttt{QUEUE\_NAME}, \texttt{NODES\_REQUESTED}, %\texttt{GPU\_CORES} \doubted{FIX THIS}, 
%\texttt{CORES\_REQUESTED}, \texttt{GPUS\_REQUESTED} and \texttt{REQUESTED\_CORE\_HOURS}. We set the target variable as the effective wall-clock runtime in days %(Core-Days)
%, as described in \Cref{sec:data_norm}.
We use the ALCF workload data ($6$ machines) as %our 
primary %case study 
for training and evaluating $\pname$~\cite{PatelEtAl2020,alcf_data}. Specifically, we aggregate the \emph{DIM\_JOB\_COMPOSITE} datasets spanning several years across %multiple 
major ALCF supercomputers~\cite{ArgonneALCF}. %, including Cooley, Polaris, Aurora, Theta, Mira, and Intrepid~\cite{ArgonneALCF}. 
%ALCF jobs' wall-time distribution has a mean of $3.06$ hours, a median of $1.00$ hour, a $75^{th}$ percentile of $2.76$ hours, and a $95^{th}$ percentile of $12.00$ hours. 
%Additionally, to evaluate the generalizability of our architecture
To test whether $\pname$ can generalize beyond the ALCF %environment
workload, we additionally evaluate it on the MIT Supercloud~\cite{MITSamsi} and {UIUC Blue Waters~\cite{UIUC_data} HPC datasets.~\Cref{tab:bucket_sizes} shows the wall-time distribution of each dataset, demonstrating %can be found in , and is shared to illustrate 
the large tail present across each of these datasets.
%\doubted{ref}
%MIT Supercloud jobs' wall-time distribution has a mean of $3.97$ hours, a median of $0.67$ hour, a $75^{th}$ percentile of $2.67$ hours, and a $95^{th}$ percentile of $15.87$ hours, while UIUC Blue Waters jobs' wall-time distribution has a mean of $4.91$ hours, a median of $0.83$ hour, a $75^{th}$ percentile of $4.79$ hours, and a $95^{th}$ percentile of $23.53$ hours.
%We select 
These datasets %because they 
represent independent HPC environments with substantially different schedulers, hardware configurations, and workload distributions.  %MIT Supercloud provides {\scshape Slurm} traces containing a substantial fraction of GPU and AI/ML workload, and Blue Waters provides {\scshape Torque} traces from a heterogeneous system. 
Each UIUC Blue %UIUC 
Waters job has %two types of log entries, 
a start record and a completion record. We consider only completion records for our study. For all %three 
datasets, we map their original scheduler fields to %the same %submission-time representation. 
a common schema. Following~\Cref{sec:methodology}, we %restrict %the XGBoost input vector to six explicit 
%base our analysis on 
use six submission-time features for analysis: \texttt{NODE\_TYPE} and \texttt{QUEUE\_NAME} (categorical features that identify the type of compute node and submission queue, respectively), \texttt{NODES\_REQUESTED}, %\texttt{GPU\_CORES} \doubted{FIX THIS}, 
\texttt{CORES\_REQUESTED}, \texttt{GPUS\_REQUESTED} and \texttt{REQUESTED\_CORE\_HOURS} (%four 
numerical resource-request features that represent the requested node, CPU-core, GPU, and aggregate core-hour allocations, respectively). We set the prediction target %variable 
as the effective wall-clock runtime in days %(Core-Days)
%, as described in \Cref{sec:data_norm}.
($\cf$, \Cref{sec:data_norm}).

\smallskip

\noindent \bem{Remark}. For ALCF, \texttt{NODE\_TYPE} identifies the name of the machine~\cite{PatelEtAl2020,alcf_data} ($\eg$, Cooley, Polaris, Aurora, Theta, Mira, and Intrepid). For MIT Supercloud, it identifies the processor type%on which the job is executed
~\cite{MITSamsi} ($\eg$, Intel\textsuperscript{\textregistered} Xeon Gold 6248, Intel\textsuperscript{\textregistered} Xeon Platinum 8260). For UIUC Blue Waters, it identifies the compute node type~\cite{UIUC_data} ($\eg$, XE: CPU-only, XK: GPU-accelerated). 

\smallskip

\noindent \textbf{Hardware Configuration}. We use Intel\textsuperscript{\textregistered} Xeon W5-3433 CPU (16 cores, 32 threads, 2.0 GHz) with 512 GB DDR5 RAM to handle all Host-side operations. %, including job orchestration, feature scaling, quantization, and online model retraining. 
For acceleration, we deploy two PCIe-connected AMD\textsuperscript{\textregistered} Xilinx\textsuperscript{\textregistered} Alveo U55C FPGAs \cite{u55c}, %connected to the Host via PCIe, 
one each for the \emph{Inference FPGA} and  \emph{Scheduling %Kernel 
FPGA}. 

\smallskip

\noindent \textbf{Baselines}. 
% \begin{comment}
% \doubted{To isolate the performance benefits of our hardware-ML co-design, we compare $\pname$ against  %against 
% a software-only prediction baseline (SW XGBoost).  This baseline executes the XGBoost inference %of XGBoost 
% \bem{entirely on the Host CPU} and leverages Advanced Vector Extensions (AVX) to maximize SIMD parallelism~\cite{intelAVX}. This comparison allows us to quantify the latency and throughput improvements achieved by spatial hardware acceleration.}
% \end{comment}
To evaluate %our 
XGBoost models in the HPC context, we establish several baselines. \bem{First}, we define a \emph{Human Estimation}, %baseline, 
representing conventional scheduler behavior relying %on user expectations. 
on the user-estimated execution times %estimated by users prior to 
at submission. %It is quantified by 
We quantify it %ting the
by reconstructing requested Core-Days (requested wall-time multiplied by the requested core count). 
\bem{Second}, we compare XGBoost against a %diverse 
set of ML predictors: Random Forest (RF \cite{Breiman2001}), AdaBoost (Ada \cite{FreundSchapire1997}), %K-Nearest Neighbors (KNN \cite{CoverHart1967}), 
and Linear Regression (LR \cite{MontgomeryPeckVining2012}). 
For the tree-based ensembles, we define static \emph{CORE} baselines using 64 and 128 estimators to evaluate baseline accuracy and to serve as the foundation for testing our additive training policy. 
\bem{Third}, %following prior work \cite{Tsafrir2007}, 
we implement a \emph{Historical Average} baseline~\cite{Tsafrir2007} that predicts runtime as the mean \texttt{USED\_CORE\_HOURS} of past jobs with the same \texttt{MACHINE\_NAME}, \texttt{QUEUE\_NAME}, and requested number of nodes. 
\bem{Finally}, %to isolate the performance benefits of our architecture, 
we compare $\pname$ against a software-only ML-Scheduling baseline to isolate its performance benefits. This baseline executes both the XGBoost inference and the SOS algorithm entirely on the Host CPU and leverages Intel\textsuperscript{\textregistered} Advanced Vector Extensions (AVX) to maximize SIMD parallelism~\cite{intelAVX}. This comparison allows us to quantify the latency and throughput improvements of spatial, application-specific, hardware acceleration.

\smallskip

\noindent \textbf{Evaluation Metrics}. We use MAE %Mean Absolute Error (MAE) 
%as the %our primary evaluation 
%metric 
to test the predictors' EPT prediction accuracy with respect to RPT. %, defined as: 
% For a set of
% $N_J$ evaluated jobs, MAE is defined as:
% $
% \mathrm{MAE}
%     =
%     \frac{1}{N_J}
%     \sum_{j=1}^{N_J}
%     \left|
%         J_j.\hat{\epsilon}_{i_j}
%         -
%         p_{i_j}^{J_j}
%     \right|,
% $
% where $J_j.\hat{\epsilon}_{i_j}$ is the predicted EPT of job $J_j$
% on its assigned machine $M_{i_j}$, and $p_{i_j}^{J_j}$ is the
% RPT ($\cf$, \Cref{sec:background_definitions}).
We prefer MAE over Root Mean Square Error (RMSE) because HPC job runtimes are often heavy-tailed and include extreme outliers. %, which 
RMSE can magnify such outliers due to its squared penalty~\cite{Hodson2022}. %In addition, 
Additionally, RMSE can be more affected by a few large deviations, causing models with similar average error magnitudes to %may still 
receive %very 
different RMSE scores. For our setting, MAE %thus %offers 
provides a more stable and interpretable comparison across predictors \cite{Hodson2022}. To quantify the computational efficiency and resource overhead of the $\pname$ %hardware-ML co-design %of $\pname$ 
against the AVX baseline, we measure \bem{end-to-end scheduling latency in seconds}, %(total execution time in seconds), 
\bem{system throughput} (jobs processed/second), and \bem{Host-side memory consumption} (peak Resident Set Size, or RSS).%\ar{Was there something about MAE being a better test across different models? That feels like the most important part no?}

\section{Experimental Results}\label{sec:exp_results}

\begin{table}
\centering
\caption{Bucket sizes, job counts, wall-time distributions for the datasets of \Cref{sec:exp_setup}, for the additive training policy. {\em Bucket Size}: number of jobs used for additive updates. {\em Raw Jobs}: number of jobs before preprocessing ($\cf$, \Cref{sec:methodology}). {\em Retained Jobs}: the number of jobs remaining after preprocessing. %Distribution values in hours.}
{\em Mean}, {\em Median}, $75^{th}\%$ / $95^{th}\%$: job wall-times in hours.}
\label{tab:bucket_sizes}

\begin{tabularx}{\columnwidth}{lYYc}
\toprule

\rowcolor{blue!15}
{\textbf{Dataset}} &
{\textbf{Bucket Size}} &
{\textbf{Raw Jobs}} &
{\textbf{Retained Jobs}}
\\

% & & & & \textbf{Mean} & \textbf{Median} & \textbf{$75^{th}\%$} & \textbf{$95^{th}\%$}\\
\midrule

\rowcolor{blue!4}
ALCF
    & 45,000  & 4,003,802 & 3,978,553\\

\rowcolor{blue!8}
MIT Supercloud
    & 125,000 & 395,914   & 309,037\\

\rowcolor{blue!12}
UIUC Blue Waters
    & 35,000  & 8,750,131 & 5,379,998\\

\bottomrule
\end{tabularx}
\begin{tabularx}{\columnwidth}{lYYYYl}
\toprule

\rowcolor{blue!15} 
 &
\textbf{Mean} & 
\textbf{Median} & 
\textbf{$75^{th}\%$} & 
\textbf{$95^{th}\%$} & {\bf Scheduler}\\
\midrule

\rowcolor{blue!4}
ALCF
    & 3.06  & 1.00 & 2.76 & 12.00 & {\scshape Slurm}\\

\rowcolor{blue!8}
MIT Supercloud
    & 3.97 & 0.67 & 2.67 & 15.87 & {\scshape Slurm}\\

\rowcolor{blue!12}
UIUC Blue Waters
    & 4.91  & 0.83 & 4.79 & 23.53 & {\scshape Torque}\\

\bottomrule
\end{tabularx}

\vspace{-4mm}
\end{table}

In this section, we analyze various aspects of $\pname$ and compare its performance against various baselines. ALCF is our primary case study, for which cross-validation selects a $45,000$ job bucket size for ADD ($\cf$, \Cref{sec:async_xgboost}). %To test the generalizability of $\pname$, we use MIT Supercloud and UIUC Bluewaters, for which we repeat the bucket size cross validation evaluation ($\cf$, \Cref{sec:methodology}), which reveals an optimal bucket size of 125,000 jobs for MIT SuperCloud and 35,000 for UIUC Blue Waters. The 3.57$\times$
%To evaluate generalizability, 
We repeat the same cross-validation on MIT SuperCloud and UIUC Blue Waters, with a job bucket size of %obtaining 
$125,000$ and $35,000$ jobs, respectively %, respectively,
%as shown in 
(\cf,~\Cref{tab:bucket_sizes}).
%The difference is consistent with the shorter wall-clock duration of MIT jobs. Their mean and median wall-times of 3.97 and 0.67 hours, respectively, are 9.1\% and 19.6\% lower than Blue Waters (4.91 and 0.83 hours), respectively, and this gap becomes more pronounced in the $75^{th}$ and $95^{th}$ percentiles, where MIT runtimes are 42.4\% and 32.5\% lower (p75: 2.76 vs. 4.79 hours; p95: 15.87 vs. 23.53 hours), respectively. %Thus, under comparable job-flow conditions.
\begin{table}[t]
\centering
\caption{Runtime MAE (Core-Days), reported as
mean $\pm$ standard deviation across 5 folds. Lower is better.
Bold denotes the best result for each dataset. Green cells highlight tree-based models and KNN, which consistently achieve lower MAE. Red cells indicate historical average, linear regression, and human estimation, with higher MAE. XGB = XGBoost, RF = Random Forest, Ada = AdaBoost, LR = Linear Regression, Hist. Avg. = Historical Average; 64 and 128 denote the number of trees.}
\label{tab:model-comparison-all}

\scriptsize
\setlength{\tabcolsep}{2.5pt}
\renewcommand{\arraystretch}{0.96}

\resizebox{\columnwidth}{!}{
\begin{tabular}{lccc}
\toprule
\textbf{Predictor}
& \textbf{ALCF}
& \textbf{MIT Supercloud}
& \textbf{UIUC Blue Waters} \\
\midrule

XGB (128)
& \cellcolor{green!15}468.22 $\pm$ 3.50
& \cellcolor{green!15}\textbf{1.77 $\pm$ 0.12}
& \cellcolor{green!15}48.29 $\pm$ 1.61 \\

XGB (64)
& \cellcolor{green!15}474.66 $\pm$ 3.80
& \cellcolor{green!15}1.79 $\pm$ 0.12
& \cellcolor{green!15}50.42 $\pm$ 1.63 \\

RF (128)
& \cellcolor{green!15}\textbf{439.04 $\pm$ 1.11}
& \cellcolor{green!15}2.24 $\pm$ 0.10
& \cellcolor{green!15}34.81 $\pm$ 0.91 \\

RF (64)
& \cellcolor{green!15}439.09 $\pm$ 1.29
& \cellcolor{green!15}2.24 $\pm$ 0.10
& \cellcolor{green!15}\textbf{34.78 $\pm$ 0.92} \\

Ada (128)
& \cellcolor{green!15}457.83 $\pm$ 5.78
& \cellcolor{green!15}2.69 $\pm$ 0.09
& \cellcolor{green!15}65.05 $\pm$ 3.36 \\

Ada (64)
& \cellcolor{green!15}457.83 $\pm$ 6.02
& \cellcolor{green!15}2.63 $\pm$ 0.09
& \cellcolor{green!15}65.05 $\pm$ 3.36 \\

%KNN
%& \cellcolor{green!15}468.27 $\pm$ 7.89
%& \cellcolor{green!15}2.47 $\pm$ 0.30
%& \cellcolor{green!15}38.60 $\pm$ 1.03 \\

Hist. Avg.
& \cellcolor{red!12}884.89 $\pm$ 8.43
& \cellcolor{red!12}3.02 $\pm$ 0.09
& \cellcolor{red!12}78.91 $\pm$ 1.09 \\

LR
& \cellcolor{red!12}1,238.33 $\pm$ 10.11
& \cellcolor{red!12}3.56 $\pm$ 0.24
& \cellcolor{red!12}75.41 $\pm$ 4.02 \\

Human
& \cellcolor{red!12}1,295.33 $\pm$ 1,368
& \cellcolor{red!12}2,376.36 $\pm$ 47.48
& \cellcolor{red!12}112.39 $\pm$ 2.41 \\

\bottomrule
\end{tabular}
}
\end{table}

\subsection{Appropriateness of a ML Predictor for Scheduling}\label{sec:predictor_requriement}
%USE: XGBOOST, RAND FOREST, DECISION TREE, ADA BOOST
%and maybe linear regression or non-tree predictors (just like 2 to make results generalizable)
\emph{To validate the effectiveness of ML predictors in the HPC context}, we 
study whether they can provide the low-variance EPTs required for the $\stannic$ %systolic 
scheduling engine. \Cref{tab:model-comparison-all} compares the MAE %of several predictive models -- XGBoost (XGB), Random Forest (RF), AdaBoost (Ada), K-Nearest Neighbors (KNN), and Linear Regression (LR) -- against a baseline of Human Variance. 
for the predictors defined in~\Cref{sec:exp_setup}, %for 
the {\em human estimation}, and %for 
the {\em historical average} on the three datasets.
%The baseline represents the default behavior in conventional schedulers, which rely heavily on static user expectations. %For this evaluation, the human baseline 
%and is quantified as the requested compute core-days (calculated as the requested wall-time multiplied by the requested core count) compared against the actual consumed core-days. 
As shown, human estimates are highly unreliable, exhibiting an enormous MAE, especially for ALCF and MIT Supercloud (mean MAE: $1,295$ core-days for ALCF, $2,376.36$ for MIT Supercloud) and \bem{extreme variance} (up to $1,368$ for ALCF). %Similarly, LR and historical average perform poorly (mean MAE: $1,238$ for LR, $884$ for historical average). %This suggest that the the relationship between submission-time telemetry features and actual job duration is \bem{non-linear}. 
Relative to the best ML predictor (XGBoost for MIT Supercloud, Random Forest for the other two datasets), the historical average incurs $2.02\times$, $1.71\times$, and $2.27\times$ higher MAE on ALCF, MIT Supercloud, and UIUC Blue Waters, respectively, while LR is $2.82\times$, $2.01\times$, and $2.17\times$ higher.
This suggests that job duration cannot be accurately captured by either linear relationships between submission-time features or historical averages, and motivates the use of models that capture richer non-linear feature interactions.
In contrast, tree-based models drastically reduce both the absolute prediction error and the variance across the cross-validation folds. In particular, their variance is minimal, making them \bem{highly robust predictors in the HPC domain.} %While Random Forest (RF) achieves the lowest overall error (mean MAE $\approx$ $439$, $128$ trees), XGBoost provides a comparably high leap in accuracy (mean MAE $\approx$ $468$, $128$ trees). 
While Random Forest (RF) achieves the lowest MAE on ALCF and UIUC Blue Waters, XGBoost performs best on MIT Supercloud with an MAE of 1.77 Core-Days and provides a comparable improvement in prediction accuracy on the other two datasets.
Coupled with the ADD policy improvements seen in \Cref{sec:xgboost_rationale}, XGBoost is a good candidate for inferencing in this context. 
\bem{These results show that replacing user-provided estimates with tree-based ML models yields superior accuracy in forecasting actual job runtimes. When %By dividing the 
predicted core-days is divided by the requested core count, the models provide the hardware accelerator with reliable low-variance EPTs needed for competitively bounded scheduling.}

\definecolor{addred}{HTML}{D62728}
\definecolor{coregray}{HTML}{555555}

\begin{figure*}
    \centering

    \begin{subfigure}{0.32\textwidth}
        \centering
        \includegraphics[width=\linewidth]{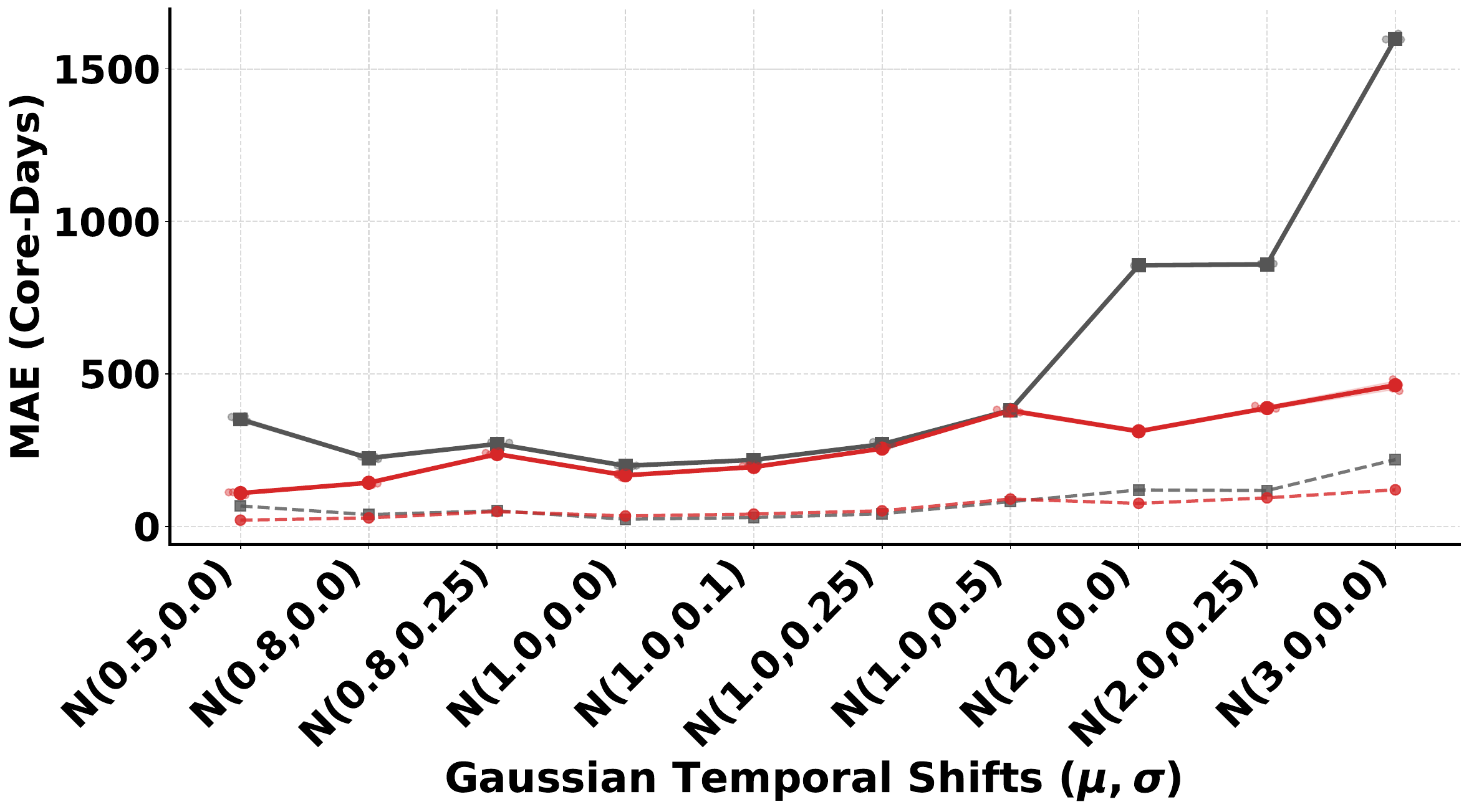}
        \vspace{-2mm}
        \caption{ALCF}
        \label{fig:fig7_a}
    \end{subfigure}
    \hfill
    \begin{subfigure}{0.32\textwidth}
        \centering
        \includegraphics[width=\linewidth]{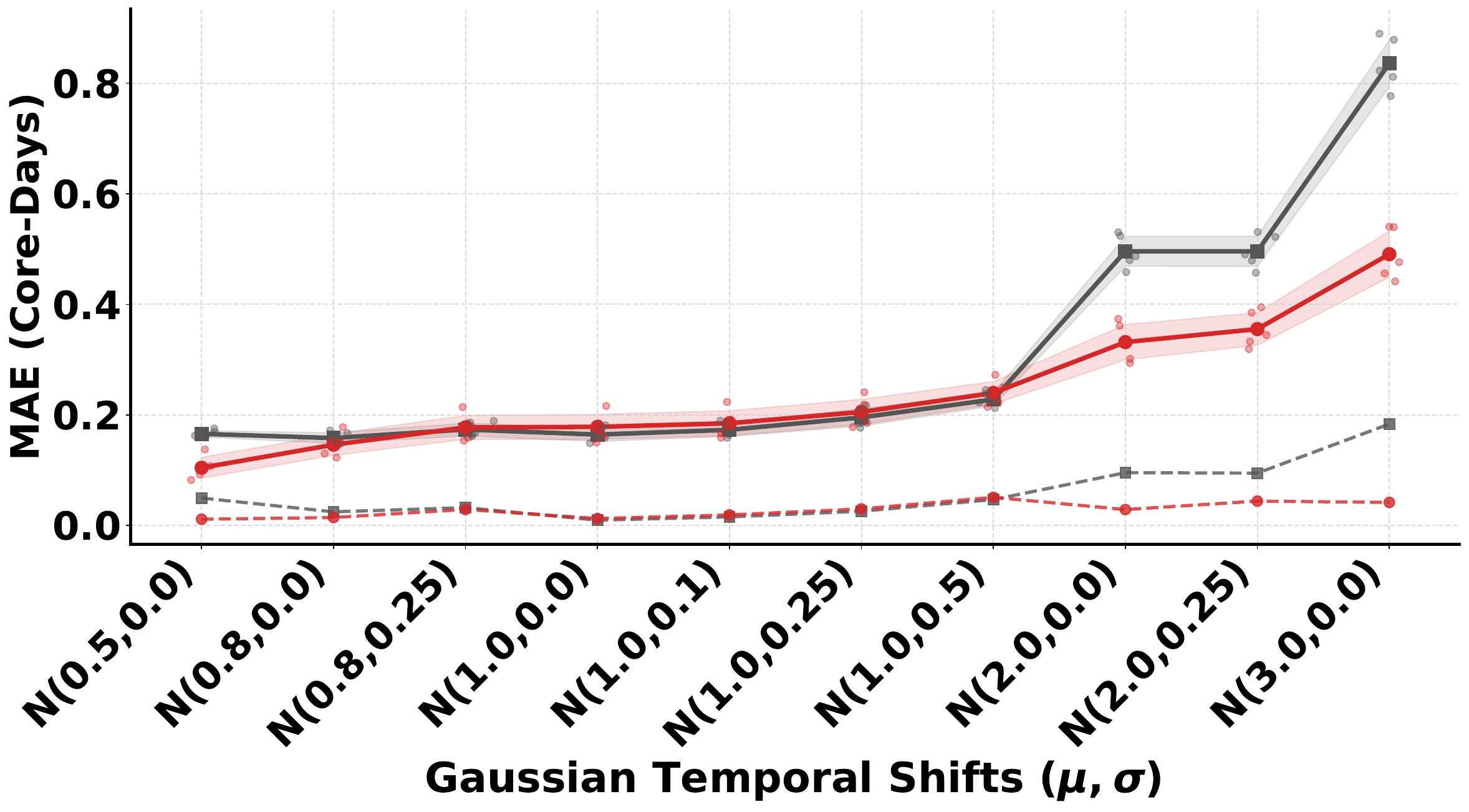}
        \vspace{-2mm}
        \caption{MIT Supercloud}
        \label{fig:fig7_b}
    \end{subfigure}
    \hfill
    \begin{subfigure}{0.32\textwidth}
        \centering
        \includegraphics[width=\linewidth]{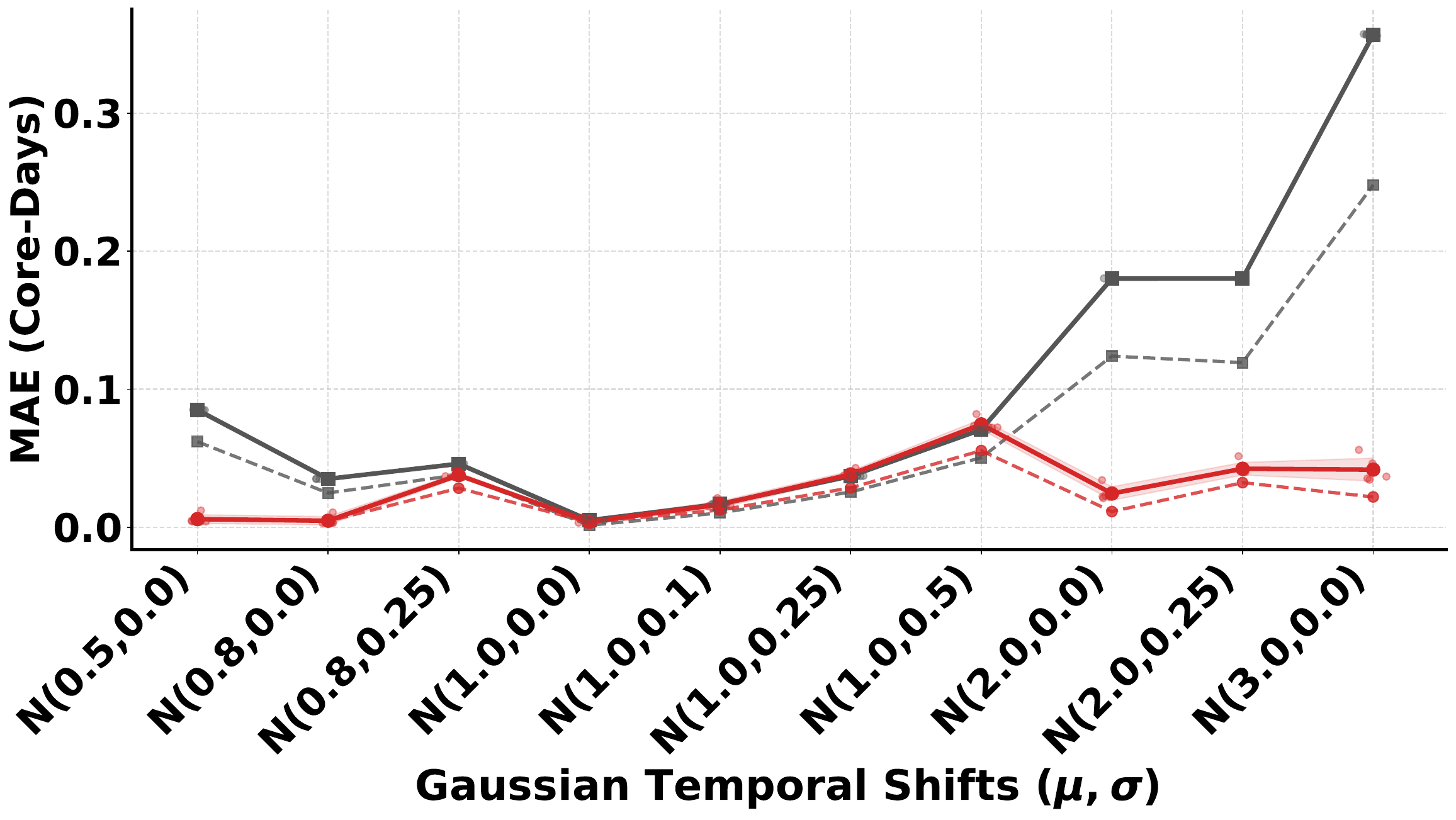}
        \vspace{-2mm}
        \caption{UIUC Blue Waters}
        \label{fig:fig7_c}
    \end{subfigure}

    \vspace{2mm}

    % Shared legend
    \begin{tikzpicture}
        % CORE MAE
        \draw[coregray!80, line width=1.5pt] (0,0) -- (0.9,0);
        \node[rectangle, fill=coregray!80, inner sep=2.3pt] at (0.45,0) {};
        \node[anchor=west] at (1.05,0) {\small\textbf{CORE (MAE)}};

        % ADD MAE
        \draw[addred!80, line width=1.5pt] (3.5,0) -- (4.4,0);
        \node[circle, fill=addred!80, inner sep=2.3pt] at (3.95,0) {};
        \node[anchor=west] at (4.55,0) {\small\textbf{ADD (MAE)}};

        % CORE MAE p95
        \draw[coregray!65, dashed, line width=1.2pt] (6.9,0) -- (7.8,0);
        \node[rectangle, fill=coregray!65, inner sep=1.8pt] at (7.35,0) {};
        \node[anchor=west] at (7.95,0) {\small\textbf{CORE (MAE p95)}};

        % ADD MAE p95
        \draw[addred!65, dashed, line width=1.2pt] (11.0,0) -- (11.9,0);
        \node[circle, fill=addred!65, inner sep=1.8pt] at (11.45,0) {};
        \node[anchor=west] at (12.05,0) {\small\textbf{ADD (MAE p95)}};
    \end{tikzpicture}

    \vspace{-1mm}

    \caption{Effectiveness of ADD under workload distribution shifts on ALCF, MIT SuperCloud, and UIUC Blue Waters. The three charts show the MAE for ADD (red, dashed for MAE p95) and CORE (grey, dashed for MAE p95) under varying Gaussian temporal shifts ${\cal N}(\mu, \sigma)$ to the resource demands (used Core-Days) within the test set. Lower MAE and MAE p95 are better.}
    \label{fig:exp_res_additive_shifts}
    \vspace{-4mm}
\end{figure*}
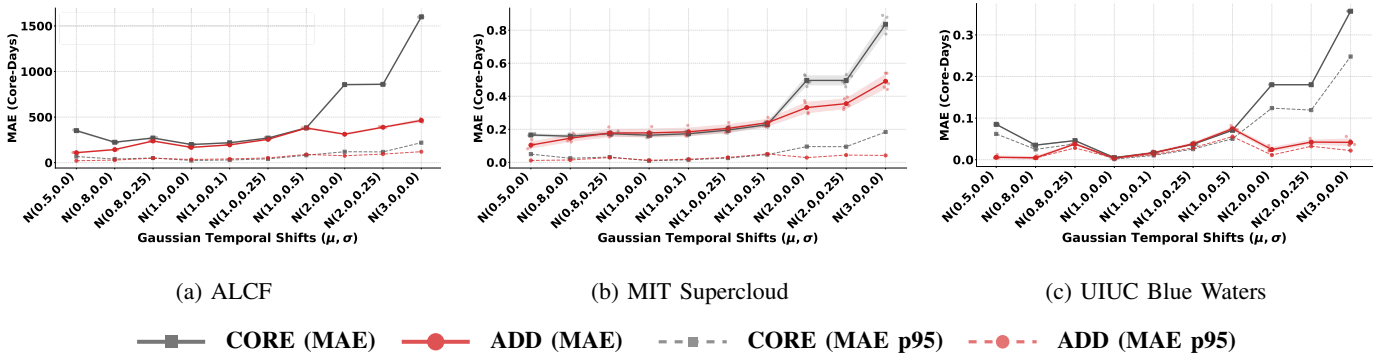

\definecolor{lightgreen}{RGB}{225,245,225}
\begin{table}
  \centering
  \caption{Relative change of ADD with respect to CORE
  for mean MAE and MAE p95 (in \%). Negative values indicate
  lower MAE (more negative values = better). The largest reduction in each metric/dataset pair is
  highlighted in %light 
  green.}
  \label{tab:add-generalization}
  \small
  \setlength{\tabcolsep}{5pt}
  \renewcommand{\arraystretch}{1.08}

  \resizebox{\columnwidth}{!}{
  \begin{tabular}{lrrrrrr}
    \toprule
    & \multicolumn{2}{c}{\textbf{ALCF}} &
      \multicolumn{2}{c}{\textbf{MIT Supercloud}} &
      \multicolumn{2}{c}{\textbf{UIUC Blue Waters}} \\
    \cmidrule(lr){2-3}\cmidrule(lr){4-5}\cmidrule(lr){6-7}
    \textbf{Shift} & MAE & MAE p95 & MAE & MAE p95 & MAE & MAE p95 \\
    \midrule

    $\mathcal{N}(0.5,0.0)$
      & -68.8 & \cellcolor{green!50}\textbf{-69.0}
      & -37.0 & \cellcolor{green!50}\textbf{-77.3}
      & \cellcolor{green!50}\textbf{-93.1} & -91.0 \\

    $\mathcal{N}(0.8,0.0)$
      & -36.0 & -27.7
      & -7.6 & -41.7
      & -86.6 & -82.3 \\

    $\mathcal{N}(0.8,0.25)$
      & -12.3 & -5.2
      & +2.2 & -10.4
      & -17.9 & -25.6 \\

    \addlinespace[2pt]

    $\mathcal{N}(1.0,0.0)$
      & -15.7 & +45.0
      & +8.5 & +30.2
      & -21.2 & +148.5 \\

    $\mathcal{N}(1.0,0.1)$
      & -10.6 & +38.5
      & +6.8 & +23.5
      & -2.1 & +20.3 \\

    $\mathcal{N}(1.0,0.25)$
      & -5.3 & +23.9
      & +5.0 & +15.9
      & +3.4 & +11.7 \\

    $\mathcal{N}(1.0,0.5)$
      & -0.3 & +10.7
      & +5.2 & +8.6
      & +5.4 & +10.2 \\

    \addlinespace[2pt]

    $\mathcal{N}(2.0,0.0)$
      & -63.5 & -36.7
      & -33.1 & -69.9
      & -86.4 & -90.8 \\

    $\mathcal{N}(2.0,0.25)$
      & -54.8 & -20.4
      & -28.4 & -53.5
      & -76.5 & -72.9 \\

    $\mathcal{N}(3.0,0.0)$
      & \cellcolor{green!50}\textbf{-71.0} & -45.1
      & \cellcolor{green!50}\textbf{-41.3} & -77.2
      & -88.3 & \cellcolor{green!50}\textbf{-91.1} \\

    \midrule
    \rowcolor{black!5}
    \textbf{Mean}
      & \textbf{-33.8} & \textbf{-8.6}
      & \textbf{-12.0} & \textbf{-25.2}
      & \textbf{-46.3} & \textbf{-26.3} \\
    
    \rowcolor{black!5}
    \textbf{Median}
      & \textbf{-25.9} & \textbf{-12.8}
      & \textbf{-2.7} & \textbf{-26.1}
      & \textbf{-48.9} & \textbf{-49.3} \\
    
    \rowcolor{black!5}
    \textbf{Improved Shifts}
      & \textbf{10/10} & \textbf{6/10}
      & \textbf{5/10}  & \textbf{6/10}
      & \textbf{8/10}  & \textbf{6/10} \\
    
    \bottomrule
  \end{tabular}
  }
  \vspace{-4mm}
\end{table}

\subsection{Effectiveness of Additive Scheduling Policy}\label{sec:additive}

%%%%HOW DO THESE BUCKETS PERFORM? IN THESE SETTINGS:
%%% With runtime variation 
%%% With switching policy
%%% Comparison of software vs. hardware

%%% SOftware only %% How often -- not too often -- not too slow not to lose temporal information

%check th replacement policy, like how often should we call that, if we slide windows too little at a time, do we learn too little? we wanna measure how often we sample the 45k bucket

%How do I simulate the jobs occurring over time in a manner that is similar to real hw setting??

%mention policy replacement in methodogy as well -> every T jobs we sample a bucket, just in one/two lines

%put figure 4 + switching policy
\emph{To evaluate the robustness and adaptability of the ADD policy %additive scheduling policy (ADD) 
against stationary baselines (CORE $64$/$128$) when subjected to workload distribution shifts}, we apply varying Gaussian temporal shifts %, denoted as 
${\cal N}(\mu, \sigma)$ to the resource demands (used Core-Days) within the test set. {\em This simulates period of higher or lower resource contention within the machines}. Because CORE ($64$) and CORE ($128$) %reveal 
show negligible performance differences, we average them into a single baseline (CORE). \Cref{fig:exp_res_additive_shifts} shows the MAE, MAE p95 and their respective distributions for ADD and CORE on the three datasets, after a $5$-fold cross validation. \Cref{tab:add-generalization} %compares the models using MAE and 95$^{th}$-percentile MAE (MAE p95). %This dual-metric approach is critical for HPC workloads, as it disentangles the outsized impact of heavy-tailed, massive tasks ("whales") from the scheduling accuracy of typical, high-frequency tasks. -> THIS GOES IN EXP SETUP
reports the relative change in MAE and MAE p95 of ADD with respect to CORE, together with the mean and median relative changes across all shifts and the number of shifts in which ADD improves over CORE, for all three datasets.
%for clarity. 
On ALCF, our primary use case, both CORE and ADD models have comparable performance under mild temporal shifts ($\eg$, ${\cal N}(1.0, 0.0)$ to $\mathcal{N}(1.0, 0.25)$). However, under more pronounced distribution shifts on either tail -- whether toward lighter queue periods ($\eg$, ${\cal N}(0.5,0.0)$, ${\cal N}(0.8,0.0)$) or toward highly concurrent node-heavy periods ($\eg$, ${\cal N}(2.0,0.25)$, ${\cal N}(3.0,0.0)$) -- ADD consistently outperforms CORE across both metrics. Under the extreme ${\cal N}(3.0,0.0)$ shift, CORE %suffers severe degradation, with MAE rising to approximately 1600 Core-Days. In contrast, ADD is able to adapt to the shifted setting, keeping Total MAE near 450 Core-Days %(\bem{70\% reduction) 
degrades sharply to an MAE of $\sim 1600$ Core-Days, whereas ADD adapts and keeps MAE near $\sim 450$ Core-Days %, corresponding to a reduction of over 
(reduction of over $\mathbf{70\%}$, \Cref{fig:fig7_a})
%and reducing 
and reduces MAE p95 to nearly half of the baseline ($120$ vs. $219$ Core-Days). Across all ten shifts, ADD reduces MAE in $10$/$10$ settings, with mean and median reductions of $33.8$\% and $25.9$\%, respectively. This improvement is significant under a Wilcoxon signed-rank test ($p=0.002$, \cite{wilcoxon1945}), with a $95$\% confidence interval (CI) of [-$54.1$,-$13.6$]\% ($\cf$,~\Cref{tab:add-generalization}).

To test generalizability beyond ALCF, we repeat the same evaluation on MIT Supercloud and UIUC Blue Waters ($\cf$, \Cref{fig:fig7_b,fig:fig7_c,tab:add-generalization}). \Cref{fig:fig7_b,fig:fig7_c} show the same overall trend observed on ALCF, as ADD provides larger MAE reductions with respect to CORE toward the tails of the workload distribution. On Blue Waters, ADD reduces MAE in $8$/$10$ settings, with mean and median reductions of $46.3$\% and $48.9$\% (the improvement is significant ($p=0.020$, $95$\% CI [-$77.1$,-$15.6$]\%). In the remaining two, more central settings ($\ie$, $\mathcal{N}(1.0,0.25)$ and $\mathcal{N}(1.0,0.50)$), CORE is slightly more accurate (by 4.4\% more on average), as these settings exhibit limited workload shift. On MIT Supercloud, ADD reduces MAE in $5$/$10$ settings, with a mean reduction of $12.0$\% and a median reduction of 2.7\% ($p=0.275$, 95\% CI $[-26.6,2.7]\%$). \bem{These results show that ADD preserves comparable accuracy under mild changes while providing the largest and most consistent benefits when workloads shift substantially, with the strongest evidence on ALCF and Blue Waters.}

%\bem{These results suggest that stationary models may remain competitive under stable conditions, but deteriorate under workload shifts. A continuous additive updating policy helps contain large errors without sacrificing nominal accuracy.}

%\input{pic/additive_results_shifts}

\subsection{Real Chronological Workload-Drift Validation}\label{sec:chrono_drift}

%\begin{figure}
%    \centering
%    \includegraphics[scale=0.25]{pic/Chronological_Exp.pdf}
%    \caption{MAE of Core-Days across chronological windows of the ALCF workloads.}
%    \label{fig:chronological_windows}
%    %\vspace{-5mm}
%\end{figure}

\definecolor{addred}{HTML}{D62728}
\definecolor{recentblue}{HTML}{1F77B4}
\definecolor{coregray}{HTML}{555555}

\begin{figure*}
    \centering

    \begin{subfigure}{0.32\textwidth}
        \centering
        \includegraphics[width=\linewidth]{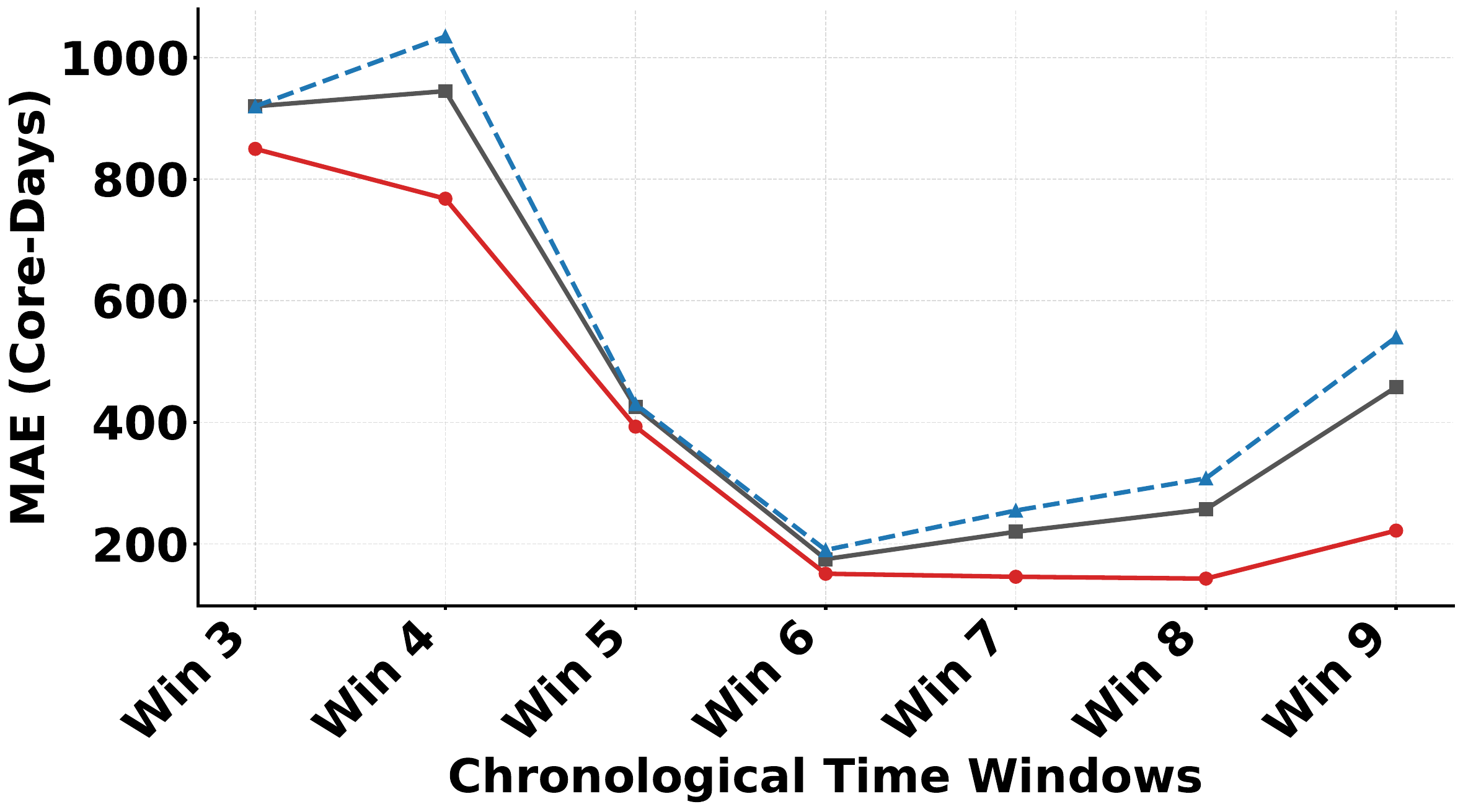}
        \vspace{-2mm}
        \caption{ALCF}
        \label{fig:fig8_a}
    \end{subfigure}
    \hfill
    \begin{subfigure}{0.32\textwidth}
        \centering
        \includegraphics[width=\linewidth]{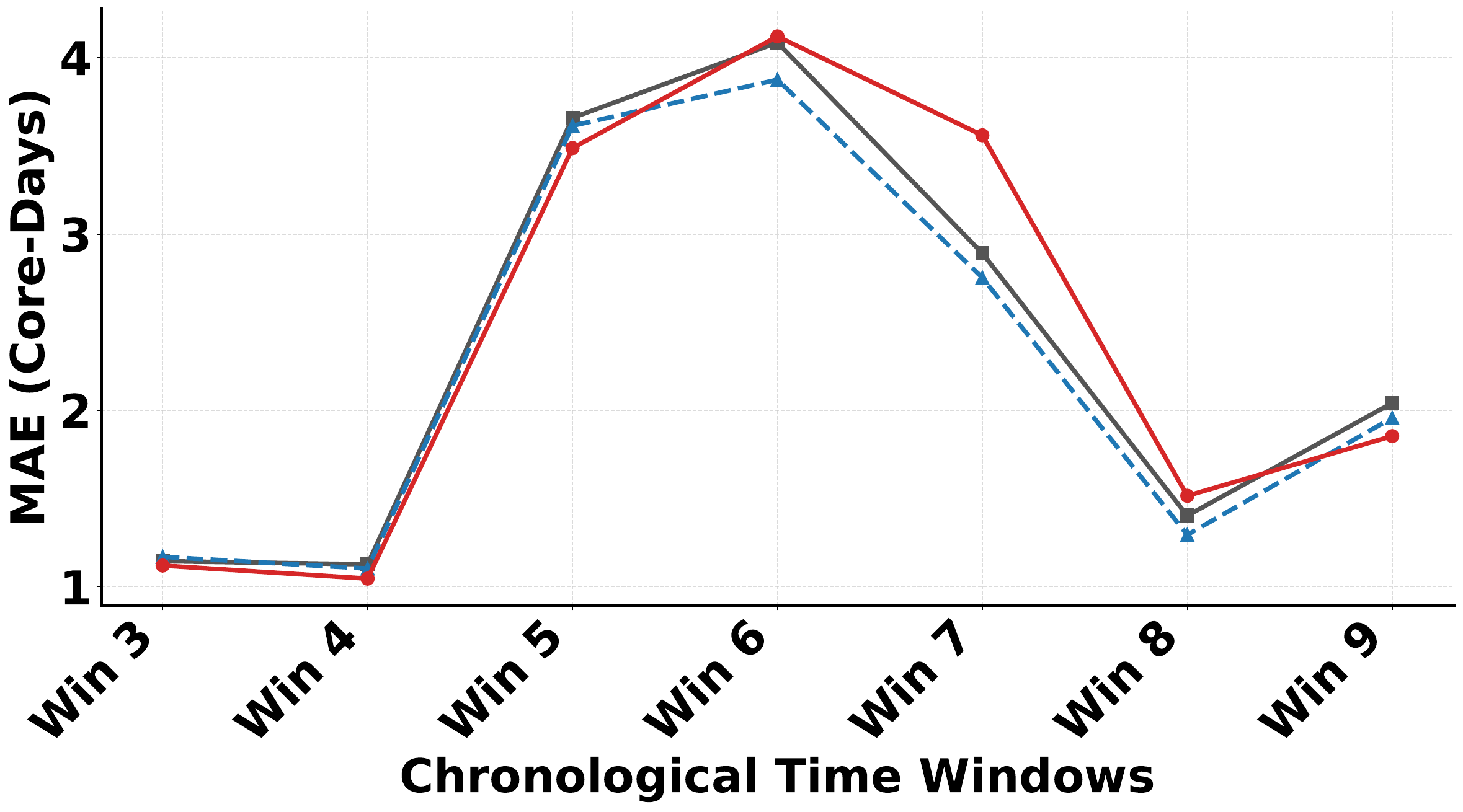}
        \vspace{-2mm}
        \caption{MIT Supercloud}
        \label{fig:fig8_b}
    \end{subfigure}
    \hfill
    \begin{subfigure}{0.32\textwidth}
        \centering
        \includegraphics[width=\linewidth]{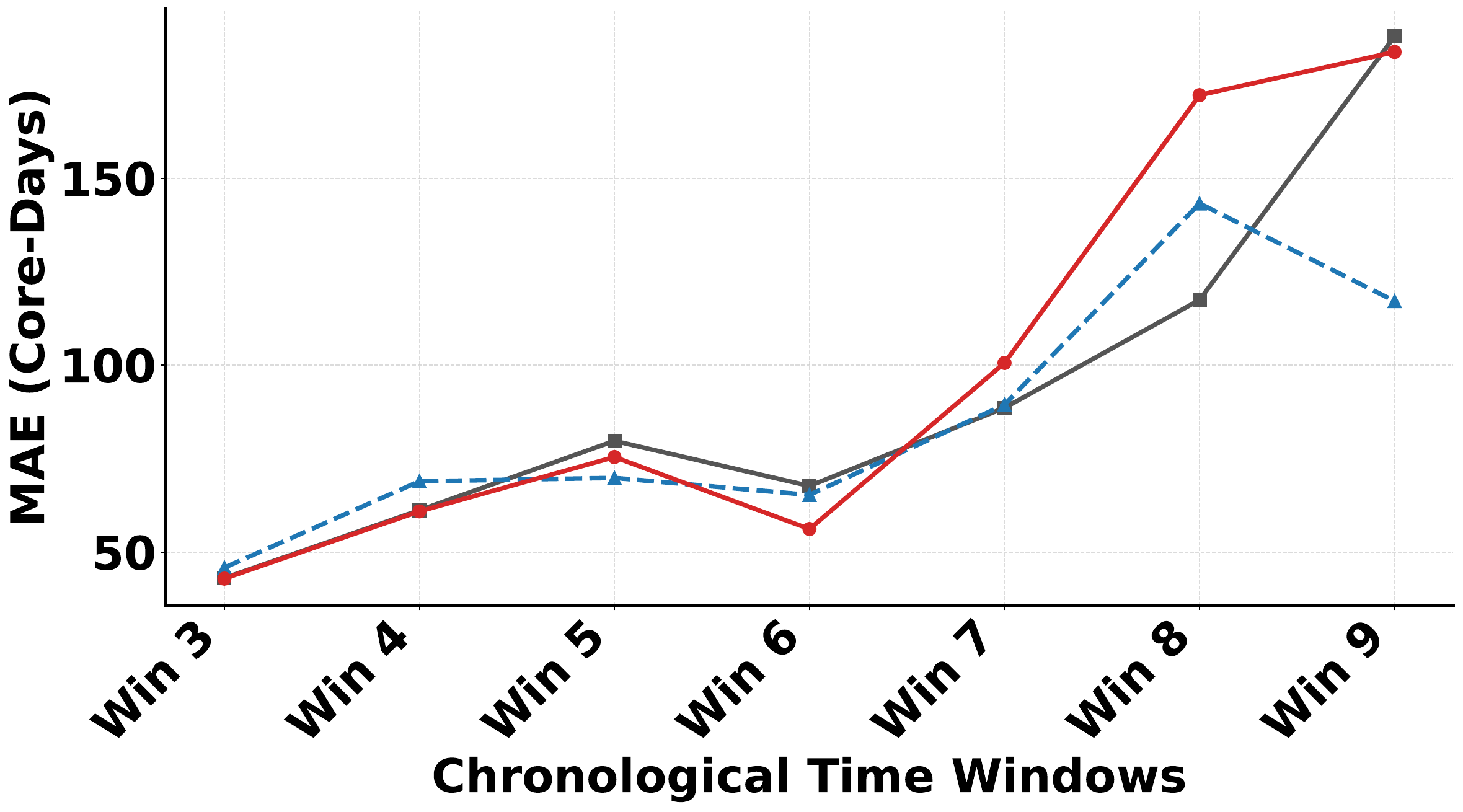}
        \vspace{-2mm}
        \caption{UIUC Blue Waters}
        \label{fig:fig8_c}
    \end{subfigure}

    \vspace{2mm}

   \begin{tikzpicture}

    % CORE
    \draw[coregray, line width=1.5pt]
        (0,0) -- (1.15,0);
    \node[
        rectangle,
        fill=coregray,
        inner sep=2.2pt
    ] at (0.575,0) {};
    \node[anchor=west, font=\small\bfseries]
        at (1.35,0) {CORE (MAE)};

    % RECENT-ONLY
    \draw[
        recentblue,
        line width=1.5pt,
        dash pattern=on 7pt off 2pt on 1.5pt off 2pt
    ] (4.6,0) -- (5.75,0);
    \fill[recentblue]
        (5.175,0.12) --
        (5.055,-0.12) --
        (5.295,-0.12) -- cycle;
    \node[anchor=west, font=\small\bfseries]
        at (5.95,0) {RECENT-ONLY (MAE)};

    % ADD
    \draw[addred, line width=1.5pt]
        (10.2,0) -- (11.35,0);
    \node[
        circle,
        fill=addred,
        inner sep=2.2pt
    ] at (10.775,0) {};
    \node[anchor=west, font=\small\bfseries]
        at (11.55,0) {ADD (MAE)};

\end{tikzpicture}

    \vspace{-1mm}

    \caption{Chronological workload-drift validation of predictors on the three datasets. We report MAE of the CORE, ADD, and RECENT-ONLY policies across ten sequential workload windows. Overall, ADD generally improves upon baselines across datasets, with temporary degradation only under abrupt workload shifts.}
    \label{fig:chrono_all}
    \vspace{-5mm}
\end{figure*}
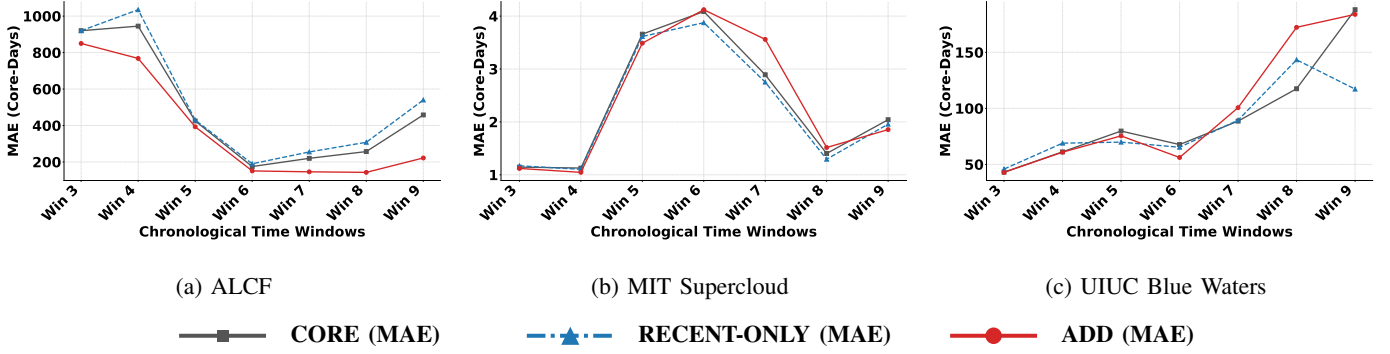

\emph{To further demonstrate the ADD policy's resilience to temporal data shifts, we partition the %ALCF 
three datasets chronologically by queue submission timestamp into ten sequential windows following the Starburst methodology}~\cite{Luo2024Starburst}. 
While \Cref{fig:exp_res_additive_shifts} demonstrates the ADD policy's resilience to synthetic Gaussian temporal shifts, using this alternative methodology allows us to additionally validate its behavior under real chronological workload evolution. %Following the Starburst methodology \cite{Luo2024Starburst}, we partition the ALCF dataset chronologically by queue submission timestamp into ten sequential windows. 
CORE is trained on the first three windows; starting from the fourth, we iteratively update ADD using window $T-1$ (\cf,~\Cref{sec:bifurcated}) %following the methodology of
and evaluate CORE and ADD on the unseen window $T$. We additionally evaluate RECENT-ONLY, a baseline that discards the existing model and retrains from scratch on only the most recent window, representing a naive alternative to ADD.

~\Cref{fig:fig8_a} reports, for the ALCF dataset, the per-window MAE for CORE and ADD, and available data points for RECENT-ONLY. ADD consistently matches or outperforms CORE from Window $5$ onward, and the gap widens as the chronological horizon extends: by Window 9, ADD attains an MAE of approximately $222$ Core-Days, compared to roughly $460$ for CORE and roughly $539$ for RECENT-ONLY. RECENT-ONLY's degradation illustrates a %specific 
failure mode of naive retraining: by discarding all prior model structure at each update, it overfits to each window's local statistics and loses the stable long-term baseline that the frozen CORE component of ADD retains. We assess the statistical significance of the CORE/ADD gap using a Wilcoxon signed-rank test %\cite{wilcoxon1945} 
over the per-window MAE values, obtaining $p = 0.05$. %Together with the synthetic drift results of \Cref{fig:exp_res_additive_shifts}, this confirms that the additive policy's robustness generalizes from controlled perturbations to the chronological drift patterns present in real HPC job traces.

\Cref{fig:fig8_b,fig:fig8_c} 
mirror the %report the
chronological evaluation %MAE of CORE, ADD, and RECENT-ONLY evaluation 
on MIT Supercloud and UIUC
Blue Waters. %, respectively, mirroring the ALCF evaluation in \Cref{sec:chrono_drift}. 
ADD generally matches or improves upon CORE
and returns to the CORE trend by Window~9. The only clear consecutive losses
occur in Windows~7 and~8 for both datasets (their identical indices are purely
coincidental as the two figures report results for the two different datasets). On Blue Waters, Windows~7--8 show a sharp increase in job length. Relative to the three windows in which we train CORE, median job wall-time for windows~7--8 increases by $10.4\times$ and $9.2\times$, respectively (p95 job wall-time increases by $15.4\times$ and $26.2\times$). MIT Supercloud shows the opposite trend. From Window~6 to~7, median, and p95 job wall-times decrease by 40.3\%, and 30.4\%, respectively. In both settings, {\em MAE increases because the unseen window present a significant distribution shift}, different from what is learned by the models. %up to that time. 
%ADD %is affected 
%degrades more because its newest trees encode residual corrections learned from the immediately preceding window. After a sudden shift, these corrections become outdated and are still added to CORE, whose prediction does not already capture the newer distribution, amplifying the error. The frozen CORE model does not carry this additional bias and therefore degrades less. %Once the shifted window is observed, resetting the adaptive component removes the outdated corrections and allows ADD to return to the CORE trend by Window~9. \bem{These results suggest that ADD remains effective across heterogeneous HPC workloads, with degradation limited to abrupt distribution shifts. ADD recovers once the adaptive component is reset.}
We observe that under these abrupt distribution shifts, ADD can temporarily overcorrect because its most recent trees continue to apply residual corrections learned from the preceding workload distribution. We refer to this phenomenon as \emph{inertia}, in the sense that the model continues to reflect previous workload distributions, until enough data from the new distribution is observed and the additive component is updated. As shown in \Cref{fig:fig8_b,fig:fig8_c}, ADD returns toward the CORE trend after the model is updated and again provides the stability observed under more gradual shifts. This inertia under abrupt changes is a limitation of the current ADD training policy. \bem{These results suggest that ADD remains effective across heterogeneous HPC workloads, with degradation limited to abrupt distribution shifts. ADD recovers once the adaptive component is reset.}

\subsection{Cost of Batch I/O Under Realistic Arrival Rates}\label{sec:batch-vs-stream}

\begin{figure}
    \centering
    \includegraphics[scale=0.15]{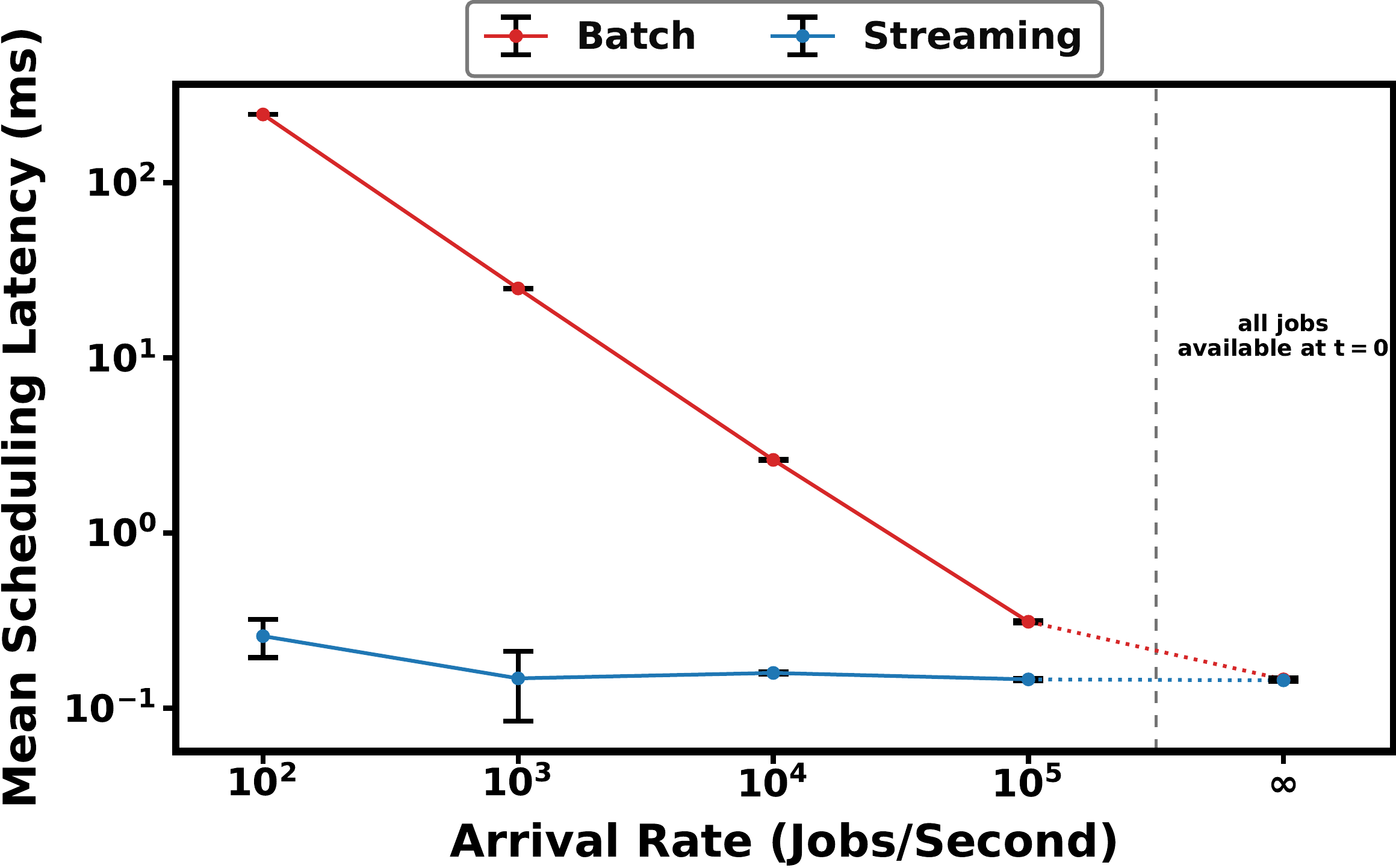}
    \caption{Mean per-job scheduling latency versus arrival rate on a 2,000-job ALCF sample, 15 runs per point; error bars give the standard deviation of per-run means. The rightmost point is unpaced, with the entire workload available at $t = 0$.}
    \label{fig:latency-vs-rate}
    \vspace{-6mm}
\end{figure}

\emph{To test the per-job scheduling latency impact of the job writing interface (\Cref{sec:streaming}),} we schedule an identical 2,000-job ALCF sample through both I/O paths on the same Scheduling FPGA, releasing jobs to the Host at a controlled arrival rate and timing on the Host clock. Each run is repeated 15 times. In \Cref{fig:latency-vs-rate}, we report per-job scheduling latency, from a job becoming visible to the Host until its scheduling decision returns, separating the wait for a batch of fifty jobs to fill from the transfer and decision itself. Accumulation of jobs governs the batch path entirely: it falls by a factor of ten for every tenfold increase in arrival rate, from $245.0$\,ms at $10^{2}$ jobs/s to $2.4$\,ms at $10^{4}$, and constitutes $99.8\%$ of total latency at the lowest rate. The streaming path, which dispatches each job on arrival, stays below $4\,\mu$s throughout. With the slower arrival rates ($10^2$ and $10^3$ jobs/sec), there is a noticeable variance in streaming's scheduling latency, due to the corresponding writing thread sleeping during a job's arrival (\Cref{sec:host_side}). Mean scheduling latency therefore differs by $951\times$ at $10^{2}$ jobs/s, narrowing to $2.1\times$ at $10^{5}$.

The difference is significant at every paced arrival rate under a Wilcoxon rank-sum test over the per-run means ($p<10^{-7}$, \cite{wilcoxon1945}), with no overlap between the two distributions at any feasible rate. When the entire workload is available at $t = 0$, (\ie, offline scheduling), the two I/O paths are statistically indistinguishable ($p=0.126$, with $95\%\ CI$ on the relative difference $[-0.2,2.6]\%$). In this case, batch scheduling could indeed offer higher throughput, but having all jobs available in advance %achieving such high arrival rates 
is simply unrealistic in an online HPC scheduling context, \bem{showing that in the context of online scheduling, a streaming I/O is crucial to reduce per-job scheduling latency.}

\subsection{Comparison of End-to-End Performance}\label{sec:runtime_memory}
%\doubted{Adam: put averages + std dev for table 3}
%% Resource Utilization
%% Memory footprint
%% Speedup software AVX vs. STANNIC based
\emph{To quantify %the computational efficiency and resource overhead of the proposed hardware-ML co-design}
$\pname$'s end-to-end performance}, we compare the execution metrics of $\pname$ against an optimized software-only AVX baseline running on the Host CPU. We compare runs across various virtual schedule $V_i$ job depth configurations \cite{stannic_2026}, running each configuration 15 times. As the focus of this experiment is to test the \emph{architecture's} throughput and memory utilization, we limit these experimetns to a singular dataset. We use a random 45k job sample from the ALCF dataset as the input and perform repeat measurements.

The %most 
primary %significant 
advantage of $\pname$ is a significant %dramatic 
reduction in end-to-end inference and scheduling latency. As shown in ~\Cref{fig:exp_res_time_speedup}, across varying job depths (ranging from $10$ to $40$), the AVX software baseline requires between $445.63\pm2.18$ and $451.91\pm12.61$ seconds to process the workloads consisting of $45$K jobs. {\em This is due to the irregular traversal patterns of the prediction trees limiting the benefits of AVX SIMD instructions}. In contrast, \textsc{BoostedSOSA} processes the exact same workloads in $26.29\pm0.11$ to $26.45\pm0.10$ seconds, respectively. %\bem{This is a consistent latency reduction of over 10$\times$ (averaging approximately 13$\times$ acceleration).}
This represents an average speedup of up to \bem{17$\times$} in total scheduling time compared to the AVX baseline. %\ar{We also 

%{\em Note that \pname\ maintains a consistent runtime regardless of $V_i$'s depth, while the AVX baseline trends upwards, %though it is a noisy trend, 
%as the AVX implementation is %greatly 
%susceptible to other concurrently running background host processes}. %(such as background OS processes) 
%running simultaneously}. 
%This time reduction directly increases system throughput by an %equivalent 
%order of magnitude, as shown in ~\Cref{tab:throughput_memory}. The AVX implementation reaches at most 129.22 jobs/sec and it drops to 86.16 jobs/sec at high schedule depths. Meanwhile, $\pname$ consistently processes between 1,217.47 jobs/sec and 1,508.57 jobs/sec across all job depths. 

Note that both implementations maintain a consistent mean runtime regardless of $V_i$'s depth: across the depths evaluated, mean total time varies by $1.4\%$ for the AVX baseline and by $0.6\%$ for $\pname$. The two differ instead in run-to-run stability. $\pname$'s coefficient of variation stays between $0.28\%$ and $0.49\%$ at every depth, whereas the AVX baseline ranges from $0.39\%$ up to $2.79\%$, as the AVX implementation is susceptible to other concurrently running background host processes. {\em This time reduction directly increases system throughput by more than an order of magnitud}e, as shown in \Cref{tab:throughput_memory}. The AVX implementation sustains between $99.65\pm2.66$ and $100.98\pm0.50$ jobs/sec across all job depths. Meanwhile, $\pname$ consistently processes between $1,701.54\pm6.33$ and $1,711.74\pm7.04$ jobs/sec.

\begin{figure}
    \centering
    \includegraphics[scale=0.15]{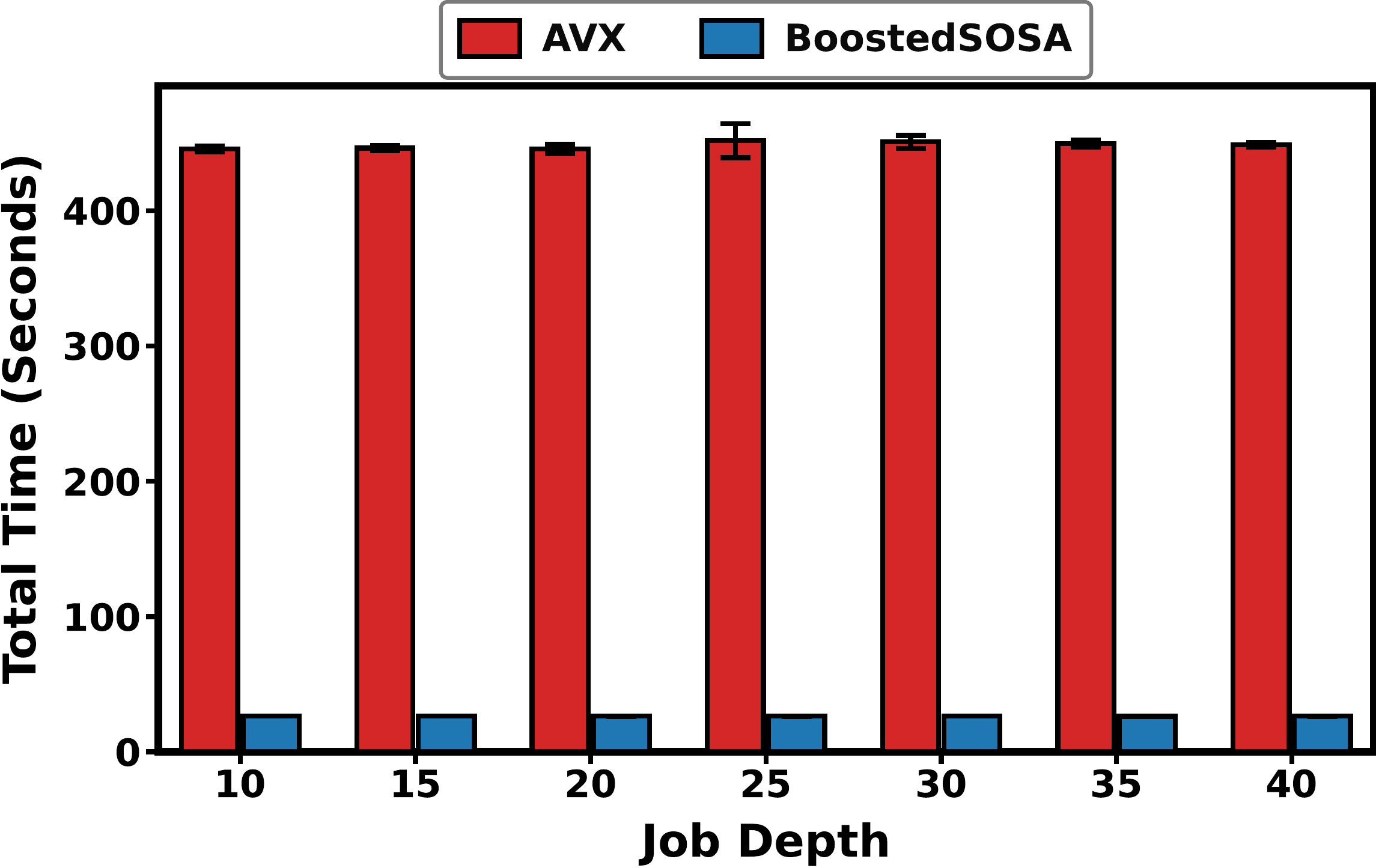}
    \caption{Execution Time vs. Job Depth. $\pname$ maintains a nearly stable latency ($\sim$26.3s on average), %across all schedule depths, 
    achieving an average $17\times$  speedup over the AVX baseline.}
    \label{fig:exp_res_time_speedup}
    \vspace{-3mm}
\end{figure}

It is worth clarifying the source of this speedup within the pipeline. The PCIe-based streaming integration between the Host and the two FPGAs does not itself contribute to the speedup; if anything, it introduces a small communication overhead at each stage of the pipeline. In \cite{stannic_2026}, it is shown that a software AVX implementation of the SOS algorithm alone slightly outperforms $\stannic$ for sufficiently small virtual-schedule configurations, including the six-machine configurations used in this evaluation. Taken together, these results indicate that the speedup reported here is attributable primarily to FPGA-accelerated XGBoost inference rather than to the scheduling hardware or the streaming integration layer, neither of which would by itself explain the observed gains.

However, the primary limitation of $\pname$ is the increase in Host-side memory use. The AVX baseline is more memory-efficient, reaching a peak %Resident Set Size (RSS) 
RSS of %only 
$\sim 35$ MB ($35{,}968$ KB). In contrast, $\pname$ reaches a peak RSS of $\sim 253$ MB ($253{,}296$ KB). This corresponds to about a $7\times$ increase, driven by the fixed allocations necessary to manage several asynchronous PCIe data streams, maintain the FPGAs' relative modules, and operate the multi-threaded Host-side orchestration. %\bem{The memory trade-off of $\pname$ is highly favorable. While our architecture requires a $\sim$250 MB memory footprint, this overhead is easily manageable by the high memory typical of modern HPC host nodes. In exchange, achieving an average 13$\times$ speedup in total scheduling time demonstrates the operational advantage of $\pname$ against in-software baselines.}
\bem{The memory trade-off of $\pname$ is favorable. Our architecture uses about 250 MB of host memory, a fraction of main memory of %which is easily supported by 
modern HPC nodes. In return, it delivers an average 17$\times$ speedup in total scheduling throughput over software baselines enabling near-real time scheduling}.

\begin{table}
\centering
\caption{Comparison of throughput, coefficient of variance ($CV$), and memory footprint. {\bf Job Depth}: Depth of $V_i$.}
\label{tab:throughput_memory}
\resizebox{\columnwidth}{!}{%
\begin{tabular}{c|ccc|ccc|c}
\toprule
\textbf{Job} & \multicolumn{3}{c|}{\textbf{AVX}} & \multicolumn{3}{c|}{\textbf{$\pname$}} & \multirow{2}{*}{\textbf{Speed-Up}}\\
\cline{2-7}
{\bf Depth} & \textbf{Jobs} & \textbf{$CV$} & \textbf{Peak Mem } & \textbf{Jobs} & \textbf{$CV$}  & \textbf{Peak Mem } \\
& {\bf /seconds} & & (KB) & {\bf /seconds} & & (KB) & \\
\midrule
10 & 100.98 & $\pm$ 0.49\% & 32,824 & 1,703.94 & $\pm$ 0.36\% & 250,832 & \textbf{16.87$\times$}\\
15 & 100.79 & $\pm$ 0.45\% & 35,216 & 1,705.93 & $\pm$ 0.28\% & 252,640 & \textbf{16.93$\times$}\\
20 & 100.98 & $\pm$ 0.79\% & 33,096 & 1,707.54 & $\pm$ 0.36\% & 250,652 & \textbf{16.91$\times$}\\
25 & 99.65 & $\pm$ 2.79\% & 33,132 & 1,708.33 & $\pm$ 0.44\% & 251,508 & \cellcolor{green!50}\textbf{17.16$\times$}\\
30 & 99.78 & $\pm$ 1.04\% & 30,824 & 1,701.54 & $\pm$ 0.37\% & 250,880 & \textbf{17.05$\times$}\\
35 & 100.04 & $\pm$ 0.61\% & 35,868 & 1,711.74 & $\pm$ 0.41\% & 253,296 & \textbf{17.11$\times$}\\
40 & 100.24 & $\pm$ 0.39\% & 35,968 & 1,706.71 & $\pm$ 0.49\% & 250,968 & \textbf{17.03$\times$}\\
\bottomrule
\end{tabular}%
}
\vspace{-2mm}
\end{table}

\subsection{Resource Utilization and Operating Frequency
}\label{sec:resource_util}

\emph{To assess the hardware cost and scalability of $\pname$}, we report the post-implementation resource utilization of each pipeline module on the target AMD Alveo U55C FPGAs in \Cref{tab:FPGA_util}. Both boards operate at 371.47 MHz. On the Inference FPGA, the FPU and Zipper modules together occupy a small fraction of available resources (at most $4\%$ of LUTs). On the {\em Scheduling FPGA}, the $\stannic$, Reader, and Writer modules similarly occupy a modest footprint (at most $2\%$ of LUTs). DSP slices are unused across all modules because the pipeline performs exclusively fixed-point arithmetic; Vitis HLS accordingly maps all arithmetic onto LUT/FF logic rather than DSP blocks. URAM is likewise unused, as on-chip storage requirements are modest (at most 64 node entries per tree engine \Cref{sec:inf_fpga} and small per job packets in the Zipper, Reader, and Writer Modules).

\begin{table}
\centering
\caption{Post-implementation resource utilization of different sub-modules of $\pname$. Numbers are raw utilization values, and percentages are the proportion of available resources used on the AMD Alveo U55C %target 
boards.}
\label{tab:FPGA_util}
\resizebox{\columnwidth}{!}{%
\begin{tabular}{|c|ccccc|}
\hline
{\bf Module} & {\bf BRAM} & {\bf URAM} & {\bf DSP} & {\bf FF} & {\bf LUT} \\
\hline\hline
Conifer FPU & $8$ ($\sim0\%$) & $0$ ($0\%$) & $0$ ($0\%$) & $73,961$ ($2\%$) & $54,859$ ($4\%$)\\
Zipper & $15$ ($\sim0\%$) & $0$ ($0\%$) & 0 ($0\%$) & $3,144$ ($\sim0\%$) & $3,389$ ($\sim0\%$)\\
$\stannic$ & $4$ ($\sim0\%$) & $0$ ($0\%$) & 0 ($0\%$) & $15,366$ ($\sim0\%$) & $37,219$ ($2\%$)\\
Reader & $15$ ($\sim0\%$) & $0$ ($0\%$) & $0$ ($0\%$)) & $3,295$ ($\sim0\%$) & $2,731$ ($\sim0\%$)\\
Writer & $29$ ($\sim0\%$) & $0$ ($0\%$) & $0$ ($0\%$) & $10,395$ ($\sim0\%$) & $9,119$ ($\sim0\%$)\\
\hline
\end{tabular}%
}
\vspace{-5mm}
\end{table}

Because both pipeline stages are lightly loaded, a single FPGA could in principle host a condensed version of the full $\pname$ design. We retain the split, two-FPGA organization for %three 
two main reasons. {\em First}, it preserves scalability: in \cite{stannic_2026}, the authors show that $\stannic$ can scale to support up to 140 machines in a single configuration, but %that 
at such a configuration, routing requirements become a significant bottleneck. Dedicating a separate FPGA to scheduling keeps sufficient on chip resources available for %this routing headroom available to STANNIC 
$\stannic$ rather than sharing it %device resources 
with inference logic, letting the design scale toward larger machine counts without contention between the two components. {\em Second}, it preserves modularity: while the currently utilized inference method (XGBoost via the {\em Conifer} FPU) fits easily within a single card's resources, splitting the pipeline across two FPGAs keeps the {\em Inference} and {\em Scheduling FPGA}s communicating only through the fixed packet interface described in \Cref{sec:inf}, so alternative inference engines can be substituted on the {\em Inference FPGA} without modifying the {\em Scheduling FPGA}, or vice versa. %{\em Third}, distributing the pipeline across two independently operating boards avoids a single point of failure in the overall system.
\bem{Overall, $\pname$ achieves a low hardware footprint while preserving scalability and modularity across inference and scheduling stages.}

\subsection{ML Predictor Ablation with Random Forest}\label{sec:rf_exp}

\emph{To test the stability of ADD to %the choice of 
the ML predictor}, we substitute XGBoost with Random Forest (RF), and repeat the experiments of \Cref{sec:additive,sec:chrono_drift} for all three datasets. 

\Cref{tab:rf_xgb_delta} reports the relative difference, in percentages, in ADD MAE between Random Forest and XGBoost across the temporal shift settings of \Cref{sec:additive}. Positive values favor XGBoost, while negative values favor Random Forest. Using XGBoost consistently results in lower MAE under pronounced temporal shifts. Across ${\cal N}(0.5,0.0)$, ${\cal N}(2.0,0.0)$, ${\cal N}(2.0,0.25)$, and
${\cal N}(3.0,0.0)$, respectively, it achieves lower MAE in all 12 dataset-shift pairs. 
RF performs better only under milder ALCF shifts, where it achieves lower MAE in six settings (${\cal N}(0.8,0.0)$, ${\cal N}(0.8,0.25)$, ${\cal N}(1.0,0.0)$, ${\cal N}(1.0,0.1)$, ${\cal N}(1.0,0.25)$, and ${\cal N}(1.0,0.5)$). XGBoost outperforms RF across
all ten MIT SuperCloud shifts, and in nine Blue Waters shifts. \bem{Overall, ADD can adapt to both ML predictors, but XGBoost is
more robust to pronounced temporal shifts}.

\Cref{fig:chrono_RF_ALL} shows, for each dataset, the per-window MAE for CORE and ADD, and available data points for RECENT-ONLY, when we substitute XGBoost (whose results are shown in \Cref{fig:chrono_all}), with Random Forest. Replacing XGBoost with Random Forest %(RF) 
does not degrade the
chronological behavior of ADD.  On the contrary, the consecutive
Window~7--8 losses observed with XGBoost on MIT Supercloud and
Blue Waters ($\cf$, \Cref{fig:fig8_b,fig:fig8_c}) disappear with Random Forest. On ALCF, ADD
also remains below CORE from Window~5 onward and reaches 
$150$ Core-Days at Window~9, compared with $318$ for CORE and
$222$ previously obtained with XGBoost ADD. By using Random Forest, however, RECENT-ONLY becomes the model which achieves the lowest MAE in some later windows, such as Window~9 on ALCF and Blue Waters. We attribute this to how Random Forest combines its trees. ADD improves over CORE because part of the trees are trained on the most recent window, and introduces information from the newer workload. However, unlike XGBoost, Random Forest trees are trained independently rather than sequentially ADD therefore does a majority voting among trees trained on the older CORE data with trees trained on the newest window, while RECENT-ONLY does a majority voting only on trees trained on the newest data. Under sufficiently large workload changes, this can favor RECENT-ONLY because all of its trees reflect the current workload, and majority voting reaches a more concistent consensus.
\bem{These results suggest that replacing XGBoost with RF preserves the improvement of ADD over
CORE on the chronological experiment and, on MIT and Blue Waters, it avoids the temporary degradation
observed with XGBoost under abrupt workload changes}.

\begin{table}
    \centering
    \caption{Relative difference in ADD MAE between Random Forest and XGBoost
    across the temporal shift settings of \Cref{sec:additive}. We report
    $\Delta\mathrm{MAE} =
    (\mathrm{MAE}_{\mathrm{RF}}-\mathrm{MAE}_{\mathrm{XGB}})
    /\mathrm{MAE}_{\mathrm{XGB}}\times100$.
    Positive values (green) favor XGBoost, while negative values (red) favor Random Forest.}
    \label{tab:rf_xgb_delta}
    \begin{tabular}{l|ccc}
        \hline
        \textbf{Shift} &
        \textbf{ALCF} &
        \textbf{MIT Supercloud} &
        \textbf{UIUC Blue Waters} \\
        \hline
        ${\cal N}(0.5,0.0)$  & \cellcolor{green!15}$+74.9\%$ & \cellcolor{green!15}$+67.5\%$ & \cellcolor{green!15}$+687.3\%$ \\
        
        ${\cal N}(0.8,0.0)$  & \cellcolor{red!12}$-29.8\%$ & \cellcolor{green!15}$+39.5\%$ & \cellcolor{green!15}$+352.6\%$ \\
        
        ${\cal N}(0.8,0.25)$ & \cellcolor{red!12}$-20.2\%$ & \cellcolor{green!15}$+27.5\%$ & \cellcolor{green!15}$+9.2\%$ \\
        
        ${\cal N}(1.0,0.0)$  & \cellcolor{red!12}$-65.3\%$ & \cellcolor{green!15}$+28.6\%$ & \cellcolor{green!15}$+38.2\%$ \\
        
        ${\cal N}(1.0,0.1)$  & \cellcolor{red!12}$-46.5\%$ & \cellcolor{green!15}$+28.4\%$ & \cellcolor{green!15}$+10.8\%$ \\
        
        ${\cal N}(1.0,0.25)$ & \cellcolor{red!12}$-26.1\%$ & \cellcolor{green!15}$+25.6\%$ & $+0.0\%$ \\
        
        ${\cal N}(1.0,0.5)$  & \cellcolor{red!12}$-12.7\%$ & \cellcolor{green!15}$+21.3\%$ & \cellcolor{red!12}$-3.4\%$ \\
        
        ${\cal N}(2.0,0.0)$  & \cellcolor{green!15}$+43.3\%$ & \cellcolor{green!15}$+43.9\%$ & \cellcolor{green!15}$+252.8\%$ \\
        
        ${\cal N}(2.0,0.25)$ & \cellcolor{green!15}$+16.5\%$ & \cellcolor{green!15}$+34.9\%$ & \cellcolor{green!15}$+107.3\%$ \\
        
        ${\cal N}(3.0,0.0)$  & \cellcolor{green!15}$+82.5\%$ & \cellcolor{green!15}$+49.8\%$ & \cellcolor{green!15}$+304.9\%$ \\
        \hline
    \end{tabular}
    \vspace{-4mm}
\end{table}

%\begin{figure}
%    \centering
%    \includegraphics[scale=0.25]{pic/Chronological_Exp.pdf}
%    \caption{MAE of Core-Days across chronological windows of the ALCF workloads.}
%    \label{fig:chronological_windows}
%    %\vspace{-5mm}
%\end{figure}

\definecolor{addred}{HTML}{D62728}
\definecolor{recentblue}{HTML}{1F77B4}
\definecolor{coregray}{HTML}{555555}

\begin{figure*}
    \centering

    \begin{subfigure}{0.32\textwidth}
        \centering
        \includegraphics[width=\linewidth]{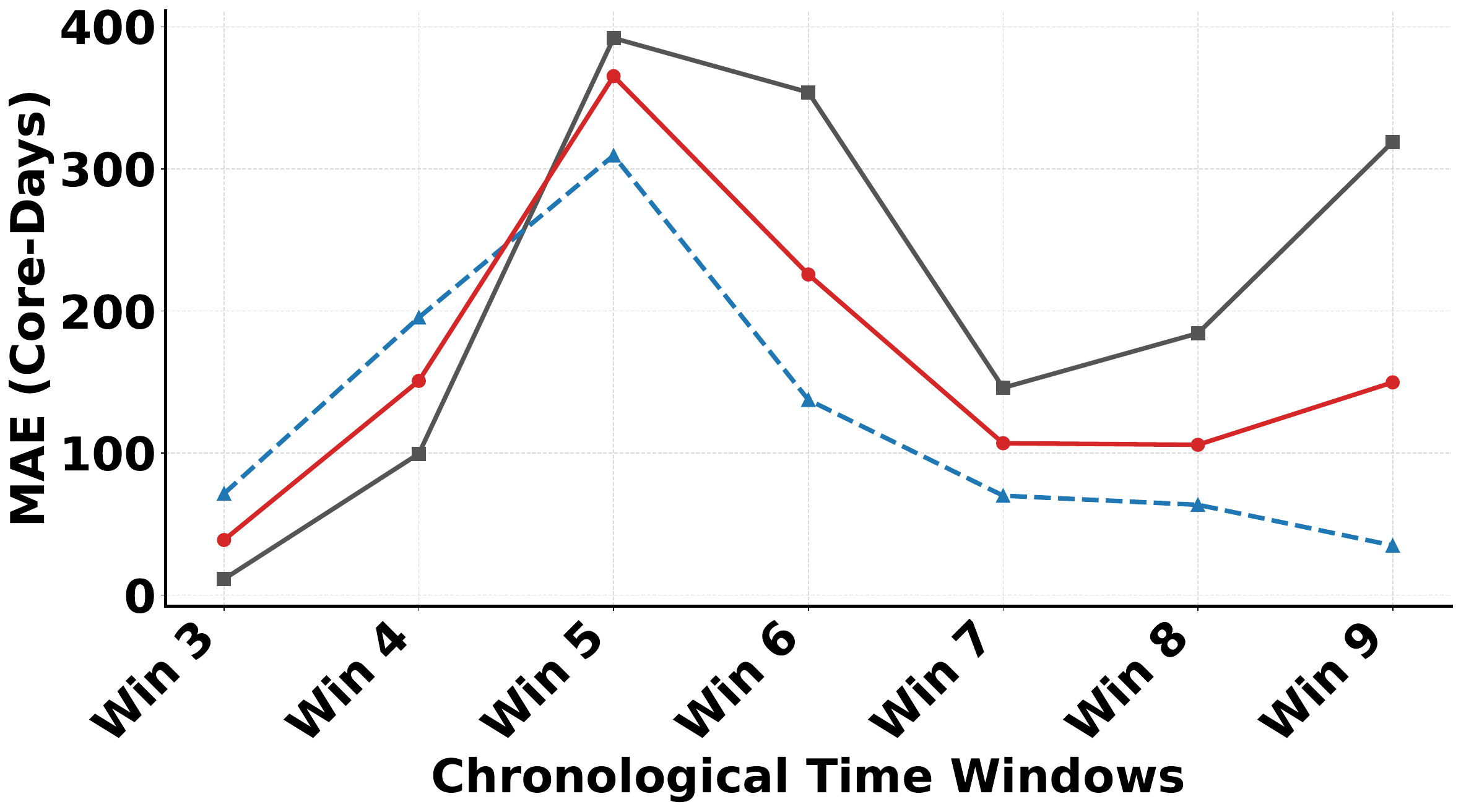}
        \vspace{-2mm}
        \caption{ALCF}
        \label{fig:RF_1}
    \end{subfigure}
    \hfill
    \begin{subfigure}{0.32\textwidth}
        \centering
        \includegraphics[width=\linewidth]{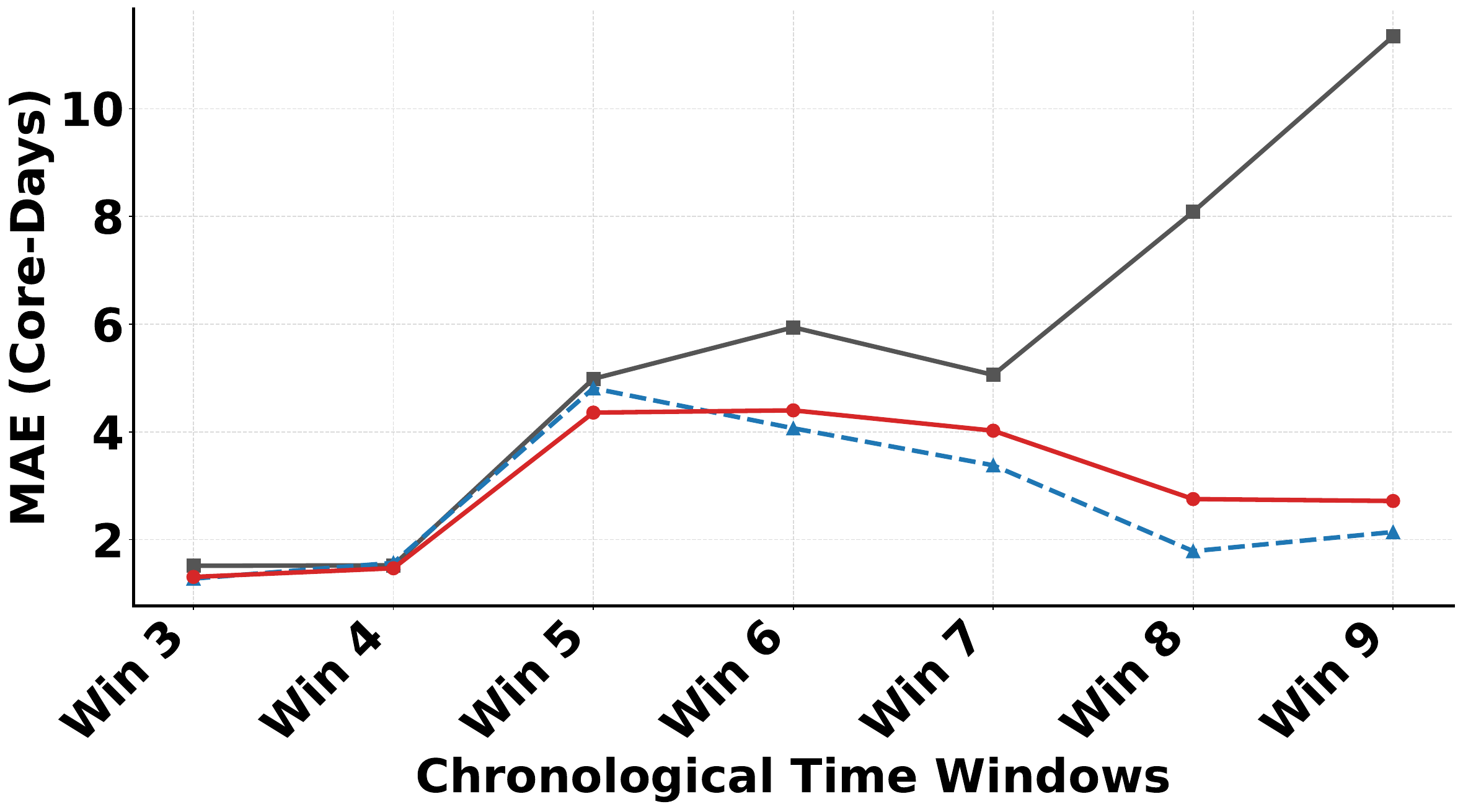}
        \vspace{-2mm}
        \caption{MIT Supercloud}
        \label{fig:RF_2}
    \end{subfigure}
    \hfill
    \begin{subfigure}{0.32\textwidth}
        \centering
        \includegraphics[width=\linewidth]{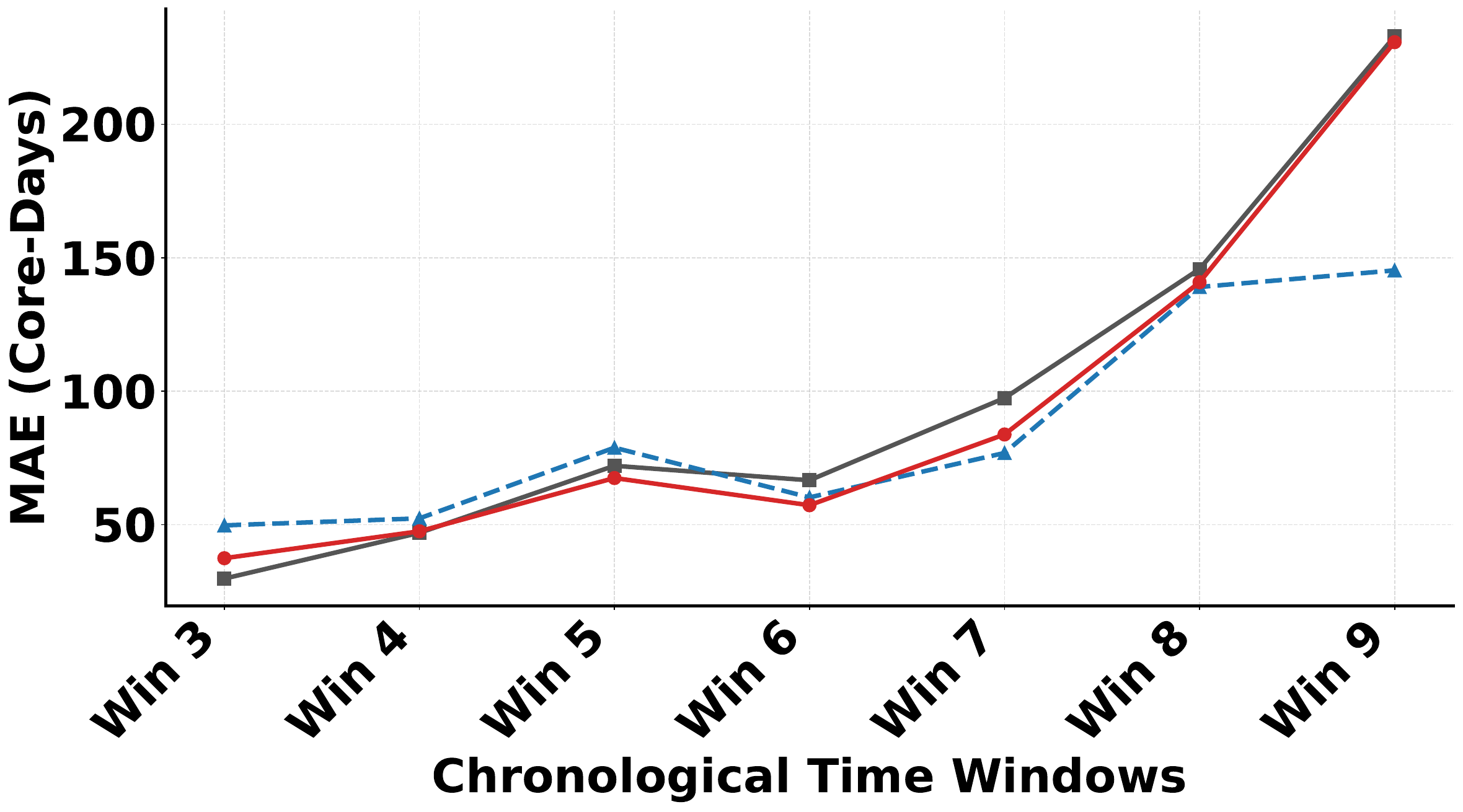}
        \vspace{-2mm}
        \caption{UIUC Blue Waters}
        \label{fig:RF_3}
    \end{subfigure}

    \vspace{2mm}

   \begin{tikzpicture}

    % CORE
    \draw[coregray, line width=1.5pt]
        (0,0) -- (1.15,0);
    \node[
        rectangle,
        fill=coregray,
        inner sep=2.2pt
    ] at (0.575,0) {};
    \node[anchor=west, font=\small\bfseries]
        at (1.35,0) {CORE (MAE)};

    % RECENT-ONLY
    \draw[
        recentblue,
        line width=1.5pt,
        dash pattern=on 7pt off 2pt on 1.5pt off 2pt
    ] (4.6,0) -- (5.75,0);
    \fill[recentblue]
        (5.175,0.12) --
        (5.055,-0.12) --
        (5.295,-0.12) -- cycle;
    \node[anchor=west, font=\small\bfseries]
        at (5.95,0) {RECENT-ONLY (MAE)};

    % ADD
    \draw[addred, line width=1.5pt]
        (10.2,0) -- (11.35,0);
    \node[
        circle,
        fill=addred,
        inner sep=2.2pt
    ] at (10.775,0) {};
    \node[anchor=west, font=\small\bfseries]
        at (11.55,0) {ADD (MAE)};

\end{tikzpicture}

    \vspace{-1mm}

    \caption{Chronological workload-drift validation of predictions using Random Forest on the three datasets. We report the MAE of CORE, ADD, and RECENT-ONLY across ten sequential workload windows. ADD generally preserves or improves upon CORE across the three datasets, showing that ADD is not specific to XGBoost.}
    
    \label{fig:chrono_RF_ALL}
    \vspace{-5mm}
\end{figure*}
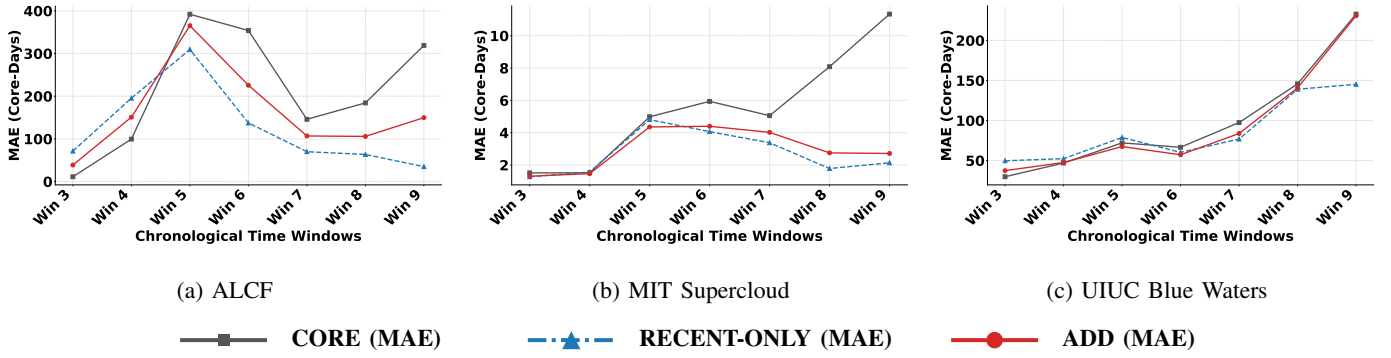
\section{Related Work}\label{sec:related_work}
%\doubted{Riccardo: put paras like in TPAMI}
%\paragraph{The Shift to Data-Centric Scheduling Paradigms} 
Modern scheduling is shifting from heuristic and %operations-research-based 
deterministic methods toward %data-centric, ML-driven paradigms 
paradigms based on data science and Machine Learning\cite{GondhiGupta2017,MaoSchwarzkopfVenkatakrishnanMengAlizadeh2019}. Machine Learning-based scheduling approaches generally follow two categories: \emph{end-to-end generation} and \emph{predictive augmentation}.

\smallskip

\noindent \hypertarget{sec:rel-work-a-a}{\textbf{Approaches Based on End-to-End Generation}}. %\emph{In the category}, H
Heuristic logic is discarded and the scheduling policy is learned directly from data. DeepRM models multi-resource scheduling as a Markov Decision Process with policy-gradient Reinforcement Learning (RL) \cite{MaoAlizadehMenacheKandula2016}, while Decima extends this idea %with GNNs to capture DAG dependencies in data-processing pipelines Decima ù
and leverages Graph Neural Networks (GNNs) to encode DAG dependencies and optimize scheduling for data-processing jobs \cite{MaoSchwarzkopfVenkatakrishnanMengAlizadeh2019}. These Deep RL approaches can discover non-obvious strategies, but their %eep 
inference pipelines introduce latency that is often \bem{incompatible %with real-time control-plane constraints. 
with real-time HPC scheduling}. %, where rapid decisions are required. %must be made immediately to maximize throughput. 
%In practice, %memory overhead may be manageable, but 
%inference delay remains the main bottleneck.

\smallskip

\noindent \hypertarget{sec:rel-work-a-a}{\textbf{Approaches Based on Predictive Augmentation}}. %\emph{In the second}, the heuristic core is preserved, while 
ML is used only to estimate unknown system variables that improve downstream decisions. Brown et al.\ train supervised models on historical HPC data %queue logs 
to predict job wait times for urgent workloads, outperforming native estimators \cite{BrownGibbBelikovNash2022}. Similarly, ML-SJF uses Support Vector Regression \cite{DruckerBurgesKaufmanSmolaVapnik1996} to predict job runtimes before applying Shortest Job First (SJF) within Apache Mesos \cite{HindmanKonwinskiZahariaGhodsiJosephKatzShenkerStoica2011}. \bem{However, executing these predictive models purely in software can introduce significant inference latency.}
These approaches expose a key gap for hardware-accelerated online schedulers, as they typically assume EPTs are already available at the hardware interface, or else rely on noisy user-provided estimates. Consequently, software-based prediction latency can negate %erase much of the 
speedup offered by the accelerator. %itself. 
This makes low-latency integration between predictive ML and hardware scheduling a \bem{central %design 
requirement as proposed in this work}.

\section{Limitations}\label{sec:limitations}
%\doubted{put validity threats}

%Despite its performance gains, $\pname$ presents some operational constraints. 
\emph{First}, $\pname$'s multi-threaded Host control and asynchronous PCIe streams raise peak Host-side memory use by about 7$\times$ %(to $\sim$253 MB) 
relative to the AVX %in-software 
baseline. \emph{Second}, to avoid temporal data leakage, XGBoost is also restricted to six submission-time features, which may miss deeper application-specific patterns. %\emph{Finally}, while we have demonstrated the temporal capabilities of the additive training policy with XGBoost, we acknowledge that other models ($\eg$, Random Forest) may also support effective temporal adaptation.
\emph{Finally},
we validate the temporal adaptation policy with XGBoost and
Random Forest, but evaluating its behavior with additional ML
predictors remains future work.

%other predictors may also support effective temporal adaptation ($\eg$, Random Forest), but we focused on XGBoost because its iterative structure %is well-suited for 
%aligns well with the FPU design and with the Core + Additive retraining strategy. 

%Added as a separate section to mimic TPAMI, we can change it though.
\section{Threats to Validity}\label{sec:validity_threats}

\noindent \hypertarget{sec:external_threats}{\textbf{External Threats}.} Our evaluation can not feasibly cover all possible HPC cluster architectures and workloads. We mitigate this by selecting varied historical datasets from unrelated systems, and testing our methodology with both chronological and synthetic workload distribution shifts, showing the methodology is transferable across systems. Nevertheless, additional dataset evaluations would further strengthen our claims.

\smallskip

\noindent \hypertarget{sec:internal_threats}{\textbf{Internal Threats}.} While we discuss an in-depth dataset quantization method to maximize parameter precision in our decision trees (\Cref{sec:categorical_encoding}), this is done using the \emph{historical} $99.9^{th}$ percentile value for that feature. In a real-world system, there is no guarantee that this value will be constant during the whole life cycle of a given system. However, $\pname$'s modular design %hot-swap flexibility
would allow such values to be re-accounted for and generate a new quantization value. %to be generated.

\smallskip

\noindent \hypertarget{sec:construction_threats}{\textbf{Construct Threats}.} %A structural weakness we have attempted to mitigate is a classic mono-method bias. For the majority of our experiments and final architecture, we utilize XGBoost as our predictor. In \Cref{fig:exp_res_models_comparison}, we show that other predictors could feasibly perform similarly as an ETP predictor, and acknowledge that other models could theoretically also perform some variation of the ADD policy. As such, the mono-method bias is partially present in the results presented in this manuscript, but is not a detriment to the overall method described.
We represent temporal workload shifts through Gaussian perturbations, which capture only a subset of the shifts that may occur in real HPC systems. We tried to mitigate this by also evaluating ADD on real chronological workload evolution across real HPC workload datasets, but these chronological windows cannot represent any possible workload change. Additionally, evaluating prediction accuracy using MAE and MAE p95, which capture the average and the upper tail of the error distribution, but may not reflect every type of prediction error relevant to scheduling.

\section{Conclusion}\label{sec:conclusion}

%We introduce $\pname$, a dual-FPGA architecture that replaces unreliable human runtime estimates with more accurate, real-time XGBoost predictions. Using a bifurcated Core + Additive training policy, the quantized model successfully adapts to temporal workload shifts. By removing the inference latency of software-only designs, $\pname$ achieves a $17\times$ speedup over the AVX baseline and demonstrates the practical value of data-centric hardware scheduling in uncertain HPC environments.
We introduce $\pname$, a dual-FPGA architecture that replaces human runtime estimates with real-time XGBoost predictions. Its Core + Additive training policy adapts to temporal workload shifts, while hardware inference achieves a $17\times$ speedup over AVX, and demonstrates the value of data-centric hardware scheduling in uncertain HPC environments.

%\medskip

%\noindent \textbf{Limitations}. 

\smallskip

\noindent {\bf AI Disclosure}. Gen AI tools were used for drafting selected text. Authors revised AI-generated content, accepting full responsibility for the intellectual content.

%\clearpage

\bibliographystyle{unsrt}
\bibliography{./bib/master,./bib/aiperformance}

\begin{IEEEbiography}
[{\includegraphics[width=1in,height=1.25in,clip,keepaspectratio]{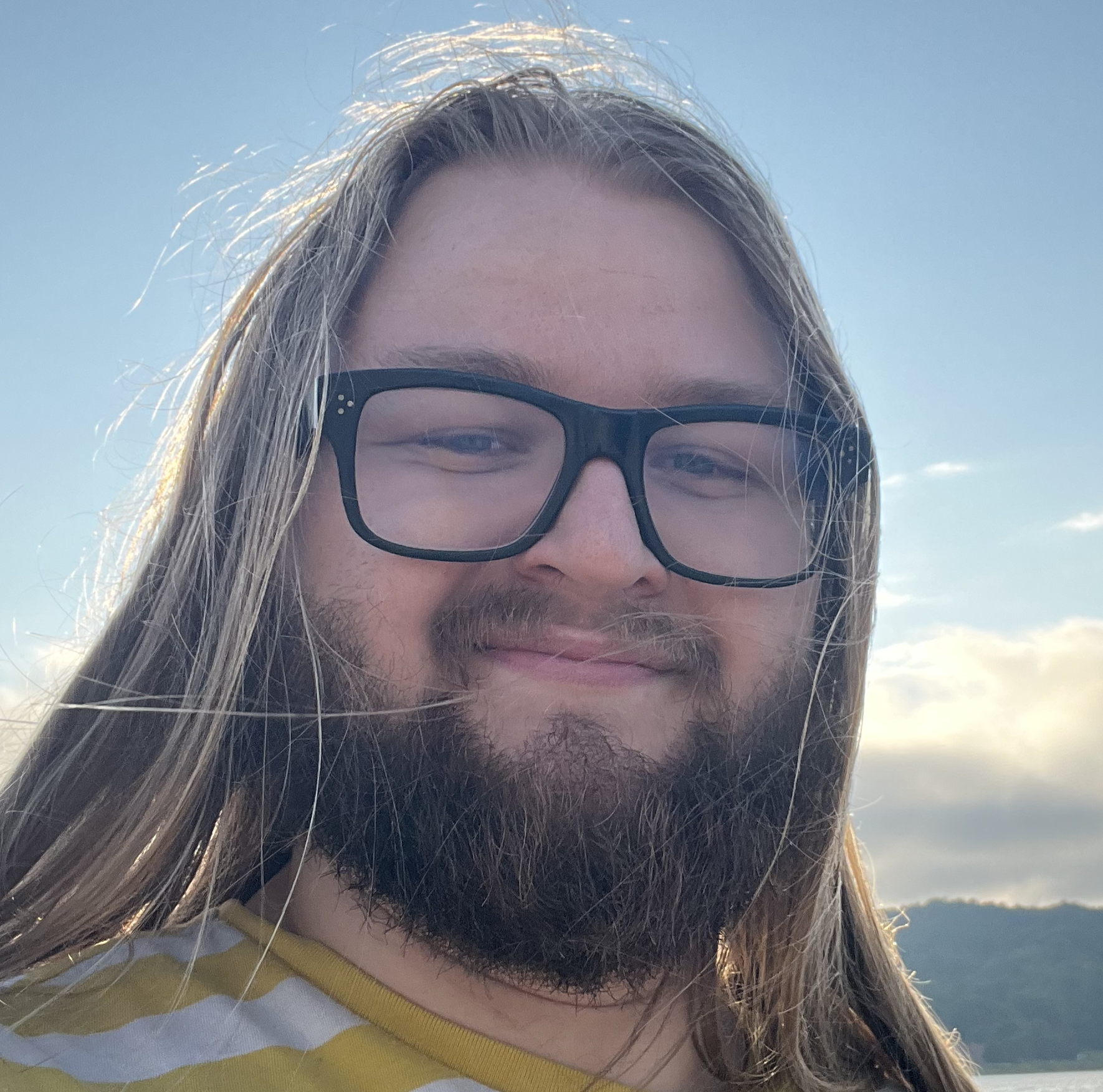}}]{Adam H. Ross} is a Ph.D. student %from the University of Illinois Chicago. 
at the Electrical and Computer Engineering Department at the University of Illinois, Chicago, and a DAC young fellow. %He received a M.S. degree from the University of Illinois Chicago, a M.S. degree from Politecnico di Torino, and a B.S. degree from Politecnico di Torino. 
He received an M.S. degree in Computer Engineering from the University of Illinois, Chicago, and a B.A. degree with a focus in Computer Science from Hampshire College.
His current research interests include hardware acceleration, SW/HW co-design, FPGA Design, and application-specific computation.
\end{IEEEbiography}

\vskip -2\baselineskip plus -1fil

\begin{IEEEbiography}
[{\includegraphics[width=1in,height=1.25in,clip,keepaspectratio]{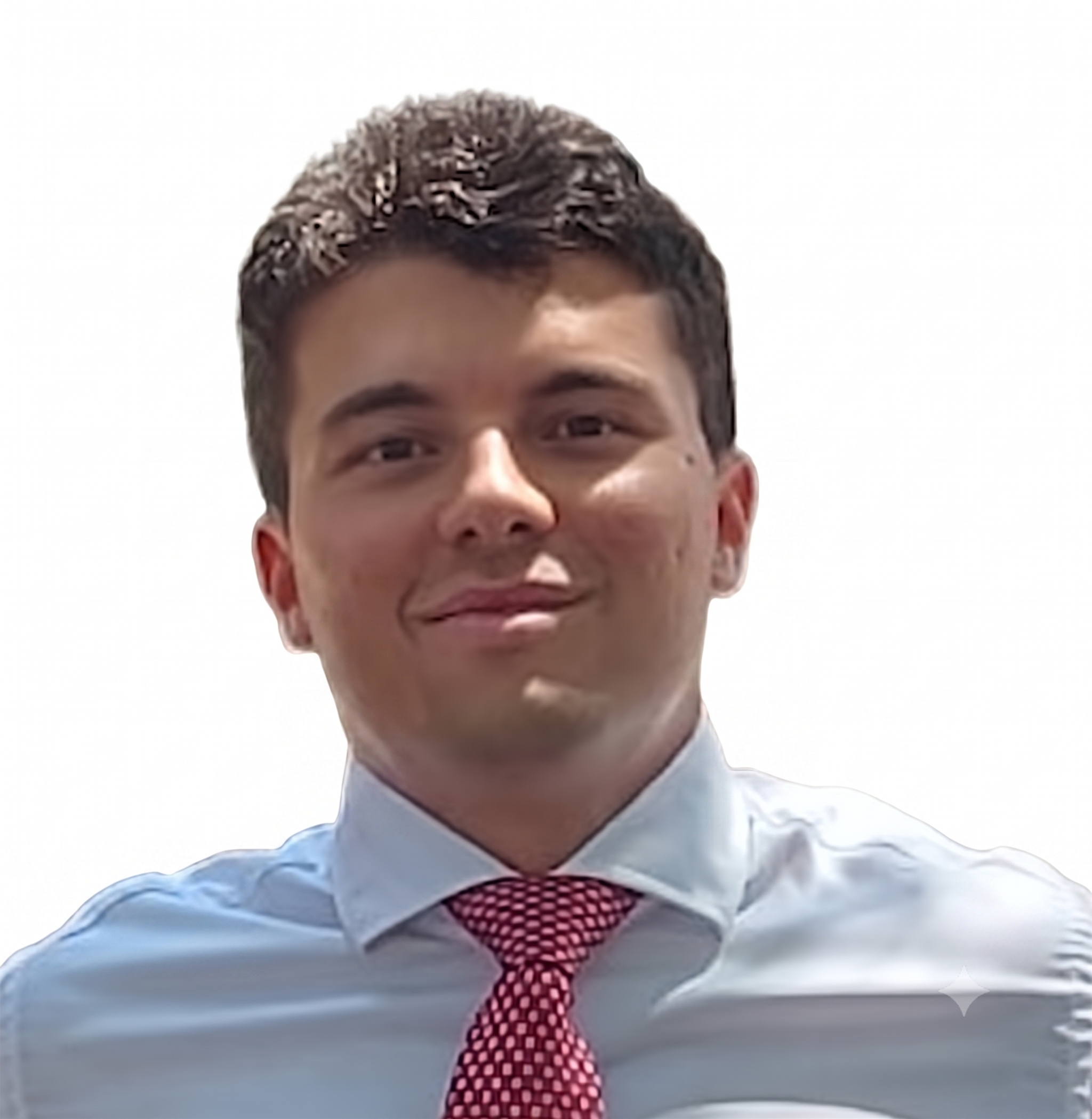}}]{Riccardo Revalor} is a Ph.D. student %from the University of Illinois Chicago. 
at the Electrical and Computer Engineering Department at the University of Illinois, Chicago. %He received a M.S. degree from the University of Illinois Chicago, a M.S. degree from Politecnico di Torino, and a B.S. degree from Politecnico di Torino. 
He received an M.S. degree in Computer Science from the University of Illinois, Chicago, and both M.S. and B.S. degrees in Computer Engineering from Politecnico di Torino, Italy.
His current research interests include neuromorphic and bio-inspired AI, AI SW/HW co-design, AI-assisted test generation and debugging of hardware designs, and LLM uncertainty estimation.
\end{IEEEbiography}

\vskip -2\baselineskip plus -1fil

\begin{IEEEbiography}[{\includegraphics[width=1in,height=1.25in,clip,keepaspectratio]{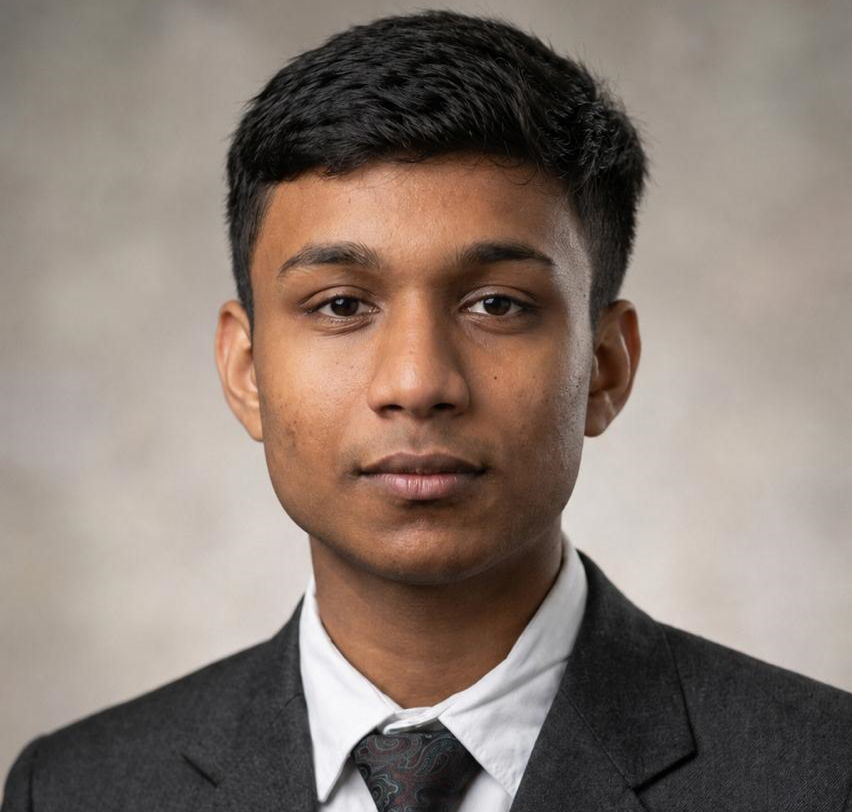}}]{Aryan Singh} is an undergraduate student pursuing a B.S. in Mathematics and Computer Science at the University of Illinois, Urbana-Champaign, expecting to graduate in May 2027. He previously conducted research on FPGA-based hardware accelerators in the Department of Electrical and Computer Engineering at the University of Illinois, Chicago. His current research interests include hardware acceleration, SW/HW co-design, %GPU computing, 
and low-latency systems.
\end{IEEEbiography}

\vskip -2\baselineskip plus -1fil

\begin{IEEEbiography}[{\includegraphics[width=1in,height=1.25in,clip,keepaspectratio]{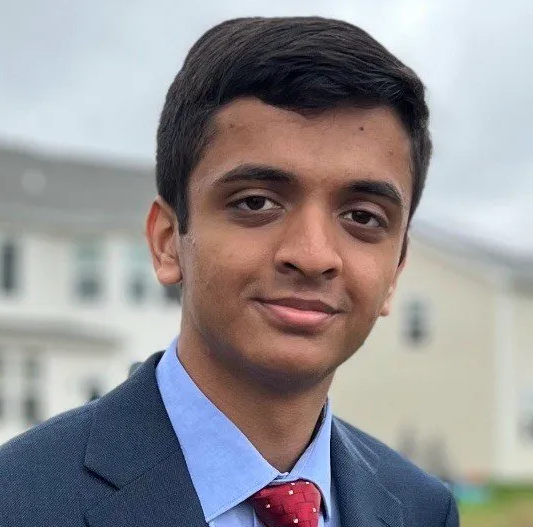}}]{Ayush Jain} is an undergraduate student, %who is 
pursuing a B.S. at the University of Illinois at Urbana Champaign in Computer Science and Astronomy. He has previously done research on hardware acceleration with Xilinx %SoC 
FPGAs in the Electrical and Computer Engineering department at University of Illinois Chicago. His research interests include computer architecture, GPU computing, VLSI design, and hardware acceleration for machine learning workloads.
\end{IEEEbiography}

\vskip -2\baselineskip plus -1fil

\begin{IEEEbiography}[{\includegraphics[width=1in,height=1.25in,clip,keepaspectratio]{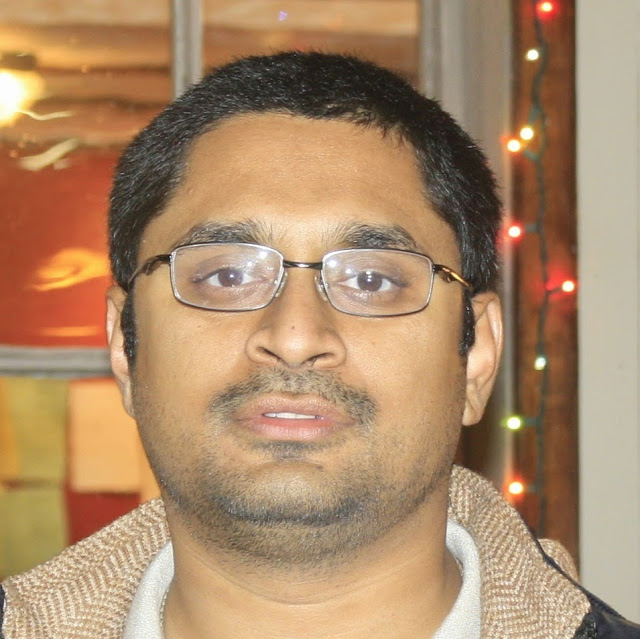}}]{Debjit Pal}
received a Ph.D. from the University of Illinois at Urbana-Champaign. He received B.Tech and M.S. degrees from Jadavpur University and IIT Kharagpur, India, respectively. He is currently an assistant professor at the Electrical and Computer Engineering Department at the University of Illinois Chicago. His current research interests include post-silicon and pre-silicon validation and debug, analog design automation, and programming paradigm for hardware accelerators. He received best paper nominations in ICCAD 2015, DAC 2018, ASP-DAC 2019, and FPGA 2024. He is the winner of the IEEE CEDA System Validation and Debug Technology Committee (SVDTC) Student Research Award 2016.
\end{IEEEbiography}

\end{document}